\documentclass[%
reprint,
superscriptaddress,
nofootinbib,
 amsmath,
 amssymb,
 aps,
prd,
floatfix,
]{revtex4-2}
\usepackage{longtable}
\usepackage{aas_macros}
\usepackage{graphicx}
\usepackage{dcolumn}
\usepackage{siunitx}
\usepackage{bm}
\DeclareSIUnit \year {yr}
\usepackage[dvipsnames]{xcolor}

\begin{document}
\graphicspath{{./}{figures/}}

\preprint{APS/123-QED}

\title{A High-Time Resolution Search for Compact Objects using Fast Radio Burst Gravitational Lens Interferometry with CHIME/FRB}

\author{Zarif Kader}
  \thanks{These authors contributed equally to this work.}
  \affiliation{Department of Physics, McGill University, 3600 rue University, Montr\'eal, QC H3A 2T8, Canada}
  \affiliation{McGill Space Institute, McGill University, 3550 rue University, Montr\'eal, QC H3A 2A7, Canada}
\author{Calvin Leung}
  \thanks{These authors contributed equally to this work.}
  \affiliation{MIT Kavli Institute for Astrophysics and Space Research, Massachusetts Institute of Technology, 77 Massachusetts Ave, Cambridge, MA 02139, USA}
  \affiliation{Department of Physics, Massachusetts Institute of Technology, 77 Massachusetts Ave, Cambridge, MA 02139, USA}
\author{Matt Dobbs}
  \affiliation{Department of Physics, McGill University, 3600 rue University, Montr\'eal, QC H3A 2T8, Canada}
  \affiliation{McGill Space Institute, McGill University, 3550 rue University, Montr\'eal, QC H3A 2A7, Canada}
\author{Kiyoshi W.~Masui}
  \affiliation{MIT Kavli Institute for Astrophysics and Space Research, Massachusetts Institute of Technology, 77 Massachusetts Ave, Cambridge, MA 02139, USA}
  \affiliation{Department of Physics, Massachusetts Institute of Technology, 77 Massachusetts Ave, Cambridge, MA 02139, USA}
\author{Daniele Michilli}
  \affiliation{MIT Kavli Institute for Astrophysics and Space Research, Massachusetts Institute of Technology, 77 Massachusetts Ave, Cambridge, MA 02139, USA}
  \affiliation{Department of Physics, Massachusetts Institute of Technology, 77 Massachusetts Ave, Cambridge, MA 02139, USA}
\author{Juan Mena-Parra}
  \affiliation{MIT Kavli Institute for Astrophysics and Space Research, Massachusetts Institute of Technology, 77 Massachusetts Ave, Cambridge, MA 02139, USA}
\author{Ryan Mckinven}
  \affiliation{Department of Physics, McGill University, 3600 rue University, Montr\'eal, QC H3A 2T8, Canada}
  \affiliation{McGill Space Institute, McGill University, 3550 rue University, Montr\'eal, QC H3A 2A7, Canada}
\author{Cherry Ng}
  \affiliation{Dunlap Institute for Astronomy \& Astrophysics, University of Toronto, 50 St.~George Street, Toronto, ON M5S 3H4, Canada}
\author{Kevin Bandura}
  \affiliation{Lane Department of Computer Science and Electrical Engineering, 1220 Evansdale Drive, PO Box 6109, Morgantown, WV 26506, USA}
  \affiliation{Center for Gravitational Waves and Cosmology, West Virginia University, Chestnut Ridge Research Building, Morgantown, WV 26505, USA}
\author{Mohit Bhardwaj}
  \affiliation{Department of Physics, McGill University, 3600 rue University, Montr\'eal, QC H3A 2T8, Canada}
  \affiliation{McGill Space Institute, McGill University, 3550 rue University, Montr\'eal, QC H3A 2A7, Canada}
\author{Charanjot Brar}
  \affiliation{Department of Physics, McGill University, 3600 rue University, Montr\'eal, QC H3A 2T8, Canada}
  \affiliation{McGill Space Institute, McGill University, 3550 rue University, Montr\'eal, QC H3A 2A7, Canada}
\author{Tomas Cassanelli}
  \affiliation{Dunlap Institute for Astronomy \& Astrophysics, University of Toronto, 50 St.~George Street, Toronto, ON M5S 3H4, Canada}
  \affiliation{David A.~Dunlap Department of Astronomy \& Astrophysics, University of Toronto, 50 St.~George Street, Toronto, ON M5S 3H4, Canada}
\author{Pragya Chawla}
  \affiliation{Department of Physics, McGill University, 3600 rue University, Montr\'eal, QC H3A 2T8, Canada}
  \affiliation{McGill Space Institute, McGill University, 3550 rue University, Montr\'eal, QC H3A 2A7, Canada}
  \affiliation{Anton Pannekoek Institute for Astronomy, University of Amsterdam, Science Park 904, 1098 XH Amsterdam, The Netherlands}
\author{Fengqiu Adam Dong}
  \affiliation{Department of Physics and Astronomy, University of British Columbia, 6224 Agricultural Road, Vancouver, BC V6T 1Z1 Canada}
\author{Deborah Good}
  \affiliation{Department of Physics, University of Connecticut, 196 Auditorium Road, U-3046, Storrs, CT 06269-3046, USA}
  \affiliation{Center for Computational Astrophysics, Flatiron Institute, 162 5th Avenue, New York, NY 10010, USA}
\author{Victoria Kaspi}
  \affiliation{Department of Physics, McGill University, 3600 rue University, Montr\'eal, QC H3A 2T8, Canada}
  \affiliation{McGill Space Institute, McGill University, 3550 rue University, Montr\'eal, QC H3A 2A7, Canada}
\author{Adam E.~Lanman}
  \affiliation{Department of Physics, McGill University, 3600 rue University, Montr\'eal, QC H3A 2T8, Canada}
  \affiliation{McGill Space Institute, McGill University, 3550 rue University, Montr\'eal, QC H3A 2A7, Canada}
\author{Hsiu-Hsien Lin}
  \affiliation{Institute of Astronomy and Astrophysics, Academia Sinica, Astronomy-Mathematics Building, No. 1, Sec. 4, Roosevelt Road, Taipei 10617, Taiwan}
  \affiliation{Canadian Institute for Theoretical Astrophysics, 60 St.~George Street, Toronto, ON M5S 3H8, Canada}
\author{Bradley W.~Meyers}
  \affiliation{Department of Physics and Astronomy, University of British Columbia, 6224 Agricultural Road, Vancouver, BC V6T 1Z1 Canada}
\author{Aaron B.~Pearlman}
  \altaffiliation{McGill Space Institute Fellow}
  \altaffiliation{FRQNT Postdoctoral Fellow}
  \affiliation{Department of Physics, McGill University, 3600 rue University, Montr\'eal, QC H3A 2T8, Canada}
  \affiliation{McGill Space Institute, McGill University, 3550 rue University, Montr\'eal, QC H3A 2A7, Canada}
\author{Ue-Li Pen}
  \affiliation{Institute of Astronomy and Astrophysics, Academia Sinica, Astronomy-Mathematics Building, No. 1, Sec. 4, Roosevelt Road, Taipei 10617, Taiwan}
  \affiliation{Canadian Institute for Theoretical Astrophysics, 60 St.~George Street, Toronto, ON M5S 3H8, Canada}
  \affiliation{Canadian Institute for Advanced Research, 180 Dundas St West, Toronto, ON M5G 1Z8, Canada; }
  \affiliation{David A.~Dunlap Department of Astronomy \& Astrophysics, University of Toronto, 50 St.~George Street, Toronto, ON M5S 3H4, Canada}
  \affiliation{Perimeter Institute of Theoretical Physics, 31 Caroline Street North, Waterloo, ON N2L 2Y5, Canada}
\author{Emily Petroff}
  \affiliation{Department of Physics, McGill University, 3600 rue University, Montr\'eal, QC H3A 2T8, Canada}
  \affiliation{McGill Space Institute, McGill University, 3550 rue University, Montr\'eal, QC H3A 2A7, Canada}
  \affiliation{Anton Pannekoek Institute for Astronomy, University of Amsterdam, Science Park 904, 1098 XH Amsterdam, The Netherlands}
\author{Ziggy Pleunis}
  \affiliation{Dunlap Institute for Astronomy \& Astrophysics, University of Toronto, 50 St.~George Street, Toronto, ON M5S 3H4, Canada}
\author{Masoud Rafiei-Ravandi}
  \affiliation{Department of Physics, McGill University, 3600 rue University, Montr\'eal, QC H3A 2T8, Canada}
  \affiliation{McGill Space Institute, McGill University, 3550 rue University, Montr\'eal, QC H3A 2A7, Canada}
\author{Mubdi Rahman}
  \affiliation{Sidrat Research, PO Box 73527 RPO Wychwood, Toronto, ON M6C 4A7, Canada}
\author{Pranav Sanghavi}
  \affiliation{Lane Department of Computer Science and Electrical Engineering, 1220 Evansdale Drive, PO Box 6109, Morgantown, WV 26506, USA}
  \affiliation{Center for Gravitational Waves and Cosmology, West Virginia University, Chestnut Ridge Research Building, Morgantown, WV 26505, USA}
\author{Paul Scholz}
  \affiliation{Dunlap Institute for Astronomy \& Astrophysics, University of Toronto, 50 St.~George Street, Toronto, ON M5S 3H4, Canada}
\author{Kaitlyn Shin}
  \affiliation{MIT Kavli Institute for Astrophysics and Space Research, Massachusetts Institute of Technology, 77 Massachusetts Ave, Cambridge, MA 02139, USA}
  \affiliation{Department of Physics, Massachusetts Institute of Technology, 77 Massachusetts Ave, Cambridge, MA 02139, USA}
\author{Seth Siegel}
  \affiliation{Department of Physics, McGill University, 3600 rue University, Montr\'eal, QC H3A 2T8, Canada}
\author{Kendrick M.~Smith}
  \affiliation{Perimeter Institute of Theoretical Physics, 31 Caroline Street North, Waterloo, ON N2L 2Y5, Canada}
\author{Ingrid Stairs}
  \affiliation{Department of Physics and Astronomy, University of British Columbia, 6224 Agricultural Road, Vancouver, BC V6T 1Z1 Canada}
\author{Shriharsh P.~Tendulkar}
  \affiliation{Department of Astronomy and Astrophysics, Tata Institute of Fundamental Research, Mumbai, 400005, India}
  \affiliation{National Centre for Radio Astrophysics, Post Bag 3, Ganeshkhind, Pune, 411007, India}
\author{Keith Vanderlinde}
  \affiliation{Dunlap Institute for Astronomy \& Astrophysics, University of Toronto, 50 St.~George Street, Toronto, ON M5S 3H4, Canada}
  \affiliation{David A.~Dunlap Department of Astronomy \& Astrophysics, University of Toronto, 50 St.~George Street, Toronto, ON M5S 3H4, Canada}
\author{Dallas Wulf}
  \affiliation{Department of Physics, McGill University, 3600 rue University, Montr\'eal, QC H3A 2T8, Canada}
  \affiliation{McGill Space Institute, McGill University, 3550 rue University, Montr\'eal, QC H3A 2A7, Canada}
\newcommand{\allacks}{
A.B.P. is a McGill Space Institute (MSI) Fellow and a Fonds de Recherche du Quebec -- Nature et Technologies (FRQNT) postdoctoral fellow.
C.L. was supported by the U.S. Department of Defense (DoD) through the National Defense Science \& Engineering Graduate Fellowship (NDSEG) Program
E.P. acknowledges funding from an NWO Veni Fellowship.
FRB research at UBC is funded by an NSERC Discovery Grant and by the Canadian Institute for Advanced Research. The CHIME/FRB baseband system is funded in part by a Canada Foundation for Innovation JELF grant to IHS.
J.M.P is a Kavli Fellow.
K.M.B. is supported by an NSF grant (2006548, 2018490)
K.S. is supported by the NSF Graduate Research Fellowship Program.
K.W.M. is supported by NSF grants 2008031 and 2018490.
M.B. is supported by an FRQNT Doctoral Research Award.
M.D. is supported by a Killam Fellowship, CRC Chair, NSERC Discovery Grant, CIFAR, and by the FRQNT Centre de Recherche en Astrophysique du Qu\'ebec (CRAQ).
P.S. is a Dunlap Fellow and an NSERC Postdoctoral Fellow. 
V.M.K. holds the Lorne Trottier Chair in Astrophysics \& Cosmology, a Distinguished James McGill Professorship, and receives support from an NSERC Discovery grant (RGPIN 228738-13), from an R. Howard Webster Foundation Fellowship from CIFAR, and from the FRQNT CRAQ.
We acknowledge the support of the Natural Sciences and Engineering Research Council of Canada (NSERC), [funding reference number RGPIN-2019-067, CRD 523638-201, 555585-20] We receive support from Ontario Research Fund—research Excellence Program (ORF-RE), Canadian Institute for Advanced Research (CIFAR), Thoth Technology Inc, Alexander von Humboldt Foundation, and the Ministry of Science and Technology(MOST) of Taiwan(110-2112-M-001-071-MY3). 
Z.P. is a Dunlap Fellow.
}

\collaboration{CHIME/FRB Collaboration}

\date{\today}

\begin{abstract}

The gravitational field of compact objects, such as primordial black holes, can create multiple images of background sources. For transients such as fast radio bursts (FRBs), these multiple images can be resolved in the time domain. Under certain circumstances, these images not only have similar burst morphologies but are also phase-coherent at the electric field level. With a novel dechannelization algorithm and a matched filtering technique, we search for repeated copies of the same electric field waveform in observations of FRBs detected by the FRB backend of the Canadian Hydrogen Mapping Intensity Experiment (CHIME). An interference fringe from a coherent gravitational lensing signal will appear in the time-lag domain as a statistically-significant peak in the time-lag autocorrelation function. We calibrate our statistical significance using telescope data containing no FRB signal. Our dataset consists of $\sim$100-ms long recordings of voltage data from 172 FRB events, dechannelized to 1.25-ns time resolution. This coherent search algorithm allows us to search for gravitational lensing signatures from compact objects in the mass range of $10^{-4}-10^{4} ~\mathrm{M_{\odot}}$. After ruling out an anomalous candidate due to diffractive scintillation, we find no significant detections of gravitational lensing in the 172 FRB events that have been analyzed. In a companion work \cite{leung2022constraining}, we interpret the constraints on dark matter from this search.

\end{abstract}

\keywords{Gravitational lensing (670) --- Radio transient sources (2008) --- Primordial black holes (1292)}


\maketitle

\section{Introduction} \label{sec:intro}
In recent years, the observation of gravitational waves from mergers of compact binaries~\citep{abbott2016observation} has renewed interest in the possibility that a significant fraction of dark matter is composed of dark compact objects~\citep{lens_pbh_constraints}, such as primordial black holes (PBHs). Gravitational lensing of transients like fast radio bursts (FRBs)~\citep{frb_review_cordes} has emerged as one of the cleanest ways to detect the presence of such dark compact objects~\citep{Katz_paper,Munoz_paper,eichler_frb,FRB_wave_low}. While the progenitor and emission mechanism of these millisecond long bursts are not yet well-understood, their cosmological distance and abundance make them particularly well-suited to time-domain searches for gravitational lensing~\citep{Katz_paper,Munoz_paper,eichler_frb,FRB_wave_low}. In this paper, we present a search pipeline for coherently detecting a gravitationally-lensed FRB. The fundamental idea of the search is that propagation of the FRB through the gravitational field of a foreground mass will coherently produce multiple images of the FRB, resolvable in time domain as an interference fringe.

Traditional searches for compact objects using gravitational lensing such as the MACHO and EROS projects~\citep{MACHO_project,EROS_2} monitor steady background sources on timescales of days to weeks and search for slow modulations of the sources' apparent brightness. These searches are able to constrain the distribution of PBHs within the Local Group. FRBs, on the other hand, have been localized to other galaxies (see e.g.,~\citep{2017Natur.541...58C, ravi_v_localized,macquart2020census}). Using FRBs as probes, it may be possible to constrain the cosmological abundance rather than the local abundance of PBHs. 

Unlike MACHO or EROS, our search detects lensing not through the gradual brightening of a background star as a lens transits in the foreground, but by the direct detection of a second image of the same FRB in the time domain. This means that for similar lens masses, we search for images on short timescales (nanoseconds to milliseconds). To search for a putative second image, we auto-correlate the FRB's phase-preserving baseband data similar to how baseband data are correlated in very long baseline interferometry (VLBI). We refer to this method as ``interferometric lensing''. Similar methods~\citep{cho2020,Farah2019} have had non-detections of a gravitational lensing signature using a correlation method. We seek to increase sensitivity to lensing detections by correlating with a matched filter and searching by modeling the noise properties of the system. 

In this paper, we present a gravitational lensing search pipeline, and apply it to 172 FRBs detected by the FRB instrument on the Canadian Hydrogen Intensity Mapping Experiment~\citep{chimefrb_overview}. We present constraints on the abundance of PBHs that compose dark matter in a companion paper~\citep{leung2022constraining}.


\section{Gravitational Lensing Model} \label{sec:lensing}
The phenomenology of gravitational microlensing is extremely rich, and different search techniques are sensitive to different signatures of lensing. In our time-domain search, as in previous works~\citep{munoz2016lensing,Katz_paper,eichler_frb,FRB_wave_low,first_constraint}, we search for compact objects which can be modeled by the simplest of lens models: a point-mass lens. In the point-mass model, a point mass $M_\mathrm{L}$ at a redshift $z_\mathrm{L}$ lies at a transverse physical distance $b = y R_E(M_\mathrm{L},z_\mathrm{L},z_\mathrm{S})$ from the direct line of sight. Here, $y$ is the impact parameter measured in units of $R_E$ (the Einstein radius), which in turn depends on $z_\mathrm{L}$ and $z_\mathrm{S}$ (the lens and source redshifts respectively). In this simple model, all the phenmenology is captured by two parameters: a characteristic delay $\tau$,
\begin{equation}\label{eq:time_delay_obs}
    \tau = \frac{4 G M_\mathrm{L}(1+ z_{\mathrm{L}})}{ c^3} \left( \frac{1}{2} y\sqrt{ y^2 +4}  + \ln \left( \frac{ y + \sqrt{ y^2 +4}} {y - \sqrt{ y^2 +4}} \right) \right) ,
\end{equation}
and a flux magnification ratio $\mu = |\varepsilon|^2$ (where $\varepsilon$ is the electric field magnification ratio) which we take by convention to be $0 < |\varepsilon| < 1$, where 
\begin{equation}\label{eq:rel_mag_ratio}
    |\varepsilon|^2 = \frac{y^2 + 2 - y \sqrt{y^2 + 4} }{y^2 + 2 + y \sqrt{y^2 + 4}} .
\end{equation}
Equation~\ref{eq:time_delay_obs} and~\ref{eq:rel_mag_ratio} predict that the brighter image arrives before the fainter one. It also allows us to translate the observables into the lens properties: its redshifted mass $M(1 + z_\mathrm{L})$ and impact parameter $b$.

The point lens model assumes the only propagation effect is gravitational lensing. Plasma lensing of FRBs~\citep{cordes2017lensing} also induces multi-path propagation, and scintillation and scattering of FRBs. This complicates the search for gravitational lensing events. We describe the effect of scintillation on our search in Sec.~\ref{sec:results}, and a two-screen model incorporating both gravitational and scattering effects is provided in ~\citep{leung2022constraining}.

In principle, the multi-path propagation characteristic of gravitational lensing is a purely geometric effect caused by differing path lengths around the lens. The phase coherence of the electric field is preserved by a gravitational lens. Hence, coherent algorithms for measuring time delays, such as those used in VLBI, should be able to detect an FRB that is coherently lensed by a foreground mass. Such a system would have profound implications for studies of gravitational lensing. First, it would be the most precise measurement of a gravitational lensing delay ever made, by orders of magnitude~\citep{li2018strongly}. Second, it would allow for the first observation of wave interference effects in gravitational lensing~\citep{jow2020wave}, which are inaccessible in other systems where gravitational lensing is traditionally studied due to large source sizes~\citep{katz2018femtolensing,oguri2019strong}. Finally, it would open up the possibility of using coherently lensed FRBs as some of the most exquisitely sensitive ``interferometers'' in the universe~\citep{dai2017probing,pearson2020searching,wucknitz2021cosmology} with the gravitational lens acting as an astronomical ``beamsplitter.''

\subsection{Recovery of the Observables}\label{sec:observables}
In this Section, we outline how a gravitationally-lensed FRB can be detected in the voltage timestream captured by a radio telescope. The details of the derivation can be found in Appendix \ref{sec:corr_appendix}. First, we consider the scenario where we have a gravitationally-lensed FRB. Our voltage timestream takes the form,
\begin{equation}\label{eq:voltform}
    V_P (t) =  S_P (t) + \varepsilon S_P (t-\tau) + N_P (t),
\end{equation}
where $P$ indexes the two telescope polarizations (X and Y), $S_P$ is amplitude-modulated white noise corresponding to the unlensed electric field signal \footnote{A gravitational lens magnifies both images, scaling each image's waveforms by a constant. The total observed flux is larger than the initial flux but it can be modeled as a scaled copy of the intrinsic waveform.}, $N_P$ is stationary telescope noise, and $\tau$ is the true time delay between the two images.
We seek to auto-correlate the voltage with itself to construct a time-lag spectrum as a function of many trial time lags $\hat{t}$, 
\begin{equation}\label{eq:timelagspec}
C_P(\hat{t}) = \frac{ \sum_{t} V_P(t+\hat{t}) V_P(t) W_P^2(t) }{\sqrt{\left( \sum_{t} V_P^2(t+\hat{t})W_P^2(t) \right) \left( \sum_{t} V_P^2(t) W_P^2(t) \right)}} ~.
\end{equation} 
In Eq.~\ref{eq:timelagspec}, the $W_P^2(t)$ are matched filters: smooth, positive functions constructed from the light curve of the burst as measured in each telescope polarization (Sec.~\ref{sec:mfilter}). They are approximations to the optimal matched filter, in which each time sample of $V_P(t)$ gets upweighted by its signal-to-noise ratio $S_P^2(t) / N_P^2(t)$. In our data, the noise is largely stationary, so taking $W_P^2(t) \propto S_P^2(t)$ is close to optimal.

In general, to interpret the $C_P(\hat{t})$ produced by time-lag correlating with a matched filter, there are two limiting cases: when the fainter image is (a) outside the support of $W^2_P(t)$, such that the filter is constructed without any information about the image, and (b) inside the support of $W^2_P(t)$, such that the filter model is affected by the image's presence. We find from numerical simulations, described in Sec.~\ref{sec:validation}, that the uncertainty in recovering $\varepsilon$ from $C_P(\hat{t})$ is dominated by the thermal noise in CHIME/FRB data rather than the differences between scenario (a) and (b) \footnote{The weighted unlensed fluence (Eq.~\ref{eq:fluxest}) will be overestimated by a factor of $(1+\varepsilon^2)$, where $|\epsilon| < 0.1$ in most realistic scenarios.}. Hence, we neglect the complications arising from scenario (b), and present the recovery of the lensing observables $\varepsilon$ and $\tau$ from $C_P(\hat{t})$ in scenario (a). First, we define the weighted unlensed fluence as
\begin{equation}\label{eq:fluxest}
    F_P = \sum_{t} S_P^2(t) W_P^2(t) ,
\end{equation}
and note that $F_P$ is translationally invariant in time. Additionally, we define the weighted signal-to-noise ratio as 
\begin{equation}\label{eq:noiseest}
    \Gamma = \frac{ F_P }{\sum_{t} N_P^2(t) W_P^2(t)} .
\end{equation}
At the lensing time delay, $\hat{t} =\tau$,
\begin{equation}\label{eq:lenslag}
    C_P(\tau) = \frac{\varepsilon \Gamma}{\sqrt{( \Gamma + 1 )( \varepsilon^2 \Gamma + 1)  }} .
\end{equation}
We can obtain $\sum_{t} N_P^2(t)  W_P^2(t) $, which is proportional to the system temperature, from our off-pulse realizations and calculate $\Gamma_P$. 
We can then recover the value of $\varepsilon$ as 
\begin{equation}\label{eq:mag_convert}
    \varepsilon^2 = \frac{C_P^2(\tau) ( \Gamma + 1)}{ \Gamma^2 - C_P^2(\tau)  \Gamma^2  - C_P^2(\tau) \Gamma} .
\end{equation}
We recover $\tau$ from the associated time-lag of $\varepsilon$. Then, we convert these quantities into parameters of the lensing system. We obtain the normalized impact parameter $y$ as ~\citep{munoz2016lensing}
\begin{equation}\label{eq:impact_conv}
    y = \sqrt{ |\varepsilon| + |\varepsilon|^{-1} - 2} ,
\end{equation}
and the redshift mass of the lens as
\begin{equation}\label{eq:mass_conv}
    M_\mathrm{L} (1+z_\mathrm{L}) = \frac{ c^3 \tau}{4 G} \left( \frac{y}{2} \sqrt{ y^2 +4}  - \ln \left( \frac{ y + \sqrt{ y^2 +4}} {y - \sqrt{ y^2 +4}} \right)  \right)^{-1} .
\end{equation}
We note that with $\tau$ and $\varepsilon$ alone, it is difficult to recover the actual mass of the gravitational lens because it is degenerate with its own redshift. Only the combination $M_\mathrm{L} (1+z_\mathrm{L})$, also known as the ``redshifted mass'' (see, e.g.,~\citep{paynter2021evidence}), can be estimated. For the remainder of this paper, we focus on the recovery and estimates for $\varepsilon$ and $\tau$. 

\section{CHIME} \label{sec:chime}
In this section, we provide a concise overview of CHIME~\citep{CHIMECosmologyOverview}, the FRB backend~\citep{chimefrb_overview}, and the baseband system~\citep{baseband_paper}, which records and stores the data used in our search. CHIME has no moving parts, and beams are digitally formed and steered. It is located near Penticton, British Columbia at the Dominion Radio Astrophysical Observatory (DRAO). CHIME consists of four cylindrical paraboloidal reflectors which are each $\approx$100-m long and 20-m wide. Each cylinder is populated with of 256 dual-polarization antenna feeds, along an 80-m-long focal line with each feed spaced 30 cm apart. The feeds are oriented such that the two polarizations are aligned with the N-S and E-W cardinal directions. The dimensions of the telescope are crucial for the lensing search, as signals can reflect and bounce along the cylinder. This creates time-delayed images in the system that would be indistinguishable from a time-delayed gravitationally lensed image. The largest dimension is 100 m, therefore any detections at time-lags less than $\frac{\SI{100}{\meter}}{c} = \SI{330}{\nano\second}$ are indistinguishable from instrumental echoes, which are likely at such small time-lags.

\begin{table}[htb!]
\centering
\begin{tabular}{l l} 
    \hline
    Parameter & Values \\
    \hline
    Collecting area & 8000 $\mathrm{m}^2$ \\
    Frequency range & 400 \textemdash 800 MHz \\
    Polarization	& Orthogonal linear\\
    E-W FOV &	$2.5^{\circ}$ \textemdash $1.3^{\circ}$ \\
    N-S FOV & $\sim 100^{\circ}$ \\
    Number of beams & 1024 \\
    Beam width (FWHM) &	40\textemdash 20'\\
    \hline
 \end{tabular}
 \caption{Specifications of CHIME relevant to the gravitational lensing search.}
\end{table}

Each antenna feed operates within a bandpass of 400-800 MHz. Analog signals from the feeds are digitized in the second Nyquist zone at a sampling rate of 800 MHz with 8-bit accuracy~\citep{FEngineOverview}. The signal is channelized through a Polyphase Filterbank (PFB, see Appendix~\ref{sec:pfb} for details). This signal processing chain turns the digitized voltage timestream, sampled every 1.25 ns, into 1024 frequency channels. Each channel is 390 kHz wide and is centered at $f_i = 400.390625, 400.78125,...\SI{800.0}{\mega\hertz}$. The channelized complex data compose the dynamic spectrum which is sampled every 2.56 \textmu s with 4 + 4 bit complex accuracy. This frequency-time data or dynamic spectrum is referred to as baseband data. The signal processing and channelization are done through a signal processing system referred to as the F-engine ~\citep{FEngineOverview}. For the gravitational lensing search, we seek to invert the channelization and recover the higher time resolution voltage timestream to search for smaller time delays.

The channelized data are passed from the F-engine to the X-engine: 256 computer nodes that process the voltage data in parallel over different frequencies for various backends~\citep{CHIMECosmologyOverview,FRBSystemOverview,CHIMEPulsarOverview}. Using Fast Fourier Transform (FFT) beamforming~\citep{CHIMEBeamforming,beamforming}, the X-engine forms a grid of 1024 beams pointing towards a fixed set of azimuths and altitudes within CHIME's primary beam~\citep{chimefrb_overview}. The CHIME Fast Radio Burst project (CHIME/FRB) performs a real-time search for FRBs in each beam and once a candidate is detected, the search backend triggers a raw baseband dump, saving the data for offline analysis~\citep{chimefrb_overview}. 

Our search relies on baseband data, which are calibrated and prepared for scientific analysis through the baseband analysis pipeline ~\citep{baseband_paper}. There are two steps which affect the gravitational lensing search: beamforming and dedispersion. 

First, once an FRB has been detected in one of the beams in the search grid, the baseband analysis pipeline finds the best-fit location of the FRB in the sky~\citep{baseband_paper}. The baseband data from each antenna can be re-beamformed towards the best-fit position by applying a phase shift to each antenna. This points all of the antennas at the best-fit location of the FRB. The co-added dynamic spectrum pointed to the best-fit location for each FRB is used in the remainder of our analysis.

After beamforming, we de-disperse the data by applying the following coherent de-dispersion kernel~\citep{hankins_pulsar_signal_process}:
     \begin{equation}\label{eq:dispersion}
  H(f+f_i) = \exp \left(-2 \pi i k_\mathrm{DM} \mathrm{DM}   \frac{f^2}{f_i^2(f+f_i)} \right) ~
\end{equation}
where the $f_i$ are chosen to be the center frequency of each channel and the DM is chosen to maximize the signal-to-noise ratio of the burst summed over all 1024 channels. Next, the entire band is dedispersed incoherently, by shifting each channel by some integer multiple of $\SI{2.56}{\micro\second}$ to align neighboring channels according to the chosen DM. These two steps sufficiently compensate for interstellar dispersion for the purposes of our pipeline's successful operation\footnote{There is an overall phase per frequency channel due to dispersion that remains uncorrected (see~\citep{lorimer2004handbook} Eq. (5.17)); however this overall phase cancels out in time-lag auto-correlation.}.

Beyond the dispersive smearing of the pulse, we consider other propagation effects which may have an influence on our pipeline's sensitivity and may require compensation. One effect that has been explored is the possibility that the DM differs between the two images due to their different propagation paths~\citep{2019MNRAS.488.2989F,2020ApJ...891L..38C}. However, the largest lensing delay we consider is 100 ms. The fractional difference in path length, considering the cosmological travel times of FRB emission, is at most of order $\SI{100}{\milli\second}/\SI{1}{\giga\year} \sim 10^{-18}$. The fractional change in the DM arising from the two different paths is therefore a negligible effect. Small-scale inhomogeneities between the two paths may be relevant, but this matters most near the gravitational lens plane, where the two image paths are maximally separated. The lens in turn is most likely to be found at a significant distance from either the source or the observer. If such plasma density fluctuations were present at significant distances from either source or observer, we would observe IGM-based scatter broadening in FRBs which would likely quench any spectral scintillation. The observation of scintillation in many FRBs~\cite{masui_scattering,macquart2019spectral,schoen2020scintillation} provides evidence against the scattering originating from the IGM or in the CGM of intervening galaxies. A similar argument applies for differential Faraday rotation between the two images.

\subsection{Dataset}
We conducted our search using the baseband data of 172 FRB events with 103 events from independent sight lines. The remaining events are repeat bursts identified with one of the first 103 sources. 19 of these bursts have been previously analyzed and are publicly available in~\citep{chimefrbcatalog1}, and the remainder have not appeared in CHIME/FRB publications. Our search algorithm uses baseband data, which is available for approximately $25\%$ of all bursts detected by CHIME/FRB. Baseband is dumped when the burst's signal-to-noise ratio (S/N) $\gtrsim 15$ and $\mathrm{DM} < \SI{1000}{\centi\meter^{-3}}$. In addition, repeating FRB sources are over-represented in our sample relative to all FRBs detected by CHIME/FRB due to CHIME/FRB's data analysis prioritization strategy. Depending on the steepness of the FRB luminosity function, there might be a slight magnification bias which makes lensed bursts over-represented in our sample~\citep{oguri2019strong}. With current numbers of FRBs, however, this effect is too small to detect. We provide a table of all FRB events used in this search in Appendix \ref{sec:bursttable}.

\section{Search Algorithm} \label{sec:search} 
Our search algorithm for detecting gravitationally lensed images in the CHIME/FRB baseband data from FRB events is comprised of five main sections. In the following, we will outline the implementation of these algorithms and their importance in detecting microlensing.
\subsection{Matched Filter}\label{sec:mfilter}
\begin{figure}[htb!]
    \centering
    \includegraphics[width=\columnwidth]{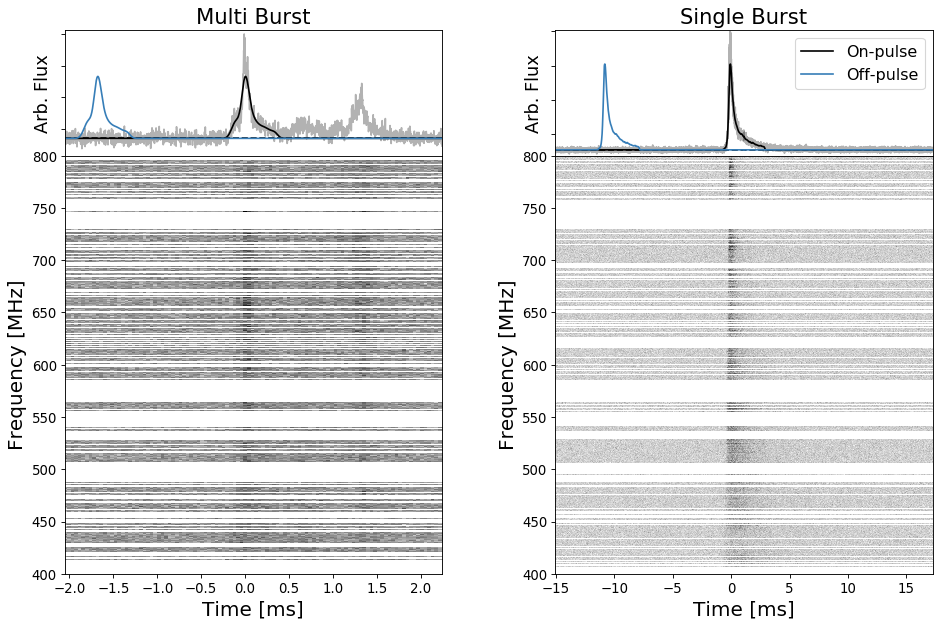}
    \caption{Two examples of FRBs used in our search: an anomalous multi-burst FRB (left) and a more common single-burst FRB (right). We compute the intensity profile (gray lines in top left and top right panels) by summing the power over all frequency channels in the dynamic spectrum of each burst (bottom two panels). The matched filter (black lines) is constructed by smoothing the intensity profile of the brightest peak and enhances sensitivity, while a translated copy (blue) serves as a null test. Channels not recorded by the X-engine or  contaminated by radio-frequency interference (RFI) are indicated by the white bands in the dynamic spectrum. We define on-pulse (top panel, black line) when the filter is centered on one FRB image (brighter burst in the left panel, lone burst in the right panel) and off-pulse (top panel, blue line) as when the filter contains no signal. Our coherent search can distinguish whether the dimmer components of multi-component FRBs (left) are images created by a lens, and it can search for temporally-unresolved images within apparently single-component FRBs (right).}
    \label{fig:filter}
\end{figure}

In order to detect gravitationally lensed signals in the voltage timestream, our search focuses on finding another copy of the same electric field waveform occurring at a delayed time. Since it could be buried in the noise, it is desirable to use a time-domain matched filter to enhance our search's sensitivity. We begin with beamformed baseband data (also referred to as the ``dynamic wavefield'' in the wave optics literature), dedispersed to a signal-to-noise maximizing DM. To understand why this metric is chosen to optimize the DM, it is helpful to consider the effect of choosing a slightly different DM. This would cause in (1) a percent-level residual dispersive smearing within each frequency channel, and (2) an uncompensated integer time delay, in units of 2.56 \textmu s, between neighboring channels. The effect of (1) is a decrease in the signal-to-noise ratio of the FRB within each channel. Similarly, the effect of (2) is that the burst is temporally smeared over the frequency band. Both effects decrease the signal-to-noise ratio of the FRB by temporally smearing some of the signal outside the support of the filter. In the correlation search for lenses, less signal is correlated while the noise contribution will remain the same. Eq.~\eqref{eq:lenslag} quantifies the height of the correlation peak as a function of $\Gamma$. In turn, Eq.~\ref{eq:noiseest} shows that if not enough signal is concentrated within the support of $W_P^2(t)$, then the correlation peak height will fall below the noise floor leading to a non-detection. Hence, the optimal DM value for this pipeline is a ``signal-to-noise maximizing'' DM, rather than, e.g., a DM that aligns temporal microstructures but results in a broader pulse profile. To maximize sensitivity, we begin by coherently dedispersing the beamformed baseband data to the DM that optimizes the signal-to-noise summed over all frequency channels.

After the pulse is aligned across the band, we construct a matched filter $W_P(t)$ in the time-lag domain following~\citep{leung2021synoptic}. We begin by summing the intensity over all frequency channels to obtain a pulse profile. This intensity profile is then smoothed by convolving it with a Gaussian whose width is set by some downsampling factor, chosen on a burst-by-burst basis by the baseband pipeline~\citep{michilli2020analysis}. When the intensity profile falls to the noise floor, we set the filter value to zero outside this region. $W_P(t)$ is therefore positive in some region, hereafter referred to as the ``on-pulse'' region, and zero everywhere else (the ``off-pulse'' region(s)). In later parts of the search, we use the off-pulse data as a null test. Additionally, we construct off-pulse filters by translating $W_P(t)$ to time ranges at least five burst widths prior to the FRB's arrival. In the point-mass lens model, the dimmer image arrives after the brighter one, so the data before the first burst are assumed to not contain any lensed images. The FRB image might lie inside or outside the on-pulse region. 

Figure \ref{fig:filter} shows the result of this process, where the matched filter outlines the intensity profile generated from the dynamic spectrum. The white bands spanning the time axis in the dynamic spectrum are frequency channels that are persistently contaminated with radio-frequency interference (RFI) or frequencies processed by correlator nodes that were not operational during the duration of the baseband dump. These channels are masked as part of the baseband pipeline.

\subsection{PFB Inversion}
In modern FX correlators, a polyphase filterbank (PFB) is used to channelize real-valued voltage timestream data into narrow-bandwidth channels of complex-valued voltage data (see Appendix~\ref{sec:pfb}). We refer to the result as baseband data; it is proportional to the underlying wavefield and from baseband data, quantities such as (intensity) dynamic spectra can be calculated. This simplifies procedures like beamforming and dedispersion but reduces the time resolution from that implied by the telescope bandwidth ($\SI{1.25}{\ns}$) to that implied by the channel bandwidths of 2.56 \textmu s.
To access timescales finer than 2.56 \textmu s, it is possible to dechannelize the data and approximately invert the channelization from a dynamic spectrum back into a single voltage timestream. This process can be thought of as turning CHIME into one effective antenna recording voltage data at a time resolution of $\SI{1.25}{\ns}$ -- CHIME's native sampling rate.

We invert the PFB by considering the time domain to be periodic about the total length of the baseband data, i.e. imposing a ``circulant'' boundary and solving a system of linear equations. A derivation of the inversion can be found in Appendix~\ref{sec:pfb_appendix} and further details can be found in~\citep{kadermscthesis}.
We summarize the method to invert the PFB in the following computational steps:
\begin{enumerate}
    \item We take the baseband data represented as a complex array, $\mathbf{V}(k,m)$. Here, $k$ is the ``channel'' axis, which represents frequency in units of $\SI{390}{\kilo\hertz}$ per channel. $m$ is the time axis (in units of frames, where each frame has a duration of 2.56 \textmu s).
    \item The PFB can be thought of as a linear filter which applies a different frequency response for each sub-frame time offset $m'$ ($m' \in 0,1.25,...\SI{2560}{\nano\second}$) and sub-channel frequency $k'$ ($k' \in [-\SI{195.3125}{\kilo\hertz},...,+\SI{195.3125}{\kilo\hertz})$). The frequency response is time-dependent and has a period of 2.56 \textmu s. This period is formed as the PFB is applied to digitized voltage data every 2.56 \textmu s. By imposing a circulant boundary on $m$, the PFB operator is akin to a convolution and therefore easy to invert in $m'$, $k'$ space using only Fourier transforms. In the language of linear algebra, the PFB can be represented in $m'$, $k'$ space as a diagonal operator and, $m'$ and $k'$ jointly specify a unique eigenmode of the PFB. To transform the data into this space, we first apply a discrete inverse Fourier transform to the channel axis. The frequency channel axis is transformed to the sub-frame time axis (i.e., $k \to m'$).
    \item $\mathbf{V}(m',m)$ is real-valued and solely in the time domain. This is a timestream with the PFB filter convolved throughout. We deconvolve the filter by first reconstructing the PFB filter in this domain. We place the PFB window coefficients (a sinc-hamming window for CHIME) at the start of a zero-padded array with the same dimensions as $\mathbf{V}(m',m)$, capturing both the sub-frame and frame information of the filter.
    \item The PFB is applied every 2.56 \textmu s; we deconvolve it along the frame ($m$) axis. We apply a discrete forward Fourier transform to the frame axis to transform it to the sub-channel frequency axis (i.e., $m\to k'$) for both the zero-padded PFB filter and $\mathbf{V}(m',m)$. The result is the diagonal matrix $\mathbf{P_c}$, which approximates the frequency response of the PFB as a function of time $m'$ and sub-channel frequency $k'$.
    \item The PFB is not perfectly invertible; therefore some elements of $\mathbf{P_c}(m',k')$ (the PFB's frequency response) are close to zero. Additionally, voltage signals arrive at the correlator at different times, scrambling sub-frame structure. To mitigate this, we average $\mathbf{P_c}(m',k')$ over all values of $m'$, $0,1.25,2.50,...\SI{2560}{\nano\second}$. This averaging procedure can be thought of as a way to average over unknown cable delays incurred prior to digitization. $$\overline{\mathbf{P_c}}(k') = \dfrac{1}{2048}\sum_{m'} \mathbf{P_c}(m',k')$$ then is independent of $m'$.
    \item We divide the data $\mathbf{V}(m',k')$ by $\overline{\mathbf{P_c}}(m',k')$, deconvolving the PFB filter from the data. Note that this is an imperfect deconvolution because of the averaging operation but gives more stationary autocorrelation functions as compared with $\mathbf{P_c}(m',k')$.  
    \item Finally, we perform a discrete inverse Fourier transform over the sub-frame frequency axis to transform $k'$ back to $m$. Our data are now represented as a function of the sub-frame time offset $m'$ and the frame delay $m$. Flattening the array $\mathbf{V}(m,m')$ results in $\mathbf{V}(m+m')$, a one dimensional function in the time-domain with a time resolution of $\SI{1.25}{\nano\second}$. This represents a good approximation to the reconstructed timestream that would be observed if CHIME were a single-dish telescope pointed precisely in the direction of the FRB, with the effects of dispersion removed.    
\end{enumerate}

In principle it is possible to invert the PFB before beamforming and dedispersing the FRB signal. This is the better method if one wants to accurately reconstruct the FRB in the time domain rather than the time-lag domain. However, the phase introduced per frequency channel by beamforming and dedispersion is not relevant for the purposes of our gravitational lensing search, as this only introduces an overall phase error per frequency channel in the dynamic spectrum. This overall phase error is redundant when performing a time-lag auto-correlation as it would be common to both images. The phase between neighboring channels, however, does affect the PFB inversion. Specifically, the PFB has a different frequency response as a function of sub-integer delay. However, since the sub-frame delay of each channel after summing over antennas is not straightforward to model robustly, we instead average the PFB's eigenspectrum as a function of sub-integer delay.

There are inversion artifacts that appear at regularly-spaced intervals in the time-lag domain, every integer multiple of 2.56 \textmu s due to imperfect inversion of the PFB. We refer to these artifacts as correlation leakage. In real data, we are able to remove the correlation leakage for off-pulse data. Unfortunately, this procedure is not able to completely remove correlation leakage for some extremely bright FRBs. However, we are able to replicate this feature by simulating similar bursts having similar widths, S/N ratios, and DMs to the observed burst (Section \ref{sec:validation}). This correlation leakage is understood and can be modeled exactly with our simulation framework. If these correlation leakage features are present in data, we will always find them at integer multiples of 2.56 \textmu s in the time-lag domain. As the locations of these correlation leakage spikes are known and they are able to be replicated in simulations, we chose to account for this in our pipeline by only searching for lensing correlations at lags corresponding to non-integer multiples of a frame (2.56 \textmu s). This reduces our exposure to lenses by about 1 part in 2048.

\subsection{RFI Flagging}\label{sec:rfi_flag}

Narrowband RFI can cause false positives in our search. Therefore, it is necessary to aggressively remove RFI. We apply RFI filtering at two different stages in our pipeline. Our first flagging occurs before inverting the channelization. This removes any strong narrowband RFI that auto-correlates. We take the matched filter, move it to a region without any FRB signal, and perform our time-lag correlation (defined in section \ref{sec:correlation}) per frequency channel. We normalize all the frequency lag spectra by the zero-lag peak and obtain the RMS per frequency channel. We calculate the mean and standard deviation $\sigma$ across all frequency channels and flag based on whether a channel is larger than 3-$\sigma$ from the mean. We move the off-pulse to five different regions in total, repeating this process, in order to generate a RFI mask from the sum of all the trials. The off-pulse realizations should not contain any FRB signal but even in that scenario a lensing signal would only exist at one frame delay and be common in amplitude across all frequencies. This algorithm, which removes channels that autocorrelate strongly with themselves across time-lag, should not mask such a non-local signal.

Our second round of RFI flagging occurs after we generate the high frequency resolution spectrum, which occurs after we have inverted the PFB. Here, we flag by applying a median absolute deviation (MAD) filter and masking any excursions in the spectra which exceed 3-$\sigma$, where $\sigma = 1.4826 \times \sigma_{MAD}$ and where $\sigma_{MAD}$ is the median absolute deviation of the nearest 15 neighbors. We additionally remove the 2048 (out of $10^6$ values) highest peaks in the frequency spectra to further ensure that there are no outliers remaining. This is sufficient to removing all narrowband RFI signals without removing a significant fraction of the FRB's spectral content. Because the lensing signal is non-local in frequency space, our narrowband RFI flagging removes less than $1\%$ of the lensing signal in a typical run. The final RFI mask is the union of all frequencies flagged in both polarizations, and results in a significant reduction in the non-Gaussian statistics in the time-lag domain.

In order to claim a statistically significant excursion, we require a thorough model of the underlying statistics of noise and other contaminations such as RFI, such that we can reject the null hypothesis. We use the off-pulse realizations to capture the instantaneous noise environment and sample the RFI conditions at the time of each event. We find this noise follows Gaussian statistics in the time-lag domain with the largest source of non-Gaussianity in the distribution of correlation values coming from strong narrowband RFI. We find our RFI flagging method results in an average removal $\sim 20 \%$ of the total bandwidth and is able to significantly reduce non-Gaussian contributions.



\subsection{Time-Lag Correlation}\label{sec:correlation}
 \begin{figure}[htb!]
    \centering
    \includegraphics[width=\columnwidth]{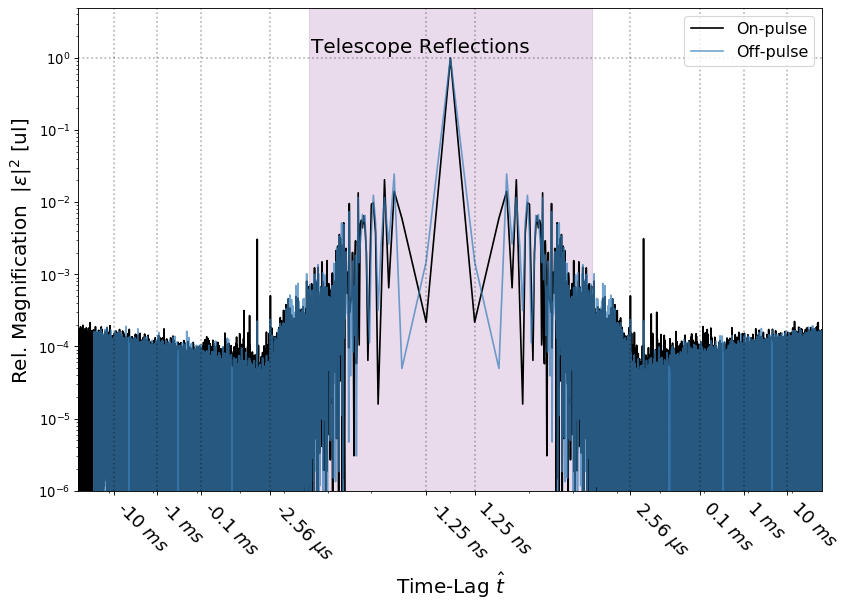}
    \caption{Time-lag correlation of both telescope polarizations, shown on a log-log scale. The on-pulse (black) is the time-lag correlation of the matched filters aligned with the FRB in each polarization. The off-pulse (blue) is the time-lag correlation with the matched filter moved to a region containing no burst. Telescope reflections dominate at lag timescales $< 300$ ns (shaded in pink). Peaks from PFB correlation leakage are visible at certain integer multiples of 2.56 \textmu s. A statistically significant correlation at any other time-lags might be a gravitational lensing signature. }
    \label{fig:autocorr}
\end{figure}
For our correlation algorithm, we implement Eq.~\ref{eq:timelagspec} through FFTs. This allows us to generate the time-lag spectrum using the discrete Fourier transformed spectra of the reconstructed voltage timestream in a computationally efficient manner. 
We generate two time-lag correlation functions; $C_X(\hat{t})$ and $C_Y(\hat{t})$, one from each antenna polarization. For shorthand, we introduce the vector, 
\begin{equation}\label{eq:corr_vec}
\vec{C} (\hat{t}) = [ C_X(\hat{t}) , C_Y(\hat{t}) ] ~.
\end{equation} 
In principle, a statistically-significant outlier in the time-lag domain could be interpreted as a signature of coherent gravitational lensing with some time delay $\tau$ and the relative image magnification ratio. 
We can convert the independent antenna polarization components of $\vec{C}(\hat{t})$ correlation values to the corresponding field amplitude ratios, giving $\vec{\varepsilon}(\hat{t})$, using the square root of Eq.~\ref{eq:mag_convert} and conserving the sign of the components of $\vec{C}(\hat{t})$.

To assess the statistical properties of $\vec{\varepsilon}(\hat{t})$, we generate off-pulse time-lag spectra by shifting the matched filter to a region without the burst and repeating the correlation. We average several off-pulse realizations into an estimate of the mean off-pulse time-lag spectra, which we denote as $\vec{\mu_{\varepsilon}}(\hat{t})$. We also measure the standard deviation of off-pulse spectra realizations from the mean time-lag spectrum, denoted as $\vec{\sigma_{\varepsilon}}(\hat{t})$. Our final data product consists of $\vec{\mu_{\varepsilon}}(\hat t)$, $\vec{\sigma_{\varepsilon}}(\hat t)$, and a single off-pulse realization left out of the calculation of $\vec{\sigma_{\varepsilon}}(\hat{t})$.


Fig.~\ref{fig:autocorr} shows the time-lag spectrum of on-pulse data containing an FRB (shown in black) and off-pulse data (shown in blue). The similarity of the two curves reflects the fact that the telescope data are dominated by thermal noise. We expect a coherent lensing event to appear as a peak in the time-lag spectrum in both telescope polarizations, with the same magnification ratio. One complication that arises is that at delays less than $\sim \SI{300}{\nano\second}$, internal reflections within the telescope dominate the on-pulse time-lag spectrum and can cause it to deviate from the off-pulse time-lag spectrum, even in the absence of a lensed image. Additionally, the frequency channel mask introduces spectral structure between $\pm \SI{2.56}{\micro\second}$ for both on-pulse and off-pulse data. This leads to non-zero structure for the time-lag spectrum.

In the absence of a detection, our measurements of $|\varepsilon|^2$ can still be interpreted as an upper limit on the relative magnification ratio as a function of lag, presented in our companion paper (\citet{leung2022constraining}).

\subsection{Outlier Detection}\label{sec:compression}
\begin{figure}[htb!]
    \centering
    \includegraphics[width=\columnwidth]{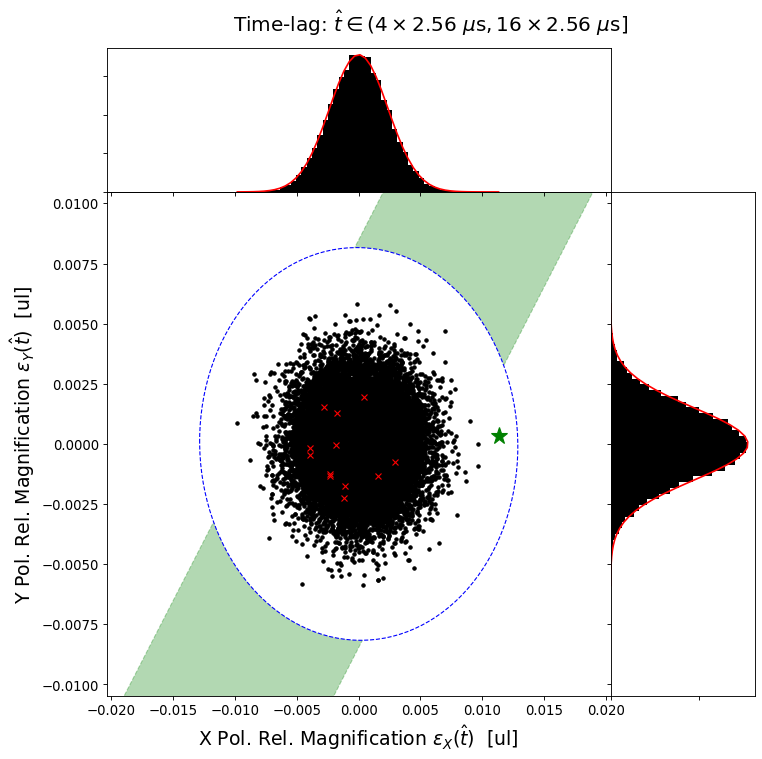}
    \caption{A graphical representation of off-pulse spectra for both antenna polarizations within one time-lag bin, and the vetoes that we use to reject noise candidates. We show the joint distribution of $\varepsilon_X$ and $\varepsilon_Y$ (black points), within a time-lag bin, for off-pulse data from one FRB event. The polarizations each follow a Gaussian distribution (red) as highlighted by the top and right histograms. The vetoed frame integer lags are the red crosses (condition 1, see text). The 2D significance threshold (blue) indicates which candidates are considered to be significant (condition 2), with the largest excursion as quantified by its $\chi^2$ value highlighted as the green star. A 99\% confidence region (green), derived from the local 2D Gaussian distribution and the requirement $\varepsilon_X \approx \varepsilon_Y$, indicates the region consistent with a gravitational lens (polarization condition, see Sec.~\ref{sec:results} and Tab.~\ref{tab:local_sig}); the region in which there are no candidates for this time-lag bin.}    
    \label{fig:2d_offstats}
\end{figure}

Lensing events are expected to be quite rare; optimistic estimates~\citep{Munoz_paper} vary from 1 in 100 to 1 in 1000 FRB sightlines for our mass range of interest.  Detecting FRB lensing therefore requires a search through all time-lags from a large sample of FRBs. Additionally, each $\vec{\varepsilon}(\hat{t})$ contains $\approx 10^9$ time-lags. It is difficult to search for lenses in such a large volume of data. However, the data volume can be reduced. We use the fact that instrumental reflections and frequency masking dominate only over a short range of time-lag scales. At short time-lag scales ($|\hat{t}| \leq 4\times \SI{2.56}{\micro\second}$), these systematics can significantly affect the distribution of time-lag spectrum values. However, at larger time-lag scales ($|\hat{t}| > 4\times \SI{2.56}{\micro\second}$), the majority of time-lag spectrum values can be modeled as realizations of a Gaussian random variable. Therefore, we can divide the time-lag spectrum at large time-lag scales into logarithmically-spaced bins. The bin edges are defined by $\pm \SI{2.56}{\micro\second} \times 4^i$ for integer values of $i=0,1,...$. For large time-lag scales, we characterize the statistics of each time-lag bin separately such that the systematics and statistics of smaller time-lag bins likely do not affect those of the larger time-lag bins and vice versa. Then, we save only the outliers in each time-lag bin.

We quantify outliers as follows. For each time-lag bin we calculate and save the mean over lags within a time-lag bin, denoted $\vec{\mu}_i$ (not to be confused with the mean over off-pulse time-lag spectra $\vec{\mu_{\varepsilon}}(\hat{t})$). We also calculate the $2\times2$ covariance matrix $\mathbf{G}_i$ from the time-lag spectrum values within that time-lag bin, for every lag bin. The matrix elements of $\mathbf{G}_i$ are computed by empirically estimating the moments  $\langle \varepsilon_X^2\rangle$, $\langle \varepsilon_Y^2 \rangle$, and $\langle \varepsilon_X \varepsilon_Y \rangle$, $\langle \varepsilon_X \rangle$, and $\langle \varepsilon_Y \rangle$ from the time-lag spectrum values within lag bin $i$.

\begin{figure}[htb!]
    \centering
    \includegraphics[width=\columnwidth]{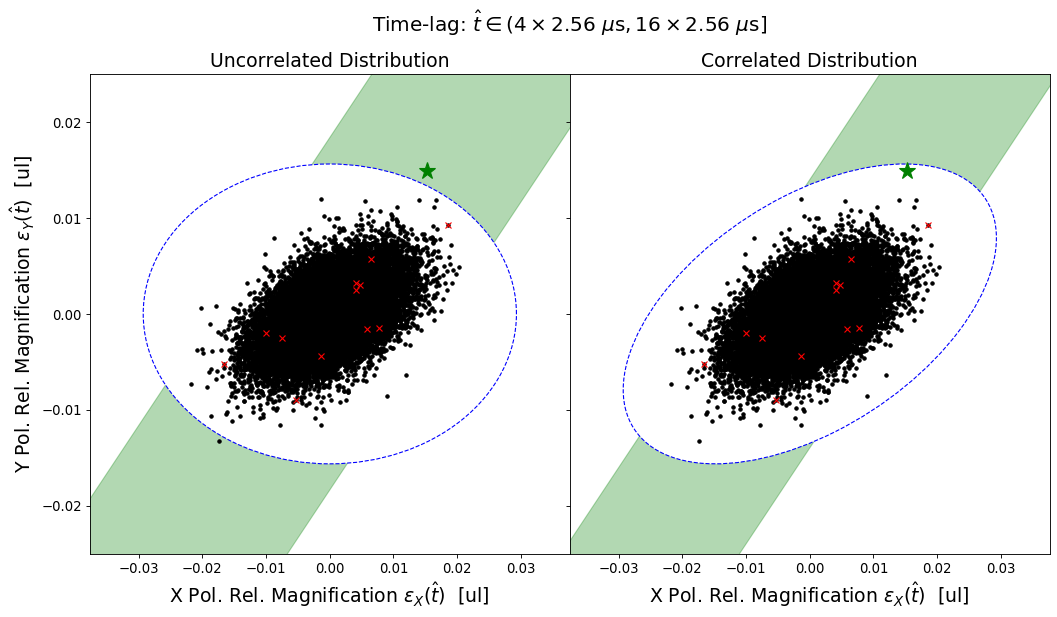}
    \caption{The joint distribution of $\varepsilon_X$ and $\varepsilon_Y$ (black points) for the on-pulse realization for an FRB event with excess correlation present. The criteria for a candidate are defined in Tab.~\ref{tab:local_sig}. \textbf{Left:} A graphical representation of the excursion significance not taking into account correlations between feed polarizations. The candidate event (green star) lies slightly outside the threshold contour (blue) and within expected range of a gravitational lensing signal (green region). \textbf{Right:} However, an improved estimate taking into account polarization correlations, shows the significance of the excess is consistent with the null hypothesis. } 
    \label{fig:corr_reanalyze}
\end{figure}

\begin{figure}[htb!]
    \centering
    \includegraphics[width=\columnwidth]{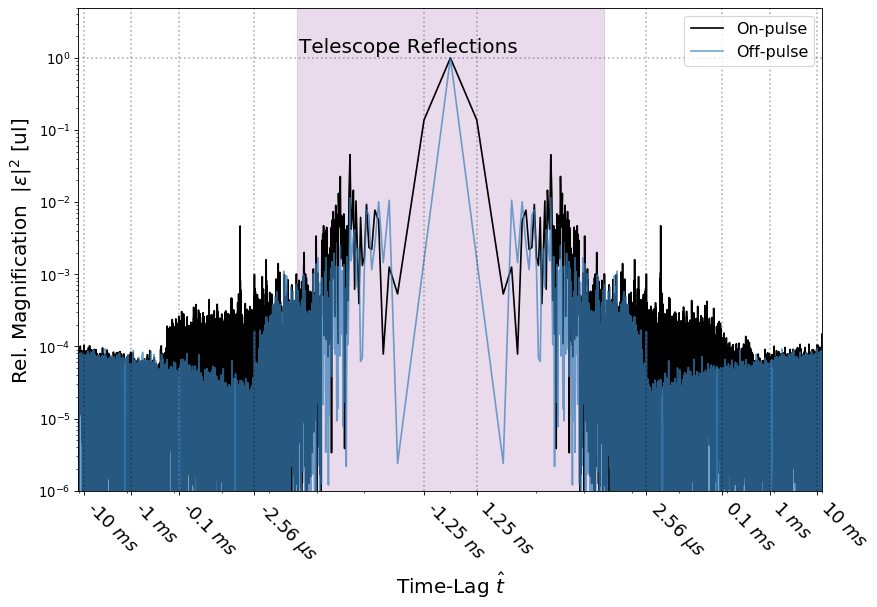}
    \caption{ Comparison of the time-lag spectrum for the on-pulse data (black) to that of the off-pulse data (blue) for an event with a correlated polarization. Time-lags that contain telescope reflections are indicated as the shaded purple region. There is evidence of excess correlation structure compared to the instrumental response extending to $ \hat{t} < 1 ~\mathrm{ms}$. At larger time-lags, on-pulse and off-pulse time-lag spectra become nearly identical. This excess structure appears across a broad range of time-lags, which is inconsistent with the expectation for a single gravitationally lensed image. }
   \label{fig:full_spec_compare}
\end{figure}

We find that the $\mathbf{G}_i$ differ significantly between the off-pulse and on-pulse data. In particular, Fig.~\ref{fig:2d_offstats} shows that for off-pulse data, the X and Y components of the time-lag spectrum vector $\vec{\varepsilon}(\hat{t})$ can be modeled as two independent Gaussian random variables. In contrast, for on-pulse data (Fig.~\ref{fig:corr_reanalyze}), the two spectra often exhibit a high degree of correlation between polarizations. For instance, we see that the correlation between the two polarizations form a tilted ellipse in Fig.~\ref{fig:corr_reanalyze}. The source of the correlation between polarizations is likely diffractive scintillation of the FRB. In Fig.~\ref{fig:full_spec_compare}, we see that the excess correlation in the time-lag spectrum itself in the on-pulse deviates significantly from the off-pulse/instrumental response suggesting an astrophysical origin. When comparing the fine spectral structure of the burst in a dynamic spectrum to the time-lag spectrum (Fig.~\ref{fig:full_spec_compare}), we see a clear relationship between the frequency bandwidth of all the sub-bursts and the width of the excess correlation structure in our time-lag spectrum analysis. We interpret the excess in Fig.~\ref{fig:full_spec_compare} as the signature of diffractive scintillation as seen in the time-lag spectrum.

A vector $\vec{\varepsilon}(\hat{t})$ belonging to time-lag bin $i$ has a  $\chi^2$ value of
\begin{equation}\label{eq:corr_r}
    \chi^2 = (\vec{\varepsilon}(\hat{t}) - \vec{\mu}_i)^{T} \cdot \mathbf{G}_i^{-1} \cdot (\vec{\varepsilon}(\hat{t}) - \vec{\mu}_i) .
\end{equation}
To keep the data volume manageable, we keep the 2048 most significant vectors, i.e. the vectors with the largest $\chi^2$ values. We find that keeping the 2048 top candidates within each lag bin is sufficient to preserve the tails of the distribution of vectors $\vec{\varepsilon}$ in which a lensing signal may be present. In this scheme, we keep a larger fraction of peaks at short time-lags than at large time-lags. While this property may seem undesirable, the logarithmic binning is natural for an unbiased search. The lensing time delay that we expect to observe is proportional to the characteristic mass scale of the lenses. Without a preferred mass scale over our mass range ($10^{-4}-10^4 M_{\odot}$), all decades in mass (and therefore in time-lag) should be treated as equally likely to produce a lensing event. From this perspective, the fact that a smaller fraction of time-lags are saved in larger bins can be thought of as a look-elsewhere effect. At our fixed time resolution of $\SI{1.25}{\nano\second}$ there are more lag trials at larger lag scales.

We refer to the remaining set of vectors as an ``excursion set'' containing $\approx 2048$ vectors, or ``excursions,'' per time-lag bin per burst. Any significant lensing event should be part of the excursion set, and should stand out in the excursion set. We analyze this possibility in Sec.~\ref{sec:candidate}. Alternatively, a large number of sub-threshold lensing events, may still measurably distort the distribution of excursions. This is considered in Sec.~\ref{sec:results}. To assess this latter possibility, we also generate an excursion set for the off-pulse data.

\section{Simulations}\label{sec:validation}

\begin{figure}[htb!]
    \centering
    \includegraphics[width=0.8\columnwidth]{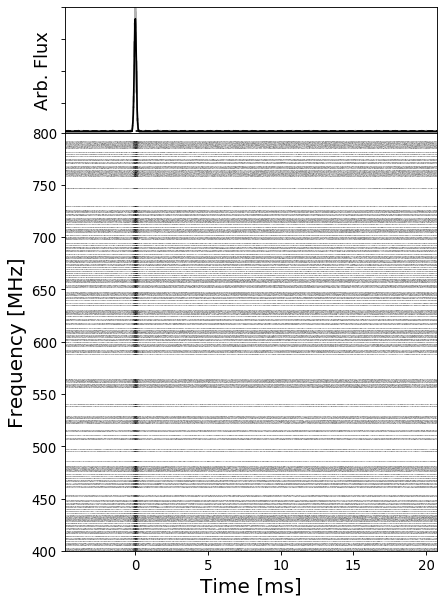}
    \caption{A simulated gravitationally lensed FRB injected into telescope noise data. The second image is injected at $\tau= 1.53$ ms and with $\varepsilon=0.1$. The image itself is not visible by eye but the second image is still detected by the search pipeline, shown in Fig~\ref{fig:sim_cand}. Both images are dispersed to the same DM, have the PFB channelization applied, and are then coherently dedispersed. Channels not recorded by the X-engine or contaminated by radio-frequency interference (RFI) are indicated by the white bands in the dynamic spectrum. } 
    \label{fig:sim_lensed}
\end{figure}

\begin{figure}[htb!]
    \centering
    \includegraphics[width=\columnwidth]{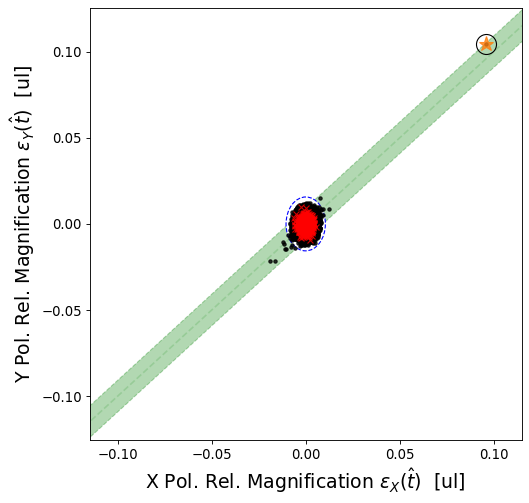}
    \caption{Simulated candidate detection with our selection criteria. The injected image had $\varepsilon = 0.1$ and the detection of the event can be seen as the circled orange star in the expected region for lensing (see section \ref{sec:candidate}). The other candidates in the green region result from trials which are correlated with the brightest candidate; they either differ by $<\SI{5}{\nano\second}$ (covariance introduced by masking parts of the band), or exactly 2.56 \textmu s (covariance introduced by correlation leakage). The red crosses are integer multiples of 2.56 \textmu s, which might be PFB inversion artifacts. The simulated baseband data is shown in Fig~\ref{fig:sim_lensed}.} 
    \label{fig:sim_cand}
\end{figure}

To check whether our pipeline can detect true gravitationally-lensed signals, we inject a coherent gravitationally-lensed FRB signal into noise samples from all the events in our dataset. The noise data is selected from a region of baseband data without a detected FRB. This tests our search pipeline in a realistic noise environment that includes RFI and frequency channel masking. 

Using this simulation framework, we may test our selection criteria and whether our pipeline recovers the parameters of the injected lensing event. We simulate a high-resolution voltage timestream at 800 Megasamples per second (the CHIME sampling rate) and simulate an FRB signal, which we model as white noise modulated with a Gaussian-shaped pulse profile. We do not consider any multi-path propagation effects other than the coherent gravitational lensing signal. We synthesize a lensing event by delaying the signal in time using Fourier methods and multiplying the signal by $\varepsilon$ as in Eq.~\ref{eq:voltform}. Both signals are dispersed with the same DM. The signal is sent through a PFB mimicking that of CHIME to channelize the voltage timestream into baseband data. The signals are dedispersed with the true DM and then injected into noise data. The noise data realizations are constructed from the off-pulse region of FRB events captured by CHIME. With this framework, we are able to capture instrumental effects and validate the recovery of lensing signals in a realistic noise environment, which includes effects such as masked channels, RFI, and telescope reflections. 

In Fig.~\ref{fig:sim_lensed}, we show the baseband data containing a simulated gravitationally lensed FRB injected into noise data recorded by CHIME. The second, delayed image has a $\tau=1.53$ ms with $\varepsilon=0.1$. In Fig.~\ref{fig:sim_cand}, we show the simulated lensing event which passes all of our selection criteria discussed in Sec.~\ref{sec:candidate}. We note that while the second image may not be visible by eye and exist within the noise, that with the matched filter and phase correlation, we are able to detect and recover the lensing signal. 

\begin{figure}[htb!]
    \centering
    \includegraphics[width=0.9\columnwidth]{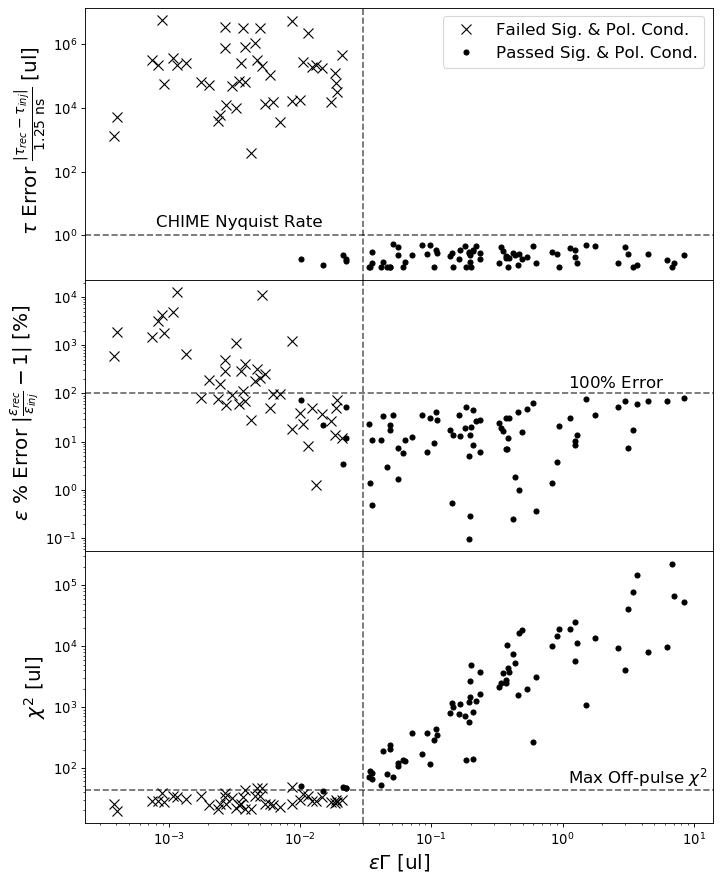}
    \caption{Magnitude of the residuals between the input gravitational lensing observable and the pipeline recovered values in fractional error, for simulated gravitational lensing events. $\varepsilon \Gamma$ is the fractional increase of the system temperature due to the second, delayed image. Both the time delay (top) and relative magnification (middle) are recovered by our pipeline when $\chi^2 \gtrsim 40$ and $\varepsilon \Gamma \gtrsim 0.03$. The $\chi^2$ (bottom) is a normalized measure of the height of the correlation peak in relation to the noise environment of the associated time-lag bin. Black dots highlights points that satisfy all our veto conditions while black crosses are simulations that did not pass the veto conditions. The largest $\chi^2$ observed from the off-pulse simulations is shown to represent the largest noise excursion observed. At $\varepsilon \Gamma = 0.03$ we indicate, with a vertical line, where the lensing signal is classified as a signal rather than a noise fluctuation. This validates our search pipeline and confirms our ability to reliably recover lensing parameters when their second images are sufficiently bright compared to the noise. The average error in recovery of the relative magnification, $\varepsilon$, is $\sim 23\%$.}
    \label{fig:sim_recover}
\end{figure}
For each noise dataset in our sample of FRBs, we randomly choose S/N values between 9 and 30, a DM between 10 and 50 pc $\mathrm{cm}^{-3}$, and fix the pulse width at 256 \textmu s. Low DMs are chosen to reduce the computational cost of manipulating extremely long data streams. The lensing parameters are chosen as follows; we take a relative field amplitude ratio, $\varepsilon \in [0.001, 0.9)$, and a lensing time delay, $\tau$, between $7 \times 2.56$ \textmu s and $5000 \times 2.56$ \textmu s. All simulation parameter values are drawn from a flat distribution between the ranges listed. In Fig.~\ref{fig:sim_recover}, we highlight how our pipeline is able to recover the input lensing parameters as a function of $\varepsilon \Gamma$, the fractional increase of the system temperature due to the second, delayed image (see Eqs.~\ref{eq:ucorr_tt} \& \ref{eq:gammadef}). We additionally highlight the relationship of $\varepsilon \Gamma$ to $\chi^2$ (Eq.~\ref{eq:corr_r}), where $\chi^2$ is a standardized measurement of the height of the correlation peak in relation to the noise statistics of the time-lag bin. The dashed line at $\chi^2 = 43$ is the largest $\chi^2$ observed from the corresponding off-pulse realizations and represents the noise floor. Excursions close to this line are more likely to be noise fluctuations. We indicate a turnover point at $\varepsilon \Gamma = 0.03$ where the lensing signal is classified as a signal rather than a noise fluctuation. We find the pipeline is able to detect the lensing signal above these thresholds, indicated by the recovery of the time delay. For these detected excursions, we find an average error of $\sim 23\%$ in the recovery of the relative magnification. There are three reasons for this error.

First, we assume that the matched filter is a template of one image rather than two images. This assumption fails strongly if the second image is comparable in brightness to the first one and close in arrival time to the first image such that the matched filter is constructed with the contribution of both images. In this scenario, the estimate of the relative magnification is incorrect as the unlensed fluence, $F$ (Eq.~\ref{eq:corrwnorm}), becomes $(1+\varepsilon^2 )F$. This does not affect the constraints on PBHs as the contribution of one image is approximate when $\varepsilon \rightarrow 0$, near the noise floor. The other bound is known physically, $\varepsilon \rightarrow 1$ as $\tau \rightarrow 0$. For a real detection, this effect can be corrected for after confirmation.

Second, the peak height is reduced by $13\%$ on average as a result of discretely sampling the phase delay between the two images at 1.25 ns. The arrival time between the two images is vanishingly unlikely to be an integer multiple of 1.25 ns. The result of this effect is to generate a sinc response rather than a delta function and smear power over neighbouring samples. This effect can be accounted for in future work by considering these neighbouring samples, however for this current work, this effect results in an average error of $13\%$ in the recovery of $\varepsilon$.

Finally, we see a decoherence within the sub-frame separate from the previous effect related to the PFB and our inversion method. This is similar to the previous effect, as it relates to the phase delay between the two images on the scale of 2.56 \textmu s rather than 1.25 ns. We can fully characterize this response, which is periodic every 2.56 \textmu s, by empirically modelling it through simulations. We then correct it for both the simulations and the data. After correcting for this effect, our recovery of $\varepsilon$ is limited by noise variance (a $<10\%$ error).

As a final note, this simulation framework is only meant to test our search pipeline and understand our sensitivity to gravitational lensing parameters in real noise environments. Additional astrophysical effects, such as the spatial distribution of PBHs in the large-scale structure along the line of sight, could be added in future work.

\begin{figure*}[htb!]
    \centering
    \includegraphics[width=0.85\textwidth]{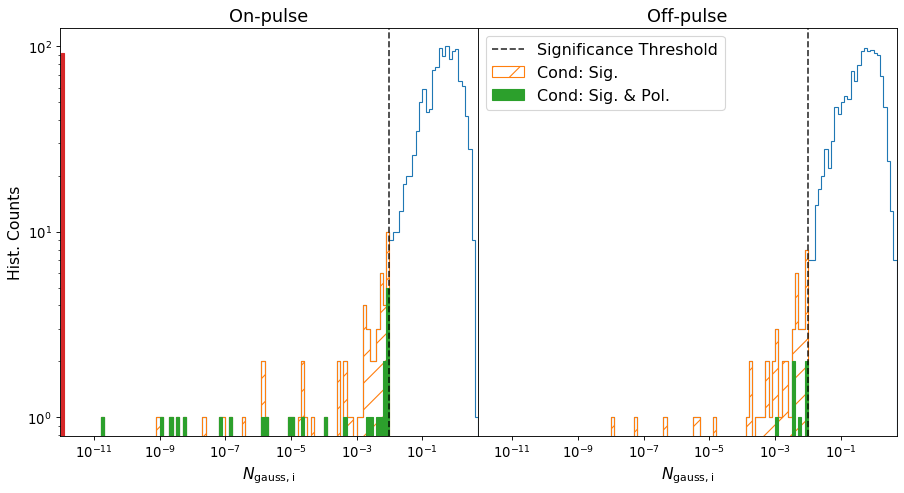}
    \caption{Histogram of $N_{\mathrm{gauss,i}}$ (see Eq.~\ref{eq:ngauss}), derived from the time-lag spectrum of simulated lensed FRBs injected into real telescope noise. The $N_{\mathrm{gauss,i}}$ values are aggregated over all lag bins $i$, by taking the largest statistical excursion in each correlation time-lag bin $i$ for all simulated events with successive veto conditions applied. There exists a lensing signature for every event. On-pulse data (left) highlights the large tail resulting from gravitational lensing. The off-pulse data (right) highlights the false-positive rate from noise. Conditions are defined in Tab.~\ref{tab:local_sig}. The red bin contains all excursions with $N_\mathrm{gauss} \leq 10^{-12}$ that satisfy both conditions.}
    \label{fig:sim_veto_stats}
\end{figure*}



\section{Veto Conditions} \label{sec:candidate}
In order to label an individual excursion as a gravitational lensing event, it must stand out as an outlier among the excursion set. We impose four conditions to quantify this, which we summarize in Table \ref{tab:local_sig}. Briefly, we require that the lensing event not occur at a frame boundary (close to an integer multiple of $\SI{2.56}{\micro\second}$), that it is unlikely to be a noise fluctuation, and that the FRB is detected in both the X and Y time-lag spectra and that they experience similar magnification ratios.

First, as discussed in Sec.~\ref{sec:pfb}, we do not consider excursions that are within \SI{0.625}{\nano\second} (set by CHIME's Nyquist limit) of an integer multiple of, \SI{2.56}{\micro\second} (our ``delay'' condition). Those excursions are likely to be non-astrophysical PFB inversion artifacts. If the largest excursion corresponds to an integer multiple of a frame, we disregard it and consider the next largest excursion. 

Second, for each lag bin, we can quantify whether the largest excursion is likely to be a noise fluctuation, and discard excursions attributable to noise (our ``significance'' condition). Since $\chi^2$ values from noise fluctuations follow a $\chi^2$ distribution with two degrees of freedom (e.g. Fig~\ref{fig:2d_offstats}), we calculate the probability of obtaining an excursion at least as large as the largest within that lag bin (whose significance we refer to as $\chi^2_{\mathrm{max},i}$. 
\begin{equation} 
\begin{split}
  p_i &= P(\chi^2 \geq \chi^2_{\mathrm{max},i} ) = 1 - \int_0^{\chi^2_{\mathrm{max},i}}~dx f(x;2)\\
  &= \exp(-\chi_{\mathrm{max},i}^2/2)~,
\end{split}
\end{equation}
where $f(\chi^2;2)$ denotes the probability density function for a $\chi^2$ random variable with two degrees of freedom, and where $p_i$ is the probability we would obtain our candidate excursion due to noise. To account for the trials factor $N_i$, the total number of time-lags in bin $i$, we multiply $p_i$ by $N_i$ to get $N_{\mathrm{gauss},i}$,
\begin{equation}\label{eq:ngauss}
    N_{\mathrm{gauss},i} = p_i N_i ,
\end{equation}
which may be interpreted as the answer to the question, ``How many excursions of size $\chi^2_{\mathrm{max},i}$ or larger are expected from lag bin $i$?'' 
If $N_{\mathrm{gauss},i}$ is $\mathcal{O}(1)$, then it is probable that the largest excursion in that lag bin was a noise fluctuation. However, if $N_{\mathrm{gauss},i} \ll 1$, then it is more likely that the excursion does not originate from statistical fluctuations in the time-lag spectrum.

We set a threshold of $N_{\mathrm{gauss},i} < 10^{-2}$ and expect 1 in $10^2$ lag bins to contain an excursion that passes this criterion purely due to statistical fluctuations. The threshold is visible as the blue ellipses in e.g., Figs.~\ref{fig:2d_offstats},~\ref{fig:corr_reanalyze},~\ref{fig:sim_cand}. A low threshold value of $N_{\mathrm{gauss},i}$ means that a false positive is unexpected when only a single lag bin is considered. As more lag bins are considered, including those from other bursts, the number of false positives will increase with this choice of threshold. The number of false positives can be estimated by tracking the total number of excursions from every lag bin and every burst that pass all our conditions using off-pulse data. 

\begin{table}[!htb]
\begin{tabular}{ r | p{0.35\textwidth} }
 Condition & Description\\
 \hline 
 Delay & $\tau \not\in N\times\SI{2.56}{\micro\second} \pm \SI{0.625}{\nano\second}$ where $N\in \mathbb{Z}$\\ 
 Significance &  $\vec{\varepsilon}$ has $N_{\mathrm{gauss},i} < 0.01$ in its lag bin \\
 Polarization &  $ |\varepsilon_X - \varepsilon_Y| $ within $99^{\mathrm{th}}$ percentile in its lag bin 
\end{tabular}
\caption{Conditions for a candidate excursion to be considered as a potential lensing event. Conditions are considered successively, with each step acting only on excursions passing all previous conditions.}
\label{tab:local_sig}
\end{table}

Third, in the absence of noise, we would expect that a gravitational lens would affect the two polarizations in precisely the same way (our ``polarization'' condition). Hence, we expect $\varepsilon_X = \varepsilon_Y$; in the presence of noise there will be some discrepancy from perfectly equal flux ratios. We therefore require that the burst be detected in both telescope polarizations, that is, $\Gamma_X > 1$ and  $\Gamma_Y > 1$ (Eq.~\eqref{eq:noiseest} in Sec.~\ref{sec:observables}). Then, for each lag bin, we compute the difference in relative magnification ratio between the two polarizations and consider this condition satisfied if $| \varepsilon_Y(\hat{t}) - \varepsilon_X(\hat{t}) | $ is within the $99^{\mathrm{th}}$ percentile of the noise fluctuations within its bin. 

Graphically, the region that is not excluded is represented as the green band in e.g. Figs.~\ref{fig:2d_offstats},~\ref{fig:corr_reanalyze},~\ref{fig:sim_cand}. This condition effectively disqualifies faint bursts where the detection is marginal in one of the telescope polarizations. While it is possible to incorrectly disqualify bright bursts which are coincidentally polarized along the other telescope polarization, it is much more likely that the observed burst is intrinsically dim. In this case, a possible lensing event would be even fainter and hard to robustly validate.

The final possibility for a candidate, once instrumental noise and diffractive scintillation are ruled out, is that the candidate is due to gravitational lensing.

The time-lag spectrum values can be used to obtain upper limits on impact parameters, $y$, and redshifted mass, $M_\mathrm{L}(1+z_\mathrm{L})$. This information is used in our separate analysis that provides upper limits on the cosmological abundance of compact objects~\cite{leung2022constraining}.

\section{Detecting Lensing}\label{sec:results}
In Sec.~\ref{sec:candidate}, we established our conditions for the detection of a candidate excursion in a single time-lag bin. In this section, we apply those conditions to excursion sets from every FRB event in our dataset to search for lensing in our sample. We aggregate the largest $N_{\mathrm{gauss},i}$ values from every lag bin from every FRB into a  ``global'' distribution of excursions. With the global distribution, we can consider two questions: ``Is there a single bright lensing event that is a distinct outlier from all excursions?'' and ``Does there lie a distribution of faint lensing events within our search data that we might detect as a statistical excess?''. 

\begin{figure*}[htb!]
    \centering
    \includegraphics[width=0.85\textwidth]{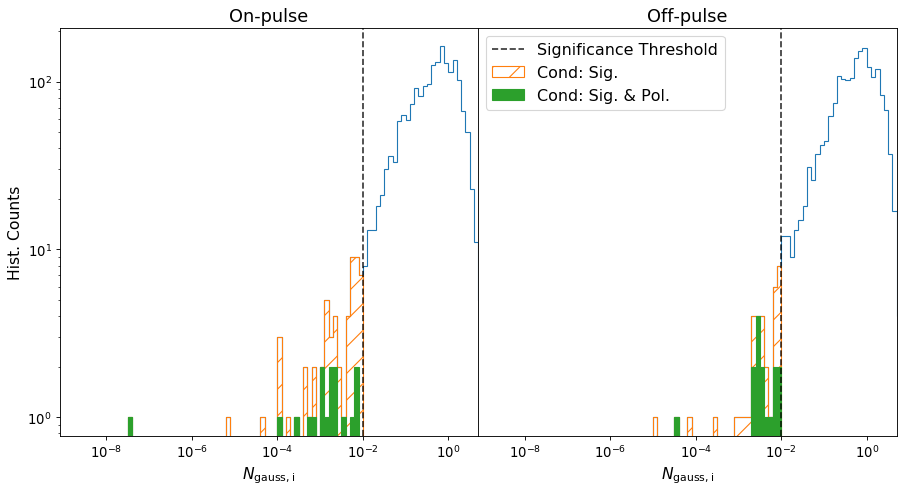}
    \caption{The distribution collecting the smallest values of $N_{\mathrm{gauss},i}$ (see Eq.~\ref{eq:ngauss}) observed (blue) in each time-lag bin for 172 FRB events. The most significant excursion in the global distribution of the on-pulse dataset lies farther below the threshold than the most significant excursion in the off-pulse dataset. On-pulse data (left) have more excursions which survive the vetoes than off-pulse data (right). $3.0~\%$ of on-pulse excursions compared to $2.0~\%$ of off-pulse excursions survive the significance condition (orange hatched). After applying all three conditions (green), $0.8~\%$ of both on-pulse and off-pulse excursions remain (green filled). Note the difference in x-axis scales from Fig.~\ref{fig:sim_veto_stats}. There are also no excursions with $N_{\mathrm{gauss},i}$ smaller than the scale shown. }
    \label{fig:veto_stats}
\end{figure*}

Our search set included all the FRB events with available beamformed baseband data and therefore contains no selection bias other than that bias imposed by which some events are given processing priority. In the future, the pipeline will become automated such that all baseband events will have time-lag correlation data available. For each of our 172 FRB events, we collect the associated probability of obtaining the largest excursion, $N_{\mathrm{gauss},i}$, from every time-lag bin. In Fig.~\ref{fig:veto_stats}, we show a histogram of these $N_{\mathrm{gauss},i}$ values compiled from our search. One value of $N_{\mathrm{gauss},i}$ for each time lag bin for each of the 172 FRB events makes a total of 1905 excursions: in the on-pulse distribution (left) and 1861 in the off-pulse distribution (right). The discrepancy comes from the inclusion of both positive and negative time-lag bins; off-pulse time-lag spectra are taken before the burst's arrival, where no lensing is expected. Since bursts may arrive close to the start of the data acquisition, there are on average fewer negative time-lag bins than for on-pulse spectra. 

If $N_{\mathrm{gauss},i} \ll 1$ (i.e. that the small chance occurrence probability is low), the corresponding $\varepsilon$ value is larger. $N_{\mathrm{gauss},i}$, therefore, acts as an indicator for the detection of a lensing signal. If there exists any lensing signals, faint or bright, they can be detected by comparing the on-pulse to the off-pulse distribution. Any excess in the on-pulse data might be due to lensed FRB events. 

\begin{figure}[htb!]
    \centering
    \includegraphics[width=0.9\columnwidth]{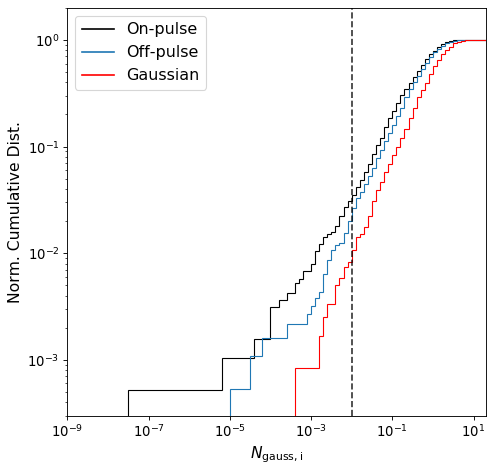}
    \caption{A normalized cumulative histogram of $N_{\mathrm{gauss},i}$ (see Eq.~\ref{eq:ngauss}), derived from the time-lag spectrum of 172 FRB events without any veto conditions applied (corresponding to the blue distributions seen in Fig.~\ref{fig:veto_stats}). The $N_{\mathrm{gauss},i}$ values are aggregated over all time-lag bins, $\mathrm{i}$. Gaussian (red) refers to sampling a Gaussian distribution with the selection effects imposed by our pipeline. We are biased and sensitive to any tail distributions, as that is where we can identify any lensing signals. Our pipeline will also observe any non-gaussianities, such as that from RFI and diffractive scintillation, which would cause deviations from the Gaussian expectation and could explain what is seen here. On-pulse (black) and off-pulse (blue) are shown to be non-Gaussian in their distribution of $N_{\mathrm{gauss},i}$.}
    \label{fig:cumdist_gauss}
\end{figure}

In Fig.~\ref{fig:cumdist_gauss}, we show the normalized cumulative distribution of $N_{\mathrm{gauss},i}$ for both on-pulse (black) and off-pulse (blue) and compare to a distribution created by sampling from a Gaussian distribution and including all selection effects of our pipeline (red). Our pipeline creates logarithmic time-lag bins and then selects the statistical excursion within that bin. If we sampled from Gaussian noise, we would observe the red distribution in this figure. However, it can be seen that both the on- and off-pulse distributions are not Gaussian in nature. The discrepancy originates from non-Gaussianity present in the tail distribution of telescope data. We are sensitive in detecting any non-Gaussian tail distributions, such as RFI for off-pulse data and RFI and diffractive scintillation for on-pulse data. As the search is expanded to more events, the tail distribution can be properly sampled, modelled, and accounted for such that the significance threshold can be accurately set. For this work, we refer only to the expected number of excursions assuming Gaussian sampling, $N_{\mathrm{gauss},i}$ and assert that this is a metric for evaluating the distributions, enough though the numeric values for $N_{\mathrm{gauss},i}$ would have the correct overall normalization for Gaussian distributions only. 

\begin{figure*}[htb!]
    \centering
    \includegraphics[width=0.85\textwidth]{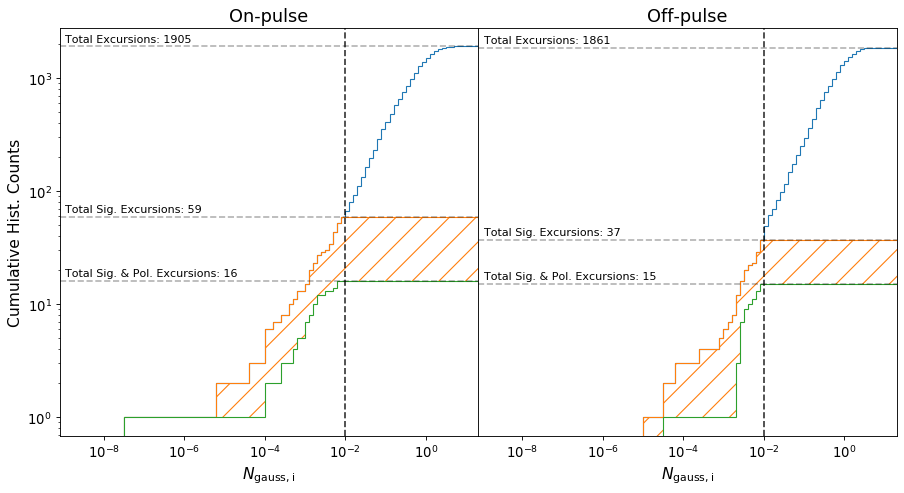}
    \caption{Cumulative distribution of expected number of excursions, $N_{\mathrm{gauss},i}$ (see Eq.~\ref{eq:ngauss}), observed for the largest statistical excursion, collected from each time-lag bin from all FRB events. After applying the significance condition (orange hatched), the on-pulse data (left) has $3.0 \% $ of excursions lying below the threshold and off-pulse data (right) has $2.0 \%$ of excursions lying below the threshold. The local significance threshold for every time-lag bin was set at $N_{\mathrm{gauss},i} = 0.01$. After applying the significance and polarization conditions (green), the on-pulse data and the off-pulse data both have effectively $0.8 \% $ of excursions that survive. }
    \label{fig:cumdist_compare}
\end{figure*}

In the global distribution, we can visualize the successive application of the candidate selection criteria in Table~\ref{tab:local_sig} and the resulting candidates that survive each condition. First, no excursion lies at an integer multiple of 2.56 \textmu s (delay condition); these events are not shown. Next, the covariance matrices $\mathbf{G}_i$ are calculated from the remaining excursions and the black dashed line indicates which excursions (those with $N_{\mathrm{gauss},i} < 0.01$) are classified as significant (significance condition). The orange hatched part of the histogram labels excursions that have a corresponding $N_{\mathrm{gauss},i}$ lower than this threshold value of 0.01. This choice is arbitrary but our results are insensitive to small changes in the threshold. This can be seen in Fig.~\ref{fig:cumdist_compare}, where the cumulative global distributions are shown. There is a deviation between the tails of the on-pulse and off-pulse distributions but it does not lie near the chosen threshold. Finally, the green part of the histogram refers to excursions that additionally have a magnification ratio that is consistent between the two polarizations (Polarization condition). After all conditions are considered, we see a single outlier event. This is a systematic outlier whose origin we discuss in Sec.~\ref{sec:eventdisc}. Hence, we exclude it from further analysis of the global distribution.

To check whether the on-pulse and off-pulse data are consistent with being drawn from the same underlying probability distribution, we perform several two-sample Kolmogorov-Smirnov (KS) tests. When comparing the two distributions with only the delay condition applied, the KS test statistic is $D = 0.075$; the associated $p$-value is $p = 3\times 10^{-5}$. If we consider only those excursions meeting the delay and significance conditions with $N_{\mathrm{gauss},i} < 0.01$ (orange hatched) we obtain $D = 0.16$ and $p = 0.52$. When applying the delay, significance, and polarization conditions (green filled), we obtain $D = 0.53$ and $p = 0.026$. Of these three cases, we reject the null hypothesis --- that the on-pulse distribution is the same as the off-pulse distribution --- in the first and the last case.

We find the cumulative fraction of statistically-significant off-pulse events is $2.0\%$. For on-pulse events we find this is $3.0\%$. The lack of excess events in the tail of the off-pulse data suggests these on-pulse excursions are not related to instrumental effects. However, these outliers are not consistent with the gravitational lensing hypothesis, as after all conditions are applied the cumulative fraction for both data sets is effectively the same ($0.8\%$). The origin of the excess is likely diffractive scintillation, not gravitational lensing.

We also propagate the simulations of Sec.~\ref{sec:validation} forward to show how the presence of lensing might distort the global distribution. 
The resulting simulated on-pulse and off-pulse distributions of $N_{\mathrm{gauss},i}$ are shown in the left and right panels of Fig.~\ref{fig:sim_veto_stats}  respectively. After all our conditions are applied, the remaining excursions are shown in green.

In the first scenario, the most probable lensing candidate is the one with the smallest $N_{\mathrm{gauss},i}$. Its significance can be assessed by comparing the smallest $N_{\mathrm{gauss},i}$ value from the on-pulse to that of the off-pulse distribution. In the latter (many, faint lensing events) scenario, when the null hypothesis is be rejected, it is possible that the excess of low-$N_{\mathrm{gauss},i}$ events originates from many faint gravitational lensing events.

\subsection{The Outlier Event}\label{sec:eventdisc}
\begin{figure}[htb!]
    \centering
    \includegraphics[width=\columnwidth]{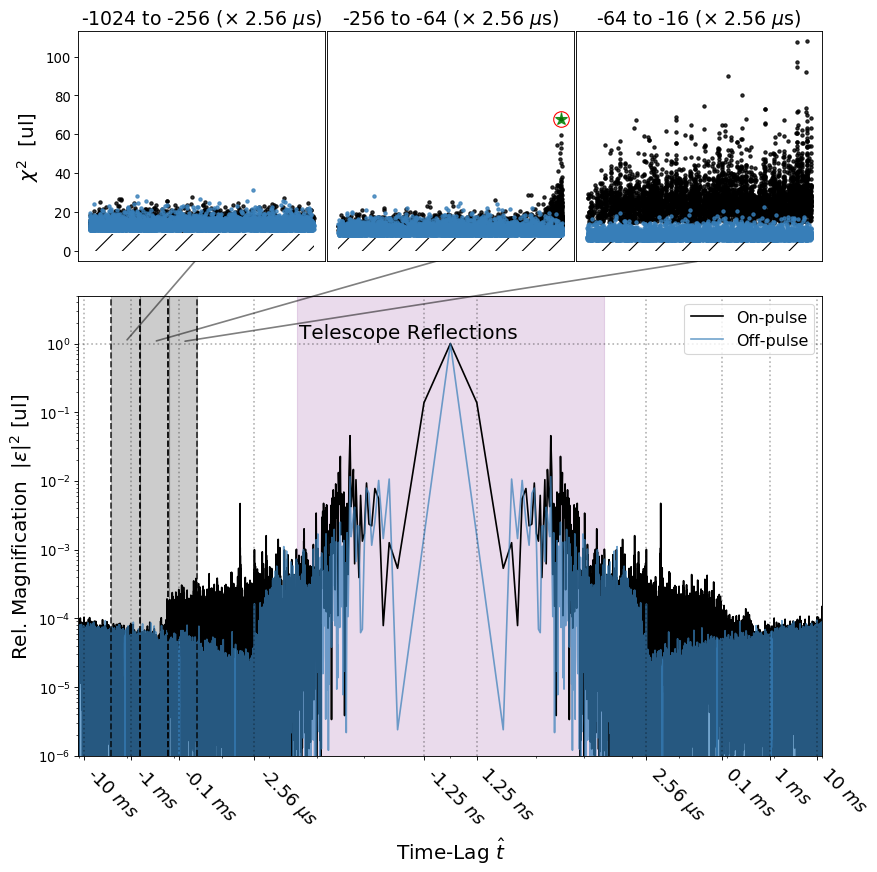}
    \caption{The time-lag spectrum of the outlier event, FRB20190624B, (circled star) from Sec.~\ref{sec:results}. The excursion was assigned an inappropriate probability due to the non-stationary noise environment within the time-lag bin. The on-pulse correlation structure, likely from scintillation, of this burst extends to large time scales (bottom). The logarithmic bin (top middle) containing the excursion overlaps the end of the correlation structure while still being noise dominated within the bin. Had the bins been chosen such that the excursion compared to the statistics of the -64 to -16 bin (top right), it would not be significant. Hatched region contains data not saved by the pipeline.} 
    \label{fig:corr_ev_acf}
\end{figure}

From the comparison of the two distributions of all time-lag bin excursions, shown in Fig.~\ref{fig:veto_stats}, we found one excursion in the on-pulse distribution that survives our three veto criteria. The excursion has $N_{exp} \sim 10^{-8}$, many orders of magnitude lower than the most significant excursion in the off-pulse distribution satisfying the same criteria (see green bars in Fig.~\ref{fig:veto_stats}).

However, the excursion is unusual for several reasons. First, it appears at a negative time lag. This is not expected in the point-mass lensing model, where the fainter image generically arrives after the brighter image. Second, its statistical significance as quantified by $N_{\text{exp}}$ is likely overestimated. This is due to our fixed binning scheme combined with the time-lag spectrum for this burst being unusually non-stationary. Upon inspection of the on-pulse time-lag spectrum for this event, the outlier was revealed to be a false positive because the time-lag spectrum for this FRB event is highly non-stationary as a function of time-lag. Fig.~\ref{fig:corr_ev_acf} shows the excess correlation in the on-pulse time-lag spectrum compared to the off-pulse time-lag spectrum. The -256 to -64 $\times\SI{2.56}{\micro\second}$ bin contains the outlier, which is highlighted as the circled star in Fig.~\ref{fig:corr_ev_acf}. In Fig.~\ref{fig:corr_ev_acf}, it is clear that the candidate is comparable to the structure seen in the -64 to -16 $\times \SI{2.56}{\micro\second}$ bin, but since the bin edge almost exactly coincides with the extent of the structure in the time-lag spectrum, just enough structure leaks into the the -256 to -64 $\times \SI{2.56}{\micro\second}$ bin from the -64 to -16 $\times \SI{2.56}{\micro\second}$ bin to generate a false positive but not enough to significantly modify the $\mathbf{G}_i$ for the latter bin.

This is a shortcoming of the pipeline, which assumes that the statistics of the time-lag spectrum values only change slowly as a function of lag. However, Fig.~\ref{fig:corr_ev_acf} shows that diffractive scintillation changes the statistics within lag bins in a way that changes rapidly as a function of lag. We emphasize that this event is not representative of our sample of FRBs. Only two FRB events, FRB20190417C and FRB20190624B, which includes the one shown here, in our sample exhibit this 0.1-ms-scale excess correlation; this represents an edge case rather than the standard for all our FRB events. We leave a more systematic treatment of time-lag dependent noise variance for future work.

\section{Final Remarks}\label{sec:discussion}
In this paper, we have presented a phase-coherent search pipeline to detect gravitational lensing of FRBs by compact objects of masses $\sim 10^{-4}$ to $10^{4} M_{\odot}$. We have conducted a search in a sample of 172 FRBs observed with CHIME. We have developed a comprehensive set of vetoes to classify and flag possible gravitational lensing candidates. We found no significant indications of a gravitational lensing signature for these events. Two bursts in our sample show statistically-significant structures in their time-lag spectra persistent in both telescope polarizations. One likely explanation for this is diffractive scintillation, which has already been observed in some FRBs~\citep{masui_scattering,macquart2019spectral,schoen2020scintillation}. A study of a handful of bursts exhibiting diffractive scintillation with CHIME~\citep{schoen2020scintillation} has shown that the scintillation is largely consistent with expectations from turbulent plasma in the Milky Way as quantified by the NE2001 model~\cite{cordes2001new}. This feature can be explained by diffractive scintillation in that FRB. If the excess correlation is from multi-path propagation through interstellar plasma it shows that many FRBs maintain phase-coherence as they propagate to Earth. Future work should characterize the feasibility of observing gravitational lensing even in the presence of plasma-related multi-path propagation effects.
In our companion paper~\citep{leung2022constraining}, we consider the constraints derived from our search on primordial black holes along our line of sight. Together these papers pioneer the use of coherent FRB lensing as a powerful cosmological tool.






\begin{acknowledgments}
We acknowledge that CHIME is located on the traditional, ancestral, and unceded territory of the Syilx/Okanagan people. We are grateful to the staff of the Dominion Radio Astrophysical Observatory, which is operated by the National Research Council of Canada.  CHIME is funded by a grant from the Canada Foundation for Innovation (CFI) 2012 Leading Edge Fund (Project 31170) and by contributions from the provinces of British Columbia, Qu\'{e}bec and Ontario. The CHIME/FRB Project is funded by a grant from the CFI 2015 Innovation Fund (Project 33213) and by contributions from the provinces of British Columbia and Qu\'{e}bec, and by the Dunlap Institute for Astronomy and Astrophysics at the University of Toronto. Additional support was provided by the Canadian Institute for Advanced Research (CIFAR), McGill University and the McGill Space Institute thanks to the Trottier Family Foundation, and the University of British Columbia.
\allacks
\end{acknowledgments}

\appendix
\section{Correlation Algorithm}\label{sec:corr_appendix}
\subsection{Time-lag Correlation}
In this section, we derive the search algorithm and how the lensing signal appears in the time-lag domain.
Let us define the real-valued voltage timestream, $V(t)$, which contains the system noise, $N(t)$, one FRB signal, $S(t)$, and a gravitational image pair of the FRB signal, $\varepsilon S(t-\tau)$,
\begin{equation}\label{eq:corr_model}
    V(t) = S(t) + \varepsilon S(t-\tau) + N(t)   .
\end{equation}

We want to find the delayed echo of this FRB signal so let us construct a matched filter, $W(t)$, from the intensity profile of the burst,
\begin{equation}
    W^2(t) = k S^2(t).     
\end{equation}
Here we are considering the case where the FRB and its echo exist as separate resolved images. We normalize $W^2(t)$ by choosing $k$ such that the weighted unlensed fluence of the FRB burst, $F_P$, agrees with the fluence of the burst in physical units.
\begin{equation}\label{eq:corrwnorm}
   \int dt W^2_P(t) S_P^2(t) =  F_P .
\end{equation}

For the search, we seek to correlate the timestream with itself to find the images. We can define the time-lag correlation as, 
\begin{equation}
    C'(\hat{t}_1,\hat{t}_2) =  \int dt V(t) W(t-\hat{t}_1 ) V(t-\hat{t}_2) W(t-\hat{t}_2 )    .
\end{equation}
Here we create two templates to correlate, $V(t-\hat{t}_2) W(t-\hat{t}_2 )$ and $ V(t) W(t-\hat{t}_1 ) $. We can change basis such that, $t' = t - \hat{t}_2$,
\begin{equation}
    C'(\hat{t}_1,\hat{t}_2) =  \int dt' V(t'+\hat{t}_2) W(t'+\hat{t}_2-\hat{t}_1 ) V(t') W(t' )    .
\end{equation}
Here we note the only non-zero contributions to this equation are when the non-zero weights of $W(t')$ and $W(t'+\hat{t}_2-\hat{t}_1 )$ overlap. We can then consider $\hat{t}_2$ to span all possible time-lag values while $\hat{t}_1$ must be bound between $ [\hat{t}_2 + T_W, \hat{t}_2 - T_W ]$, where $T_W$ is the filter width and any value outside these bounds is exactly 0. For our search, however, it is not necessary to search over two different time-lags, the other image will still be detected if we set $\hat{t}_2=\hat{t}_1$ and only have one time-lag $\hat{t}$ (which is bound between all possible time-lags),
\begin{equation}\label{eq:unnormcorr}
    C'(\hat{t}) =  \int dt' V(t'+\hat{t}) W^2(t' ) V(t') .
\end{equation}

Eq.~\ref{eq:unnormcorr} does not account for the noise variance such that, in the time-lag domain, the noise is not stationary. This is an issue for the search but is fixed by estimating the noise variance for every time-lag. If we consider there to be two random variables, $X_1 = V(t'+\hat{t}) W(t')$, and, $ X_2 = V(t') W(t') $, then the normalization of Eq.~\ref{eq:unnormcorr} is given by,
\begin{equation}\label{eq:corr2}
    C(\hat{t}) =  \frac{C'(\hat{t})}{\sqrt{C_{\sigma^2}(\hat{t})}} ,
\end{equation}
where 
\begin{equation}\label{eq:corr3}
    C_{\sigma^2}(\hat{t}) = \left( \int dt V^2(t) W^2(t)  \right) \left( \int dt V^2(t+\hat{t}) W^2(t) \right) .
\end{equation}
The two terms in Eq.~\ref{eq:corr3} represent the variance as a function of time-lag for random variables, $X_1$ and $X_2$. The term, $\int dt V^2(t) W^2(t)$, is a measurement of the variance of the quantity $V(t)W(t)$ while the term, $\int dt V^2(t+\hat{t}) W^2(t)$, can be considered as a weighted rolling variance estimator. We can define the time-lag correlation of the latter term as,
\begin{equation}\label{eq:corr4}
    \sigma^2 (\hat{t}) =  \int dt V^2(t+\hat{t}) W^2(t)  .
\end{equation}
and rewrite Eq.~\ref{eq:corr3} as 
\begin{equation}\label{eq:corr5}
    C_{\sigma^2}(\hat{t}) = \sigma^2 (0) \sigma^2(\hat{t})   .
\end{equation}
Then the variance adjusted time-lag correlation is given by,
\begin{equation}\label{eq:corr6}
    C(\hat{t}) =  \frac{C'(\hat{t})}{\sqrt{ \sigma^2 (0) \sigma^2 (\hat{t})}} ,
\end{equation}

\subsection{Observables in the time-lag domain}
Before we compute terms, let us state the assumptions made for this search. We assume the correlation between the signal and the system noise to be negligible and the correlation between the system noise and itself to also be negligible at non-zero time-lags. The system noise can correlate if RFI is present, but we consider our RFI cleaning algorithims to remove all significant correlating contributions. We will also focus on the case where the two gravitational images exist as two images separated by a minimum of the filter width. 

Let us first consider the $\hat{t}=0$. Using equations, \ref{eq:unnormcorr}, \ref{eq:corr2}, \ref{eq:corr5}, and \ref{eq:corr_model}, we obtain
\begin{equation}\label{eq:corr_t0}
    C(\hat{t}=0) = \frac{\int dt V^2(t) W^2(t) }{ \int dt V^2(t) W^2(t) }  = 1 .
\end{equation}

At the time-lag corresponding to the time delay between the gravitational images, $\hat{t}=\tau$, let us evaluate the components of Eq.~\ref{eq:corr2} separately.
First, we use Eqs.~\ref{eq:unnormcorr}, \ref{eq:corr_model}, and \ref{eq:corrwnorm} to obtain 
\begin{equation}\label{eq:ucorr_tt}
    C'(\hat{t}=\tau) = \varepsilon \int dt S^2(t)W^2(t) =\varepsilon F ,
\end{equation}
where all other terms are considered negligible; noise is assumed to not correlate with the signal or other noise at non-zero time-lags. For the other component in Eq.~\ref{eq:corr2}, we use equations \ref{eq:corrwnorm}, \ref{eq:corr_model}, and Eq.~\ref{eq:corr5} to get
\begin{equation}\label{eq:varcorr_tt}
     C_{\sigma^2}(\hat{t}) (\hat{t}=\tau) = ( F + \sigma_N^2 ) (\varepsilon^2 F + \sigma_N^2 ) ,
\end{equation}
where
\begin{equation}
     \sigma_N^2  = \int dt  N^2(t)W^2(t) = \int dt  N^2(t+\tau)W^2(t) ,
\end{equation}
as we assume the noise to be stationary. We can define a ratio, $\Gamma$, which is  similar to a signal to noise ratio as,
\begin{equation}\label{eq:gammadef}
    \Gamma = \frac{F}{ \sigma_N^2} .
\end{equation}
Then, the variance adjusted time-lag correlation, Eq.~\ref{eq:corr2}, at the lensing delay is given by
\begin{equation}
    C(\hat{t}=\tau) = \frac{\varepsilon \Gamma}{\sqrt{(\Gamma + 1)( \varepsilon^2 \Gamma + 1) } } .
\end{equation}
For our search algorithim, we compute the components of Eq.~\ref{eq:corr2} separately. Additionally, we can shift one of the voltage timestreams in Eq.~\ref{eq:corr2} to region without any signal such that we only correlate the system noise, defined as our off-pulse realization.  We can compute $\Gamma$ for every event such that there is only one unknown and we can solve for the relative magnification ratio,
\begin{equation}
    \varepsilon^2 = \frac{C(\hat{t}=\tau)^2 (\Gamma + 1)}{ \Gamma^2 - C(\hat{t}=\tau)^2 \Gamma^2  - C(\hat{t}=\tau)^2 \Gamma} .
\end{equation}

\section{List of Bursts}\label{sec:bursttable}
All FRBs used in this search. We tabulate the measured total DM of the burst, as well as the expected DM contribution from the Milky Way as determined by the NE2001 electron-density model evaluated in the direction of the FRB~\citep{cordes2001new}. All DMs are given in units of pc cm$^{-3}$. In~\citet{leung2022constraining}, we use the \texttt{fitburst} DM, DM$_{\rm MW}$, and $\tau_{\rm scatt}$ as inputs to our constraints on compact objects. The total and NE2001 DM are used to infer the distance to each FRB -- a necessary ingredient to translate the results of this search into constraints on compact dark matter. In addition, we use $\tau_{\rm scatt}$ to characterize the plasma properties and any associated decoherence related to scattering in the FRB's host environment.

\begin{longtable}{p{2.5cm}p{1.5cm}p{1.5cm}p{1.5cm}r}
\hline 
Burst & Total DM (pc cm$^{-3}$)  & DM$_{\rm NE2001}$ (pc cm$^{-3}$) & $\tau_{\rm scatt}$ ~ (ms) \\
\hline
 FRB20190110C &  222.08 &  36 & 0.28   \\ 
 FRB20190117A &  393.22 &  48 & 2.22   \\ 
 FRB20190122C &  689.97 &  30 & \ensuremath{<} 0.1  \\ 
 FRB20190202B &  464.88 &  70 & 0.08   \\ 
 FRB20190224D &  752.89 &  55 & 0.01   \\ 
 FRB20190301A &  459.79 &  82 & 2.08   \\ 
 FRB20190303B &  193.50 &  47 & 1.61   \\ 
 FRB20190320B &  489.51 &  38 & 0.08   \\ 
 FRB20190417C &  320.28 & 123 & 0.57   \\ 
 FRB20190423D &  496.68 &  67 & 5.16   \\ 
 FRB20190430C &  400.41 & 102 & 0.04   \\ 
 FRB20190606A &  552.65 &  32 & 1.60   \\ 
 FRB20190609A &  316.71 &  58 & 12.13  \\ 
 FRB20190609C &  479.87 & 113 & 0.31   \\ 
 FRB20190621A &  195.49 &  39 & 3.56   \\ 
 FRB20190624B &  213.95 &  70 & 0.18   \\ 
 FRB20190625E &  188.51 &  93 & 1.81   \\ 
 FRB20190628B &  408.03 &  47 & 0.68   \\ 
 FRB20190708A &  849.34 &  45 & 0.38   \\ 
 FRB20190712A &  682.40 &  52 & 0.13   \\ 
 FRB20190715B &  182.29 &  35 & 0.04   \\ 
 FRB20190722B &  508.47 &  56 & 0.12   \\ 
 FRB20190804B &  716.07 &  37 & 7.11   \\ 
 FRB20191018A &  301.34 &  49 & 1.24   \\ 
 FRB20191020D &  222.14 &  29 & 7.20   \\ 
 FRB20191024B &  400.28 & 102 & 0.07   \\ 
 FRB20191025D &  249.13 &  84 & \ensuremath{<} 0.1  \\ 
 FRB20191029A &  188.80 &  92 & 1.48   \\ 
 FRB20191104D &  192.16 &  56 & 5.03   \\ 
 FRB20191106C &  330.70 &  25 & 20.59  \\ 
 FRB20191107C &  222.24 &  37 & 0.48   \\ 
 FRB20191113C &  617.81 &  44 & 2.43   \\ 
 FRB20191114A &  552.93 &  99 & 10.83  \\ 
 FRB20191116A &  221.80 &  29 & 1.16   \\ 
 FRB20191215A &  222.12 &  29 & 0.06   \\ 
 FRB20191219A &  349.02 & 199 & 6.58   \\ 
 FRB20191219F &  464.56 &  49 & 0.11   \\ 
 FRB20191223A &  393.07 &  47 & 8.33   \\ 
 FRB20191223C &  604.76 &  80 & 0.01   \\ 
 FRB20191225A &  683.91 &  49 & 0.71   \\ 
 FRB20191231A &  222.42 &  30 & 3.57   \\ 
 FRB20200104E &  349.83 & 195 & \ensuremath{<} 0.1  \\ 
 FRB20200109B &  745.48 &  55 & 1.05   \\ 
 FRB20200112A &  221.16 &  29 & 1.85   \\ 
 FRB20200112D &  863.44 &  45 & 0.79   \\ 
 FRB20200118D &  625.34 &  77 & 1.33   \\ 
 FRB20200120E &   87.84 &  42 & 0.45   \\ 
 FRB20200120H &  349.80 & 200 & 1.06   \\ 
 FRB20200122A &  103.54 &  41 & \ensuremath{<} 0.1  \\ 
 FRB20200122D &  103.56 &  41 & 1.29   \\ 
 FRB20200122E &  103.56 &  41 & \ensuremath{<} 0.1  \\ 
 FRB20200122J &  103.49 &  41 & 0.16   \\ 
 FRB20200124A &  580.11 &  72 & \ensuremath{<} 0.1  \\ 
 FRB20200127B &  351.34 &  57 & 0.23   \\ 
 FRB20200128A &  439.61 &  45 & 0.25   \\ 
 FRB20200203A &  349.74 & 199 & 3.37   \\ 
 FRB20200204B &  349.30 & 200 & \ensuremath{<} 0.1  \\ 
 FRB20200204D &  350.19 & 199 & 2.79   \\ 
 FRB20200204J &  348.89 & 193 & \ensuremath{<} 0.1  \\ 
 FRB20200204K &  411.17 &  47 & 14.05  \\ 
 FRB20200207A &  506.85 &  67 & 1.08   \\ 
 FRB20200212A &  174.18 &  67 & 0.04   \\ 
 FRB20200219B &  351.27 &  54 & \ensuremath{<} 0.1  \\ 
 FRB20200220G &  313.38 &  58 & \ensuremath{<} 0.1  \\ 
 FRB20200501A &  469.51 &  51 & 6.56   \\ 
 FRB20200502A &  412.10 &  47 & 1.78   \\ 
 FRB20200503B &  674.19 &  51 & 7.12   \\ 
 FRB20200505B &  760.81 &  37 & 0.34   \\ 
 FRB20200507A &  166.92 &  51 & 0.44   \\ 
 FRB20200510A &  290.92 &  40 & 0.09   \\ 
 FRB20200512A &  349.63 & 104 & 7.17   \\ 
 FRB20200513A &  349.20 & 104 & 1.64   \\ 
 FRB20200513B &  579.72 &  72 & 3.12   \\ 
 FRB20200515A &  523.26 &  43 & 14.48  \\ 
 FRB20200520B &  351.41 &  54 & 0.19   \\ 
 FRB20200525A &  471.35 &  38 & 0.34   \\ 
 FRB20200525C &  339.63 &  34 & 0.36   \\ 
 FRB20200603B &  295.08 &  42 & 2.68   \\ 
 FRB20200606A &  723.34 &  47 & 8.22   \\ 
 FRB20200613A &  348.95 & 104 & 0.63   \\ 
 FRB20200614A &  348.79 & 104 & 1.27   \\ 
 FRB20200617A &  475.73 &  61 & 1.18   \\ 
 FRB20200621C &  364.26 &  34 & 4.77   \\ 
 FRB20200622A &  223.01 &  29 & 11.21  \\ 
 FRB20200629C &  363.79 &  43 & 0.64   \\ 
 FRB20200701A &  625.23 &  77 & 0.23   \\ 
 FRB20200702C &  201.28 &  46 & 5.41   \\ 
 FRB20200707A &  218.04 &  84 & 0.07   \\ 
 FRB20200709C &  363.56 &  43 & 1.75   \\ 
 FRB20200717A &  337.98 & 135 & 1.27   \\ 
 FRB20200725B &  302.35 &  50 & 1.39   \\ 
 FRB20200809G &  221.85 &  29 & 2.75   \\ 
 FRB20200813D &  190.80 &  53 & 0.03   \\ 
 FRB20200909A &  221.22 &  29 & \ensuremath{<} 0.1  \\ 
 FRB20200917A &  883.48 &  74 & 0.73   \\ 
 FRB20200918A &  314.44 &  26 & 0.12   \\ 
 FRB20200921A &  465.25 &  31 & 0.23   \\ 
 FRB20200921B & 1582.12 & 129 & 0.77   \\ 
 FRB20200930A &  851.58 &  49 & 0.27   \\ 
 FRB20201008A &  290.26 & 109 & 1.29   \\ 
 FRB20201015B &  173.79 &  55 & 0.82   \\ 
 FRB20201017A &  773.68 &  78 & 2.29   \\ 
 FRB20201030B &  479.80 & 114 & 0.33   \\ 
 FRB20201031B &  819.71 &  21 & 2.75   \\ 
 FRB20201125B &  413.71 &  38 & \ensuremath{<} 0.1  \\ 
 FRB20201128D &  157.88 &  38 & \ensuremath{<} 0.1  \\ 
 FRB20201129A &   87.86 &  42 & 0.28   \\ 
 FRB20201203C &  413.50 &  38 & 7.23   \\ 
 FRB20201204D &  364.25 &  34 & 4.95   \\ 
 FRB20201205B &  552.46 & 174 & 0.90   \\ 
 FRB20201219A &  322.20 &  38 & 4.09   \\ 
 FRB20201225B &  362.74 &  43 & 0.87   \\ 
 FRB20201225D &  287.96 &  57 & 5.37   \\ 
 FRB20201228A &  362.91 &  43 & 3.37   \\ 
 FRB20201230B &  256.13 & 144 & 0.05   \\ 
 FRB20210104B & 1236.69 &  22 & \ensuremath{<} 0.1  \\ 
 FRB20210105G &  288.02 &  57 & 2.48   \\ 
 FRB20210111E &  349.29 & 104 & 1.13   \\ 
 FRB20210113C &  176.85 &  57 & 1.78   \\ 
 FRB20210114B &  288.48 &  57 & 0.47   \\ 
 FRB20210115C &  200.91 &  46 & 2.34   \\ 
 FRB20210117E &  289.61 &  57 & 5.56   \\ 
 FRB20210118B &  288.24 &  57 & 9.94   \\ 
 FRB20210119B &  440.43 &  29 & \ensuremath{<} 0.1  \\ 
 FRB20210122B &  369.78 &  44 & 0.83   \\ 
 FRB20210127E &  348.88 & 104 & 2.61   \\ 
 FRB20210130H &  349.98 & 104 & 49.46  \\ 
 FRB20210130I &  349.47 & 104 & 13.75  \\ 
 FRB20210131A &  349.35 & 104 & 1.18   \\ 
 FRB20210203B &  579.94 &  72 & \ensuremath{<} 0.1  \\ 
 FRB20210203C &  221.65 &  29 & 0.16   \\ 
 FRB20210206A &  361.32 & 191 & 0.75   \\ 
 FRB20210207A &  221.87 &  29 & 4.97   \\ 
 FRB20210209B &  222.19 &  29 & 2.17   \\ 
 FRB20210209C &  382.35 &  29 & \ensuremath{<} 0.1  \\ 
 FRB20210211B & 1090.02 &  76 & 4.46   \\ 
 FRB20210213A &  482.40 &  46 & 0.05   \\ 
 FRB20210216B &  301.50 &  50 & \ensuremath{<} 0.1  \\ 
 FRB20210223A &  531.01 &  43 & 0.99   \\ 
 FRB20210302A &  349.44 & 104 & 5.85   \\ 
 FRB20210302C &  221.32 &  29 & 2.60   \\ 
 FRB20210303B &  349.26 & 104 & \ensuremath{<} 0.1  \\ 
 FRB20210303F &  510.06 &  51 & 4.41   \\ 
 FRB20210304A &  348.90 & 104 & \ensuremath{<} 0.1  \\ 
 FRB20210309D &  134.07 &  32 & 0.50   \\ 
 FRB20210310A &  135.50 &  20 & 1.55   \\ 
 FRB20210313B &  414.00 &  38 & 1.59   \\ 
 FRB20210314A &  413.72 &  38 & \ensuremath{<} 0.1  \\ 
 FRB20210326B &  413.82 &  38 & \ensuremath{<} 0.1  \\ 
 FRB20210327A &  415.92 & 140 & 15.40  \\ 
 FRB20210331A &  417.48 & 140 & 11.61  \\ 
 FRB20210331C &  414.66 & 149 & 14.13  \\ 
 FRB20210331D &  415.19 & 140 & 11.17  \\ 
 FRB20210402B &  349.86 & 193 & \ensuremath{<} 0.1  \\ 
 FRB20210410C &  301.97 &  50 & 0.74   \\ 
 FRB20210421E &   87.76 &  42 & 0.24   \\ 
 FRB20210430G &   87.76 &  41 & 0.14   \\ 
 FRB20210521C &  349.11 & 104 & 3.67   \\ 
 FRB20210523A &  348.84 & 199 & 1.65   \\ 
 FRB20210523C &  532.14 &  41 & 0.87   \\ 
 FRB20210526B &  382.39 &  29 & 4.57   \\ 
 FRB20210526D &  411.86 & 131 & 10.66  \\ 
 FRB20210610B &  694.91 &  50 & \ensuremath{<} 0.1  \\ 
 FRB20210610C &  876.40 &  21 & 6.72   \\ 
 FRB20210612B &  579.88 &  72 & 262.75 \\ 
 FRB20210624A &  413.49 &  38 & 1.00   \\ 
 FRB20210625A &  349.18 & 104 & 0.54   \\ 
 FRB20210711A &  349.33 & 199 & \ensuremath{<} 0.1  \\ 
 FRB20210712A &  348.85 & 104 & 0.01   \\ 
 FRB20210810C &  694.33 &  50 & \ensuremath{<} 0.1  \\ 
 FRB20210814B &  349.07 & 104 & \ensuremath{<} 0.1  \\ 
 FRB20210814C &  348.89 & 104 & \ensuremath{<} 0.1  \\ 

\hline
\end{longtable}

\section{Polyphase Filterbank}\label{sec:pfb}
In this section, we will outline and formalize the PFB, whose main goal is to reduce the spectral leakage when performing a Fast Fourier Transform. We will formalize the PFB as a series of linear operators. The basic operations of the PFB turn the voltages measured by the antennas into what is commonly referred to as the wavefield, or simply as baseband data.

We start with our voltage timestream, $\mathbf{v}$, which is sampled at a rate of 1.25 ns for CHIME. In general, a polyphase filterbank is constructed with a number of taps, $\alpha$, where each tap takes an input of $N$ samples from $\mathbf{v}$. For CHIME $\alpha = 4$ and $N=2048$. Each tap multiplies a set coefficient to the input value. We refer to these filter coefficients as the PFB coeffients. They are obtained from a pre-determined window function, which is a sinc-hamming window for CHIME. The outputs from all four taps are averaged together to produce an output of size $N$ which we refer to as a frame. The frame is Fourier transformed to produce a single frame of the dynamic spectrum. Each frame is 2.56 \textmu s in width. The process repeats after the voltage timestream has shifted $N$ samples, i.e., a shift of exactly one frame. It is important to note that there is no reduction in the total amount of data. Each output frame contains information from all four input frames and all input frames are used four times, once for each tap.

We can represent each tap as a square matrix, $ \mathbf{W^{(i)}}$. The matrix has $N$ diagonal elements which are the PFB coefficients for that tap.
\begin{eqnarray}
\mathbf{W^{(i)}}  & = &
 \left(
\begin{array}{cccc}
    W^{(i)}_{1}  & 0 &  \cdots & 0 \\
    0  &  \ddots  & \ddots & \vdots \\
    \vdots  &  \ddots  & \ddots  & 0 \\
    0 & \cdots & 0 & W^{(i)}_{N}  
\end{array} \right) 
     ~; ~ i \in [1,\alpha]
\end{eqnarray}

To consider how all $\alpha$ taps are averaged and how the PFB samples every $N$ samples we can construct the PFB projection matrix $\mathbf{P}$ which is a band block diagonal Toeplitz matrix.

\begin{eqnarray}
    \mathbf{P} & = &
    \left(
    \begin{array}{cccccc}
     \mathbf{W^{(1)}} & \cdots & \mathbf{W^{(\alpha)}}  & \mathbf{0} &  \cdots  & \mathbf{0} \\
     \mathbf{0} & \ddots & \ddots & \ddots  & \ddots  & \vdots \\
     \vdots & \ddots & \ddots & \ddots  & \ddots  & \vdots \\
     \vdots & \ddots & \ddots & \ddots  & \ddots  & \mathbf{0} \\
     \vdots & \ddots & \ddots & \ddots  & \ddots &  \mathbf{W^{(\alpha)}} \\
     \vdots & \ddots & \ddots & \ddots  & \ddots  & \vdots \\
    \mathbf{0} & \cdots & \cdots  & \cdots   &  \mathbf{0}  & \mathbf{W^{(1)}} \\ 
\end{array} \right) 
\end{eqnarray}
The application of the discrete Fourier transform per frame can be thought of as the block diagonal matrix $ \mathbf{F}$ where $\mathbf{F^{(N)}}$ represents a discrete fourier transform the size of a frame, $N$. 

\begin{eqnarray}
    \mathbf{F^{(N)}} & = &
    \left(
    \begin{array}{cccc}
    \mathbf{F^{(N)}} & 0 & \cdots  & 0 \\
    0  &\ddots  & \ddots  & \vdots \\
    \vdots  & \ddots & \ddots  &  0\\
    0 & \cdots &  0 & \mathbf{F^{(N)}}  \\
\end{array} \right) 
\end{eqnarray}

Then all the linear operators applied to a timestream can be thought of as
\begin{equation}\label{eq:pfb_eq}
    \mathbf{V}(k,l) = \mathbf{F} \cdot \mathbf{P} \cdot \mathbf{V}(m) ~.
\end{equation}

It is apparent that the PFB, by construction, introduces instrumental correlations in the neighbouring two frames. These are instrumental correlations that should appear at time delays of $\pm 2.56 ~\mu s$ and $\pm 2 \times 2.56 ~\mu s$. When conducting the lensing search, we therefore seek to invert the PFB to remove these systematic correlations and turn the beamformed and coherently dedispersed dynamic spectrum back into a singular voltage timestream sampling at 1.25 ns. 

With Eq.~\ref{eq:pfb_eq}, one can invert the PFB by solving the equation. There does exist a problem in trying to solve this equation; while $\mathbf{F}$ is an invertible matrix, $\mathbf{P}$ is not. $\mathbf{P}$ takes four frames of information to produce one and there is not enough information in the channelized frame to invert this. One approach to account for this issue is to use circulant or periodic boundaries.

\subsection{Circulant Polyphase Filterbank Inversion}\label{sec:pfb_appendix}
The main idea with circulant PFB inversion is to fill in the missing information with the opposite ends of the dataset, making the whole dataset cyclic or periodic, and by doing so having the eigenvectors of the PFB be approximated by Fourier modes such that a Fourier transform would diagonalize the circulant matrix and form the eigenbasis where we can invert the PFB from our dataset. Our circulant PFB matrix is $\mathbf{P_c}$ which for $\alpha =4$ is:

\begin{eqnarray}
    \mathbf{P_c} & = &
    \left(
    \begin{array}{cccccc}
     \mathbf{W^{(1)}} & \cdots & \mathbf{W^{(\alpha)}}  & \mathbf{0} &  \cdots  & \mathbf{0} \\
     \mathbf{0} & \ddots & \ddots & \ddots  & \ddots  & \vdots \\
     \vdots & \ddots & \ddots & \ddots  & \ddots  & \vdots \\
     \vdots & \ddots & \ddots & \ddots  & \ddots  & \mathbf{0} \\
     \mathbf{0} & \ddots & \ddots & \ddots  & \ddots &  \mathbf{W^{(\alpha)}} \\
     \mathbf{W^{(\alpha)}} & \ddots & \ddots & \ddots  & \ddots &  \mathbf{W^{(\alpha-1)}} \\
     \vdots & \ddots & \ddots & \ddots  & \ddots  & \vdots \\
    \mathbf{W^{(2)}} & \cdots & \mathbf{W^{(\alpha)}}  & \cdots   &  \mathbf{0}  & \mathbf{W^{(1)}} \\    

\end{array} \right) ~.
\end{eqnarray}

This should work well for constructing an inverse assuming the noise in the timestream is stochastically similar at both ends.

Next, we show how a Fourier transform diagonalizes the matrix $\mathbf{P_c}$. We define the circulant shift matrix $\mathbf{C}$ of size $M \times M$, where $M$ is the total number of frames in the recorded dataset,
\begin{eqnarray}
    \mathbf{C} & = &    
    \left(
    \begin{array}{cccccccc}
    0  & 1 & 0 & \cdots & \cdots & \cdots & \cdots & 0 \\
    0  & 0 & 1 & 0 & \cdots & \cdots & \cdots & 0 \\    
    \vdots  & \ddots  & \ddots  & \ddots &\ddots && & \vdots \\
    \vdots &   & \ddots & \ddots  & \ddots & \ddots &  & \vdots \\
    \vdots &  &  & \ddots & \ddots & \ddots & \ddots & 0 \\
    \vdots &  &  &  & \ddots & \ddots & 1 & 0 \\
    0 & \cdots &  \cdots & \cdots & \cdots & 0 & 0& 1  \\
    1 & 0 &  \cdots & \cdots & \cdots & \cdots & \cdots& 0  \\
    \end{array} \right)  ~.
\end{eqnarray}
Then we can construct $\mathbf{P_c}$ as
\begin{equation}
    \mathbf{P_c} =  \sum_{i=1}^{\alpha} (\mathbf{C})^{i-1} \otimes \mathbf{W^{(i)}} ~.
\end{equation}
Here $\otimes$ is the kronecker product that separates the dimensions of the frames and the sub frames, $(\mathbf{C})^{i-1}$ is the circulant shift matrix taken to the power of $i-1$, and $\mathbf{W^{(i)}}$ is the corresponding $i$-th tap.

The circulant shift matrices all share the same eigenvectors--complex exponentials. The DFT matrix $\mathbf{F^{(M)}}$, encoding a Fourier transform over the frame axis, diagonalizes $(\mathbf{C})^{i-1}$. Then we can define 
\begin{equation}
    \mathbf{F'} =  \mathbf{F^{(M)}} \otimes \mathbf{1} ~,
\end{equation}
where $\mathbf{F'}$ represents the fourier transform over the frame axis for our dataset with $M$ representing the total number of frames.\\
Starting from 
\begin{equation}
    \mathbf{P_c} =  \mathbf{F'}^{-1} \mathbf{F'} (\sum_{i=1}^{\alpha} (\mathbf{C})^{i-1} \otimes \mathbf{W^{(i)}}) \mathbf{F'}^{-1} \mathbf{F'} ~,
\end{equation}
we can find the circulant PFB is diagonalized as
\begin{equation}
    \mathbf{P_{c,d}} = \sum_{i=1}^{\alpha}  ( e^{-2\pi j \frac{(i-1)}{M} } \mathbf{1} )\otimes \mathbf{W^{(i)}} ~.
\end{equation}
Since $\mathbf{W^{(i)}}$ is also diagonal and the kronecker product of two diagonal matrices is a diagonal matrix, $\mathbf{P_c}$ can be diagonalized by $\mathbf{F'}$ where the diagonalized matrix is $ \mathbf{P_{c,d}}$.
From this, we can invert the PFB by

\begin{equation}\label{eq:circ_inv_eq}
    \mathbf{v_{rec}} =  \mathbf{F'}^{-1} \cdot ( \mathbf{F'} \cdot \mathbf{P_c} \cdot \mathbf{F'}^{-1} )^{-1} \cdot \mathbf{F'} \cdot \mathbf{v_{pfb}} ~.
\end{equation}
With Eq.~\ref{eq:circ_inv_eq}, we have established a procedure to invert the PFB and recover the voltage timestream from the baseband dump data at the system sampling rate of CHIME. We are now able to search for gravitational time delays at time resolutions of 1.25 ns and we have removed the PFB induced correlations in our system.

\newpage
\bibliographystyle{apsrev4-2}
\bibliography{bibfiles/frb,bibfiles/chime,bibfiles/macho,bibfiles/chimefrbpapers,bibfiles/lensing,bibfiles/scattering,bibfiles/stronglensing,bibfiles/lensingsearch,bibfiles/ligo}

\begin{thebibliography}{46}%
\makeatletter
\providecommand \@ifxundefined [1]{%
 \@ifx{#1\undefined}
}%
\providecommand \@ifnum [1]{%
 \ifnum #1\expandafter \@firstoftwo
 \else \expandafter \@secondoftwo
 \fi
}%
\providecommand \@ifx [1]{%
 \ifx #1\expandafter \@firstoftwo
 \else \expandafter \@secondoftwo
 \fi
}%
\providecommand \natexlab [1]{#1}%
\providecommand \enquote  [1]{``#1''}%
\providecommand \bibnamefont  [1]{#1}%
\providecommand \bibfnamefont [1]{#1}%
\providecommand \citenamefont [1]{#1}%
\providecommand \href@noop [0]{\@secondoftwo}%
\providecommand \href [0]{\begingroup \@sanitize@url \@href}%
\providecommand \@href[1]{\@@startlink{#1}\@@href}%
\providecommand \@@href[1]{\endgroup#1\@@endlink}%
\providecommand \@sanitize@url [0]{\catcode `\\12\catcode `\$12\catcode
  `\&12\catcode `\#12\catcode `\^12\catcode `\_12\catcode `\%12\relax}%
\providecommand \@@startlink[1]{}%
\providecommand \@@endlink[0]{}%
\providecommand \url  [0]{\begingroup\@sanitize@url \@url }%
\providecommand \@url [1]{\endgroup\@href {#1}{\urlprefix }}%
\providecommand \urlprefix  [0]{URL }%
\providecommand \Eprint [0]{\href }%
\providecommand \doibase [0]{https://doi.org/}%
\providecommand \selectlanguage [0]{\@gobble}%
\providecommand \bibinfo  [0]{\@secondoftwo}%
\providecommand \bibfield  [0]{\@secondoftwo}%
\providecommand \translation [1]{[#1]}%
\providecommand \BibitemOpen [0]{}%
\providecommand \bibitemStop [0]{}%
\providecommand \bibitemNoStop [0]{.\EOS\space}%
\providecommand \EOS [0]{\spacefactor3000\relax}%
\providecommand \BibitemShut  [1]{\csname bibitem#1\endcsname}%
\let\auto@bib@innerbib\@empty
\bibitem [{\citenamefont {{Leung}}\ and\ \citenamefont
  {{Kader}}(2022)}]{leung2022constraining}%
  \BibitemOpen
  \bibfield  {author} {\bibinfo {author} {\bibfnamefont {C.}~\bibnamefont
  {{Leung}}}\ and\ \bibinfo {author} {\bibfnamefont {Z.}~\bibnamefont
  {{Kader}}}} (\bibinfo {year} {2022}),\ \bibinfo {note}
  {submitted}\BibitemShut {NoStop}%
\bibitem [{\citenamefont {Abbott}\ \emph {et~al.}(2016)\citenamefont {Abbott},
  \citenamefont {Abbott}, \citenamefont {Abbott}, \citenamefont {Abernathy},
  \citenamefont {Acernese}, \citenamefont {Ackley}, \citenamefont {Adams},
  \citenamefont {Adams}, \citenamefont {Addesso}, \citenamefont {Adhikari},\
  and\ \citenamefont {et~al.}}]{abbott2016observation}%
  \BibitemOpen
  \bibfield  {author} {\bibinfo {author} {\bibfnamefont {B.~P.}\ \bibnamefont
  {Abbott}}, \bibinfo {author} {\bibfnamefont {R.}~\bibnamefont {Abbott}},
  \bibinfo {author} {\bibfnamefont {T.~D.}\ \bibnamefont {Abbott}}, \bibinfo
  {author} {\bibfnamefont {M.~R.}\ \bibnamefont {Abernathy}}, \bibinfo {author}
  {\bibfnamefont {F.}~\bibnamefont {Acernese}}, \bibinfo {author}
  {\bibfnamefont {K.}~\bibnamefont {Ackley}}, \bibinfo {author} {\bibfnamefont
  {C.}~\bibnamefont {Adams}}, \bibinfo {author} {\bibfnamefont
  {T.}~\bibnamefont {Adams}}, \bibinfo {author} {\bibfnamefont
  {P.}~\bibnamefont {Addesso}}, \bibinfo {author} {\bibfnamefont {R.~X.}\
  \bibnamefont {Adhikari}},\ and\ \bibinfo {author} {\bibnamefont {et~al.}}
  (\bibinfo {collaboration} {LIGO Scientific Collaboration and Virgo
  Collaboration}),\ }\href {https://doi.org/10.1103/PhysRevLett.116.061102}
  {\bibfield  {journal} {\bibinfo  {journal} {Phys. Rev. Lett.}\ }\textbf
  {\bibinfo {volume} {116}},\ \bibinfo {pages} {061102} (\bibinfo {year}
  {2016})}\BibitemShut {NoStop}%
\bibitem [{\citenamefont {Laha}(2020)}]{lens_pbh_constraints}%
  \BibitemOpen
  \bibfield  {author} {\bibinfo {author} {\bibfnamefont {R.}~\bibnamefont
  {Laha}},\ }\href {https://doi.org/10.1103/PhysRevD.102.023016} {\bibfield
  {journal} {\bibinfo  {journal} {Phys. Rev. D}\ }\textbf {\bibinfo {volume}
  {102}},\ \bibinfo {pages} {023016} (\bibinfo {year} {2020})}\BibitemShut
  {NoStop}%
\bibitem [{\citenamefont {{Cordes}}\ and\ \citenamefont
  {{Chatterjee}}(2019)}]{frb_review_cordes}%
  \BibitemOpen
  \bibfield  {author} {\bibinfo {author} {\bibfnamefont {J.~M.}\ \bibnamefont
  {{Cordes}}}\ and\ \bibinfo {author} {\bibfnamefont {S.}~\bibnamefont
  {{Chatterjee}}},\ }\href
  {https://doi.org/10.1146/annurev-astro-091918-104501} {\bibfield  {journal}
  {\bibinfo  {journal} {\araa}\ }\textbf {\bibinfo {volume} {57}},\ \bibinfo
  {pages} {417} (\bibinfo {year} {2019})},\ \Eprint
  {https://arxiv.org/abs/1906.05878} {arXiv:1906.05878 [astro-ph.HE]}
  \BibitemShut {NoStop}%
\bibitem [{\citenamefont {{Katz}}\ \emph {et~al.}(2020)\citenamefont {{Katz}},
  \citenamefont {{Kopp}}, \citenamefont {{Sibiryakov}},\ and\ \citenamefont
  {{Xue}}}]{Katz_paper}%
  \BibitemOpen
  \bibfield  {author} {\bibinfo {author} {\bibfnamefont {A.}~\bibnamefont
  {{Katz}}}, \bibinfo {author} {\bibfnamefont {J.}~\bibnamefont {{Kopp}}},
  \bibinfo {author} {\bibfnamefont {S.}~\bibnamefont {{Sibiryakov}}},\ and\
  \bibinfo {author} {\bibfnamefont {W.}~\bibnamefont {{Xue}}},\ }\href
  {https://doi.org/10.1093/mnras/staa1497} {\bibfield  {journal} {\bibinfo
  {journal} {\mnras}\ }\textbf {\bibinfo {volume} {496}},\ \bibinfo {pages}
  {564} (\bibinfo {year} {2020})},\ \Eprint {https://arxiv.org/abs/1912.07620}
  {arXiv:1912.07620 [astro-ph.CO]} \BibitemShut {NoStop}%
\bibitem [{\citenamefont {{Mu{\~n}oz}}\ \emph
  {et~al.}(2016{\natexlab{a}})\citenamefont {{Mu{\~n}oz}}, \citenamefont
  {{Kovetz}}, \citenamefont {{Dai}},\ and\ \citenamefont
  {{Kamionkowski}}}]{Munoz_paper}%
  \BibitemOpen
  \bibfield  {author} {\bibinfo {author} {\bibfnamefont {J.~B.}\ \bibnamefont
  {{Mu{\~n}oz}}}, \bibinfo {author} {\bibfnamefont {E.~D.}\ \bibnamefont
  {{Kovetz}}}, \bibinfo {author} {\bibfnamefont {L.}~\bibnamefont {{Dai}}},\
  and\ \bibinfo {author} {\bibfnamefont {M.}~\bibnamefont {{Kamionkowski}}},\
  }\href {https://doi.org/10.1103/PhysRevLett.117.091301} {\bibfield  {journal}
  {\bibinfo  {journal} {\prl}\ }\textbf {\bibinfo {volume} {117}},\ \bibinfo
  {eid} {091301} (\bibinfo {year} {2016}{\natexlab{a}})},\ \Eprint
  {https://arxiv.org/abs/1605.00008} {arXiv:1605.00008 [astro-ph.CO]}
  \BibitemShut {NoStop}%
\bibitem [{\citenamefont {{Eichler}}(2017)}]{eichler_frb}%
  \BibitemOpen
  \bibfield  {author} {\bibinfo {author} {\bibfnamefont {D.}~\bibnamefont
  {{Eichler}}},\ }\href {https://doi.org/10.3847/1538-4357/aa8b70} {\bibfield
  {journal} {\bibinfo  {journal} {\apj}\ }\textbf {\bibinfo {volume} {850}},\
  \bibinfo {eid} {159} (\bibinfo {year} {2017})},\ \Eprint
  {https://arxiv.org/abs/1711.04764} {arXiv:1711.04764 [astro-ph.HE]}
  \BibitemShut {NoStop}%
\bibitem [{\citenamefont {{Jow}}\ \emph
  {et~al.}(2020{\natexlab{a}})\citenamefont {{Jow}}, \citenamefont {{Foreman}},
  \citenamefont {{Pen}},\ and\ \citenamefont {{Zhu}}}]{FRB_wave_low}%
  \BibitemOpen
  \bibfield  {author} {\bibinfo {author} {\bibfnamefont {D.~L.}\ \bibnamefont
  {{Jow}}}, \bibinfo {author} {\bibfnamefont {S.}~\bibnamefont {{Foreman}}},
  \bibinfo {author} {\bibfnamefont {U.-L.}\ \bibnamefont {{Pen}}},\ and\
  \bibinfo {author} {\bibfnamefont {W.}~\bibnamefont {{Zhu}}},\ }\href
  {https://doi.org/10.1093/mnras/staa2230} {\bibfield  {journal} {\bibinfo
  {journal} {\mnras}\ }\textbf {\bibinfo {volume} {497}},\ \bibinfo {pages}
  {4956} (\bibinfo {year} {2020}{\natexlab{a}})},\ \Eprint
  {https://arxiv.org/abs/2002.01570} {arXiv:2002.01570 [astro-ph.HE]}
  \BibitemShut {NoStop}%
\bibitem [{\citenamefont {{Alcock}}\ \emph {et~al.}(2000)\citenamefont
  {{Alcock}}, \citenamefont {{Allsman}}, \citenamefont {{Alves}}, \citenamefont
  {{Axelrod}}, \citenamefont {{Becker}}, \citenamefont {{Bennett}},
  \citenamefont {{Cook}}, \citenamefont {{Dalal}}, \citenamefont {{Drake}},
  \citenamefont {{Freeman}}, \citenamefont {{Geha}}, \citenamefont {{Griest}},
  \citenamefont {{Lehner}}, \citenamefont {{Marshall}}, \citenamefont
  {{Minniti}}, \citenamefont {{Nelson}}, \citenamefont {{Peterson}},
  \citenamefont {{Popowski}}, \citenamefont {{Pratt}}, \citenamefont {{Quinn}},
  \citenamefont {{Stubbs}}, \citenamefont {{Sutherland}}, \citenamefont
  {{Tomaney}}, \citenamefont {{Vandehei}},\ and\ \citenamefont
  {{Welch}}}]{MACHO_project}%
  \BibitemOpen
  \bibfield  {author} {\bibinfo {author} {\bibfnamefont {C.}~\bibnamefont
  {{Alcock}}}, \bibinfo {author} {\bibfnamefont {R.~A.}\ \bibnamefont
  {{Allsman}}}, \bibinfo {author} {\bibfnamefont {D.~R.}\ \bibnamefont
  {{Alves}}}, \bibinfo {author} {\bibfnamefont {T.~S.}\ \bibnamefont
  {{Axelrod}}}, \bibinfo {author} {\bibfnamefont {A.~C.}\ \bibnamefont
  {{Becker}}}, \bibinfo {author} {\bibfnamefont {D.~P.}\ \bibnamefont
  {{Bennett}}}, \bibinfo {author} {\bibfnamefont {K.~H.}\ \bibnamefont
  {{Cook}}}, \bibinfo {author} {\bibfnamefont {N.}~\bibnamefont {{Dalal}}},
  \bibinfo {author} {\bibfnamefont {A.~J.}\ \bibnamefont {{Drake}}}, \bibinfo
  {author} {\bibfnamefont {K.~C.}\ \bibnamefont {{Freeman}}}, \bibinfo {author}
  {\bibfnamefont {M.}~\bibnamefont {{Geha}}}, \bibinfo {author} {\bibfnamefont
  {K.}~\bibnamefont {{Griest}}}, \bibinfo {author} {\bibfnamefont {M.~J.}\
  \bibnamefont {{Lehner}}}, \bibinfo {author} {\bibfnamefont {S.~L.}\
  \bibnamefont {{Marshall}}}, \bibinfo {author} {\bibfnamefont
  {D.}~\bibnamefont {{Minniti}}}, \bibinfo {author} {\bibfnamefont {C.~A.}\
  \bibnamefont {{Nelson}}}, \bibinfo {author} {\bibfnamefont {B.~A.}\
  \bibnamefont {{Peterson}}}, \bibinfo {author} {\bibfnamefont
  {P.}~\bibnamefont {{Popowski}}}, \bibinfo {author} {\bibfnamefont {M.~R.}\
  \bibnamefont {{Pratt}}}, \bibinfo {author} {\bibfnamefont {P.~J.}\
  \bibnamefont {{Quinn}}}, \bibinfo {author} {\bibfnamefont {C.~W.}\
  \bibnamefont {{Stubbs}}}, \bibinfo {author} {\bibfnamefont {W.}~\bibnamefont
  {{Sutherland}}}, \bibinfo {author} {\bibfnamefont {A.~B.}\ \bibnamefont
  {{Tomaney}}}, \bibinfo {author} {\bibfnamefont {T.}~\bibnamefont
  {{Vandehei}}},\ and\ \bibinfo {author} {\bibfnamefont {D.}~\bibnamefont
  {{Welch}}},\ }\href {https://doi.org/10.1086/309512} {\bibfield  {journal}
  {\bibinfo  {journal} {\apj}\ }\textbf {\bibinfo {volume} {542}},\ \bibinfo
  {pages} {281} (\bibinfo {year} {2000})},\ \Eprint
  {https://arxiv.org/abs/astro-ph/0001272} {arXiv:astro-ph/0001272 [astro-ph]}
  \BibitemShut {NoStop}%
\bibitem [{\citenamefont {{Tisserand}}\ \emph {et~al.}(2007)\citenamefont
  {{Tisserand}}, \citenamefont {{Le Guillou}}, \citenamefont {{Afonso}},
  \citenamefont {{Albert}}, \citenamefont {{Andersen}}, \citenamefont
  {{Ansari}}, \citenamefont {{Aubourg}}, \citenamefont {{Bareyre}},
  \citenamefont {{Beaulieu}}, \citenamefont {{Charlot}}, \citenamefont
  {{Coutures}}, \citenamefont {{Ferlet}}, \citenamefont {{Fouqu{\'e}}},
  \citenamefont {{Glicenstein}}, \citenamefont {{Goldman}}, \citenamefont
  {{Gould}}, \citenamefont {{Graff}}, \citenamefont {{Gros}}, \citenamefont
  {{Haissinski}}, \citenamefont {{Hamadache}}, \citenamefont {{de Kat}},
  \citenamefont {{Lasserre}}, \citenamefont {{Lesquoy}}, \citenamefont
  {{Loup}}, \citenamefont {{Magneville}}, \citenamefont {{Marquette}},
  \citenamefont {{Maurice}}, \citenamefont {{Maury}}, \citenamefont
  {{Milsztajn}}, \citenamefont {{Moniez}}, \citenamefont
  {{Palanque-Delabrouille}}, \citenamefont {{Perdereau}}, \citenamefont
  {{Rahal}}, \citenamefont {{Rich}}, \citenamefont {{Spiro}}, \citenamefont
  {{Vidal-Madjar}}, \citenamefont {{Vigroux}}, \citenamefont {{Zylberajch}},\
  and\ \citenamefont {{EROS-2 Collaboration}}}]{EROS_2}%
  \BibitemOpen
  \bibfield  {author} {\bibinfo {author} {\bibfnamefont {P.}~\bibnamefont
  {{Tisserand}}}, \bibinfo {author} {\bibfnamefont {L.}~\bibnamefont {{Le
  Guillou}}}, \bibinfo {author} {\bibfnamefont {C.}~\bibnamefont {{Afonso}}},
  \bibinfo {author} {\bibfnamefont {J.~N.}\ \bibnamefont {{Albert}}}, \bibinfo
  {author} {\bibfnamefont {J.}~\bibnamefont {{Andersen}}}, \bibinfo {author}
  {\bibfnamefont {R.}~\bibnamefont {{Ansari}}}, \bibinfo {author}
  {\bibfnamefont {{\'E}.}~\bibnamefont {{Aubourg}}}, \bibinfo {author}
  {\bibfnamefont {P.}~\bibnamefont {{Bareyre}}}, \bibinfo {author}
  {\bibfnamefont {J.~P.}\ \bibnamefont {{Beaulieu}}}, \bibinfo {author}
  {\bibfnamefont {X.}~\bibnamefont {{Charlot}}}, \bibinfo {author}
  {\bibfnamefont {C.}~\bibnamefont {{Coutures}}}, \bibinfo {author}
  {\bibfnamefont {R.}~\bibnamefont {{Ferlet}}}, \bibinfo {author}
  {\bibfnamefont {P.}~\bibnamefont {{Fouqu{\'e}}}}, \bibinfo {author}
  {\bibfnamefont {J.~F.}\ \bibnamefont {{Glicenstein}}}, \bibinfo {author}
  {\bibfnamefont {B.}~\bibnamefont {{Goldman}}}, \bibinfo {author}
  {\bibfnamefont {A.}~\bibnamefont {{Gould}}}, \bibinfo {author} {\bibfnamefont
  {D.}~\bibnamefont {{Graff}}}, \bibinfo {author} {\bibfnamefont
  {M.}~\bibnamefont {{Gros}}}, \bibinfo {author} {\bibfnamefont
  {J.}~\bibnamefont {{Haissinski}}}, \bibinfo {author} {\bibfnamefont
  {C.}~\bibnamefont {{Hamadache}}}, \bibinfo {author} {\bibfnamefont
  {J.}~\bibnamefont {{de Kat}}}, \bibinfo {author} {\bibfnamefont
  {T.}~\bibnamefont {{Lasserre}}}, \bibinfo {author} {\bibfnamefont
  {{\'E}.}~\bibnamefont {{Lesquoy}}}, \bibinfo {author} {\bibfnamefont
  {C.}~\bibnamefont {{Loup}}}, \bibinfo {author} {\bibfnamefont
  {C.}~\bibnamefont {{Magneville}}}, \bibinfo {author} {\bibfnamefont {J.~B.}\
  \bibnamefont {{Marquette}}}, \bibinfo {author} {\bibfnamefont
  {{\'E}.}~\bibnamefont {{Maurice}}}, \bibinfo {author} {\bibfnamefont
  {A.}~\bibnamefont {{Maury}}}, \bibinfo {author} {\bibfnamefont
  {A.}~\bibnamefont {{Milsztajn}}}, \bibinfo {author} {\bibfnamefont
  {M.}~\bibnamefont {{Moniez}}}, \bibinfo {author} {\bibfnamefont
  {N.}~\bibnamefont {{Palanque-Delabrouille}}}, \bibinfo {author}
  {\bibfnamefont {O.}~\bibnamefont {{Perdereau}}}, \bibinfo {author}
  {\bibfnamefont {Y.~R.}\ \bibnamefont {{Rahal}}}, \bibinfo {author}
  {\bibfnamefont {J.}~\bibnamefont {{Rich}}}, \bibinfo {author} {\bibfnamefont
  {M.}~\bibnamefont {{Spiro}}}, \bibinfo {author} {\bibfnamefont
  {A.}~\bibnamefont {{Vidal-Madjar}}}, \bibinfo {author} {\bibfnamefont
  {L.}~\bibnamefont {{Vigroux}}}, \bibinfo {author} {\bibfnamefont
  {S.}~\bibnamefont {{Zylberajch}}},\ and\ \bibinfo {author} {\bibnamefont
  {{EROS-2 Collaboration}}},\ }\href
  {https://doi.org/10.1051/0004-6361:20066017} {\bibfield  {journal} {\bibinfo
  {journal} {\aap}\ }\textbf {\bibinfo {volume} {469}},\ \bibinfo {pages} {387}
  (\bibinfo {year} {2007})},\ \Eprint {https://arxiv.org/abs/astro-ph/0607207}
  {arXiv:astro-ph/0607207 [astro-ph]} \BibitemShut {NoStop}%
\bibitem [{\citenamefont {{Chatterjee}}\ \emph {et~al.}(2017)\citenamefont
  {{Chatterjee}}, \citenamefont {{Law}}, \citenamefont {{Wharton}},
  \citenamefont {{Burke-Spolaor}}, \citenamefont {{Hessels}}, \citenamefont
  {{Bower}}, \citenamefont {{Cordes}}, \citenamefont {{Tendulkar}},
  \citenamefont {{Bassa}}, \citenamefont {{Demorest}}, \citenamefont
  {{Butler}}, \citenamefont {{Seymour}}, \citenamefont {{Scholz}},
  \citenamefont {{Abruzzo}}, \citenamefont {{Bogdanov}}, \citenamefont
  {{Kaspi}}, \citenamefont {{Keimpema}}, \citenamefont {{Lazio}}, \citenamefont
  {{Marcote}}, \citenamefont {{McLaughlin}}, \citenamefont {{Paragi}},
  \citenamefont {{Ransom}}, \citenamefont {{Rupen}}, \citenamefont
  {{Spitler}},\ and\ \citenamefont {{van Langevelde}}}]{2017Natur.541...58C}%
  \BibitemOpen
  \bibfield  {author} {\bibinfo {author} {\bibfnamefont {S.}~\bibnamefont
  {{Chatterjee}}}, \bibinfo {author} {\bibfnamefont {C.~J.}\ \bibnamefont
  {{Law}}}, \bibinfo {author} {\bibfnamefont {R.~S.}\ \bibnamefont
  {{Wharton}}}, \bibinfo {author} {\bibfnamefont {S.}~\bibnamefont
  {{Burke-Spolaor}}}, \bibinfo {author} {\bibfnamefont {J.~W.~T.}\ \bibnamefont
  {{Hessels}}}, \bibinfo {author} {\bibfnamefont {G.~C.}\ \bibnamefont
  {{Bower}}}, \bibinfo {author} {\bibfnamefont {J.~M.}\ \bibnamefont
  {{Cordes}}}, \bibinfo {author} {\bibfnamefont {S.~P.}\ \bibnamefont
  {{Tendulkar}}}, \bibinfo {author} {\bibfnamefont {C.~G.}\ \bibnamefont
  {{Bassa}}}, \bibinfo {author} {\bibfnamefont {P.}~\bibnamefont {{Demorest}}},
  \bibinfo {author} {\bibfnamefont {B.~J.}\ \bibnamefont {{Butler}}}, \bibinfo
  {author} {\bibfnamefont {A.}~\bibnamefont {{Seymour}}}, \bibinfo {author}
  {\bibfnamefont {P.}~\bibnamefont {{Scholz}}}, \bibinfo {author}
  {\bibfnamefont {M.~W.}\ \bibnamefont {{Abruzzo}}}, \bibinfo {author}
  {\bibfnamefont {S.}~\bibnamefont {{Bogdanov}}}, \bibinfo {author}
  {\bibfnamefont {V.~M.}\ \bibnamefont {{Kaspi}}}, \bibinfo {author}
  {\bibfnamefont {A.}~\bibnamefont {{Keimpema}}}, \bibinfo {author}
  {\bibfnamefont {T.~J.~W.}\ \bibnamefont {{Lazio}}}, \bibinfo {author}
  {\bibfnamefont {B.}~\bibnamefont {{Marcote}}}, \bibinfo {author}
  {\bibfnamefont {M.~A.}\ \bibnamefont {{McLaughlin}}}, \bibinfo {author}
  {\bibfnamefont {Z.}~\bibnamefont {{Paragi}}}, \bibinfo {author}
  {\bibfnamefont {S.~M.}\ \bibnamefont {{Ransom}}}, \bibinfo {author}
  {\bibfnamefont {M.}~\bibnamefont {{Rupen}}}, \bibinfo {author} {\bibfnamefont
  {L.~G.}\ \bibnamefont {{Spitler}}},\ and\ \bibinfo {author} {\bibfnamefont
  {H.~J.}\ \bibnamefont {{van Langevelde}}},\ }\href
  {https://doi.org/10.1038/nature20797} {\bibfield  {journal} {\bibinfo
  {journal} {\nat}\ }\textbf {\bibinfo {volume} {541}},\ \bibinfo {pages} {58}
  (\bibinfo {year} {2017})},\ \Eprint {https://arxiv.org/abs/1701.01098}
  {arXiv:1701.01098 [astro-ph.HE]} \BibitemShut {NoStop}%
\bibitem [{\citenamefont {{Ravi}}\ \emph {et~al.}(2019)\citenamefont {{Ravi}},
  \citenamefont {{Catha}}, \citenamefont {{D'Addario}}, \citenamefont
  {{Djorgovski}}, \citenamefont {{Hallinan}}, \citenamefont {{Hobbs}},
  \citenamefont {{Kocz}}, \citenamefont {{Kulkarni}}, \citenamefont {{Shi}},
  \citenamefont {{Vedantham}}, \citenamefont {{Weinreb}},\ and\ \citenamefont
  {{Woody}}}]{ravi_v_localized}%
  \BibitemOpen
  \bibfield  {author} {\bibinfo {author} {\bibfnamefont {V.}~\bibnamefont
  {{Ravi}}}, \bibinfo {author} {\bibfnamefont {M.}~\bibnamefont {{Catha}}},
  \bibinfo {author} {\bibfnamefont {L.}~\bibnamefont {{D'Addario}}}, \bibinfo
  {author} {\bibfnamefont {S.~G.}\ \bibnamefont {{Djorgovski}}}, \bibinfo
  {author} {\bibfnamefont {G.}~\bibnamefont {{Hallinan}}}, \bibinfo {author}
  {\bibfnamefont {R.}~\bibnamefont {{Hobbs}}}, \bibinfo {author} {\bibfnamefont
  {J.}~\bibnamefont {{Kocz}}}, \bibinfo {author} {\bibfnamefont {S.~R.}\
  \bibnamefont {{Kulkarni}}}, \bibinfo {author} {\bibfnamefont
  {J.}~\bibnamefont {{Shi}}}, \bibinfo {author} {\bibfnamefont {H.~K.}\
  \bibnamefont {{Vedantham}}}, \bibinfo {author} {\bibfnamefont
  {S.}~\bibnamefont {{Weinreb}}},\ and\ \bibinfo {author} {\bibfnamefont
  {D.~P.}\ \bibnamefont {{Woody}}},\ }\href
  {https://doi.org/10.1038/s41586-019-1389-7} {\bibfield  {journal} {\bibinfo
  {journal} {\nat}\ }\textbf {\bibinfo {volume} {572}},\ \bibinfo {pages} {352}
  (\bibinfo {year} {2019})},\ \Eprint {https://arxiv.org/abs/1907.01542}
  {arXiv:1907.01542 [astro-ph.HE]} \BibitemShut {NoStop}%
\bibitem [{\citenamefont {{Macquart}}\ \emph {et~al.}(2020)\citenamefont
  {{Macquart}}, \citenamefont {{Prochaska}}, \citenamefont {{McQuinn}},
  \citenamefont {{Bannister}}, \citenamefont {{Bhandari}}, \citenamefont
  {{Day}}, \citenamefont {{Deller}}, \citenamefont {{Ekers}}, \citenamefont
  {{James}}, \citenamefont {{Marnoch}}, \citenamefont {{Os{\l}owski}},
  \citenamefont {{Phillips}}, \citenamefont {{Ryder}}, \citenamefont {{Scott}},
  \citenamefont {{Shannon}},\ and\ \citenamefont
  {{Tejos}}}]{macquart2020census}%
  \BibitemOpen
  \bibfield  {author} {\bibinfo {author} {\bibfnamefont {J.~P.}\ \bibnamefont
  {{Macquart}}}, \bibinfo {author} {\bibfnamefont {J.~X.}\ \bibnamefont
  {{Prochaska}}}, \bibinfo {author} {\bibfnamefont {M.}~\bibnamefont
  {{McQuinn}}}, \bibinfo {author} {\bibfnamefont {K.~W.}\ \bibnamefont
  {{Bannister}}}, \bibinfo {author} {\bibfnamefont {S.}~\bibnamefont
  {{Bhandari}}}, \bibinfo {author} {\bibfnamefont {C.~K.}\ \bibnamefont
  {{Day}}}, \bibinfo {author} {\bibfnamefont {A.~T.}\ \bibnamefont {{Deller}}},
  \bibinfo {author} {\bibfnamefont {R.~D.}\ \bibnamefont {{Ekers}}}, \bibinfo
  {author} {\bibfnamefont {C.~W.}\ \bibnamefont {{James}}}, \bibinfo {author}
  {\bibfnamefont {L.}~\bibnamefont {{Marnoch}}}, \bibinfo {author}
  {\bibfnamefont {S.}~\bibnamefont {{Os{\l}owski}}}, \bibinfo {author}
  {\bibfnamefont {C.}~\bibnamefont {{Phillips}}}, \bibinfo {author}
  {\bibfnamefont {S.~D.}\ \bibnamefont {{Ryder}}}, \bibinfo {author}
  {\bibfnamefont {D.~R.}\ \bibnamefont {{Scott}}}, \bibinfo {author}
  {\bibfnamefont {R.~M.}\ \bibnamefont {{Shannon}}},\ and\ \bibinfo {author}
  {\bibfnamefont {N.}~\bibnamefont {{Tejos}}},\ }\href
  {https://doi.org/10.1038/s41586-020-2300-2} {\bibfield  {journal} {\bibinfo
  {journal} {\nat}\ }\textbf {\bibinfo {volume} {581}},\ \bibinfo {pages} {391}
  (\bibinfo {year} {2020})},\ \Eprint {https://arxiv.org/abs/2005.13161}
  {arXiv:2005.13161 [astro-ph.CO]} \BibitemShut {NoStop}%
\bibitem [{\citenamefont {{Cho}}\ \emph
  {et~al.}(2020{\natexlab{a}})\citenamefont {{Cho}}, \citenamefont
  {{Macquart}}, \citenamefont {{Shannon}}, \citenamefont {{Deller}},
  \citenamefont {{Morrison}}, \citenamefont {{Ekers}}, \citenamefont
  {{Bannister}}, \citenamefont {{Farah}}, \citenamefont {{Qiu}}, \citenamefont
  {{Sammons}}, \citenamefont {{Bailes}}, \citenamefont {{Bhandari}},
  \citenamefont {{Day}}, \citenamefont {{James}}, \citenamefont {{Phillips}},
  \citenamefont {{Prochaska}},\ and\ \citenamefont {{Tuthill}}}]{cho2020}%
  \BibitemOpen
  \bibfield  {author} {\bibinfo {author} {\bibfnamefont {H.}~\bibnamefont
  {{Cho}}}, \bibinfo {author} {\bibfnamefont {J.-P.}\ \bibnamefont
  {{Macquart}}}, \bibinfo {author} {\bibfnamefont {R.~M.}\ \bibnamefont
  {{Shannon}}}, \bibinfo {author} {\bibfnamefont {A.~T.}\ \bibnamefont
  {{Deller}}}, \bibinfo {author} {\bibfnamefont {I.~S.}\ \bibnamefont
  {{Morrison}}}, \bibinfo {author} {\bibfnamefont {R.~D.}\ \bibnamefont
  {{Ekers}}}, \bibinfo {author} {\bibfnamefont {K.~W.}\ \bibnamefont
  {{Bannister}}}, \bibinfo {author} {\bibfnamefont {W.}~\bibnamefont
  {{Farah}}}, \bibinfo {author} {\bibfnamefont {H.}~\bibnamefont {{Qiu}}},
  \bibinfo {author} {\bibfnamefont {M.~W.}\ \bibnamefont {{Sammons}}}, \bibinfo
  {author} {\bibfnamefont {M.}~\bibnamefont {{Bailes}}}, \bibinfo {author}
  {\bibfnamefont {S.}~\bibnamefont {{Bhandari}}}, \bibinfo {author}
  {\bibfnamefont {C.~K.}\ \bibnamefont {{Day}}}, \bibinfo {author}
  {\bibfnamefont {C.~W.}\ \bibnamefont {{James}}}, \bibinfo {author}
  {\bibfnamefont {C.~J.}\ \bibnamefont {{Phillips}}}, \bibinfo {author}
  {\bibfnamefont {J.~X.}\ \bibnamefont {{Prochaska}}},\ and\ \bibinfo {author}
  {\bibfnamefont {J.}~\bibnamefont {{Tuthill}}},\ }\href
  {https://doi.org/10.3847/2041-8213/ab7824} {\bibfield  {journal} {\bibinfo
  {journal} {\apjl}\ }\textbf {\bibinfo {volume} {891}},\ \bibinfo {eid} {L38}
  (\bibinfo {year} {2020}{\natexlab{a}})},\ \Eprint
  {https://arxiv.org/abs/2002.12539} {arXiv:2002.12539 [astro-ph.HE]}
  \BibitemShut {NoStop}%
\bibitem [{\citenamefont {{Farah}}\ \emph
  {et~al.}(2019{\natexlab{a}})\citenamefont {{Farah}}, \citenamefont {{Flynn}},
  \citenamefont {{Bailes}}, \citenamefont {{Jameson}}, \citenamefont
  {{Bateman}}, \citenamefont {{Campbell-Wilson}}, \citenamefont {{Day}},
  \citenamefont {{Deller}}, \citenamefont {{Green}}, \citenamefont {{Gupta}},
  \citenamefont {{Hunstead}}, \citenamefont {{Lower}}, \citenamefont
  {{Os{\l}owski}}, \citenamefont {{Parthasarathy}}, \citenamefont {{Price}},
  \citenamefont {{Ravi}}, \citenamefont {{Shannon}}, \citenamefont
  {{Sutherland}}, \citenamefont {{Temby}}, \citenamefont {{Krishnan}},
  \citenamefont {{Caleb}}, \citenamefont {{Chang}}, \citenamefont {{Cruces}},
  \citenamefont {{Roy}}, \citenamefont {{Morello}}, \citenamefont {{Onken}},
  \citenamefont {{Stappers}}, \citenamefont {{Webb}},\ and\ \citenamefont
  {{Wolf}}}]{Farah2019}%
  \BibitemOpen
  \bibfield  {author} {\bibinfo {author} {\bibfnamefont {W.}~\bibnamefont
  {{Farah}}}, \bibinfo {author} {\bibfnamefont {C.}~\bibnamefont {{Flynn}}},
  \bibinfo {author} {\bibfnamefont {M.}~\bibnamefont {{Bailes}}}, \bibinfo
  {author} {\bibfnamefont {A.}~\bibnamefont {{Jameson}}}, \bibinfo {author}
  {\bibfnamefont {T.}~\bibnamefont {{Bateman}}}, \bibinfo {author}
  {\bibfnamefont {D.}~\bibnamefont {{Campbell-Wilson}}}, \bibinfo {author}
  {\bibfnamefont {C.~K.}\ \bibnamefont {{Day}}}, \bibinfo {author}
  {\bibfnamefont {A.~T.}\ \bibnamefont {{Deller}}}, \bibinfo {author}
  {\bibfnamefont {A.~J.}\ \bibnamefont {{Green}}}, \bibinfo {author}
  {\bibfnamefont {V.}~\bibnamefont {{Gupta}}}, \bibinfo {author} {\bibfnamefont
  {R.}~\bibnamefont {{Hunstead}}}, \bibinfo {author} {\bibfnamefont {M.~E.}\
  \bibnamefont {{Lower}}}, \bibinfo {author} {\bibfnamefont {S.}~\bibnamefont
  {{Os{\l}owski}}}, \bibinfo {author} {\bibfnamefont {A.}~\bibnamefont
  {{Parthasarathy}}}, \bibinfo {author} {\bibfnamefont {D.~C.}\ \bibnamefont
  {{Price}}}, \bibinfo {author} {\bibfnamefont {V.}~\bibnamefont {{Ravi}}},
  \bibinfo {author} {\bibfnamefont {R.~M.}\ \bibnamefont {{Shannon}}}, \bibinfo
  {author} {\bibfnamefont {A.}~\bibnamefont {{Sutherland}}}, \bibinfo {author}
  {\bibfnamefont {D.}~\bibnamefont {{Temby}}}, \bibinfo {author} {\bibfnamefont
  {V.~V.}\ \bibnamefont {{Krishnan}}}, \bibinfo {author} {\bibfnamefont
  {M.}~\bibnamefont {{Caleb}}}, \bibinfo {author} {\bibfnamefont {S.~W.}\
  \bibnamefont {{Chang}}}, \bibinfo {author} {\bibfnamefont {M.}~\bibnamefont
  {{Cruces}}}, \bibinfo {author} {\bibfnamefont {J.}~\bibnamefont {{Roy}}},
  \bibinfo {author} {\bibfnamefont {V.}~\bibnamefont {{Morello}}}, \bibinfo
  {author} {\bibfnamefont {C.~A.}\ \bibnamefont {{Onken}}}, \bibinfo {author}
  {\bibfnamefont {B.~W.}\ \bibnamefont {{Stappers}}}, \bibinfo {author}
  {\bibfnamefont {S.}~\bibnamefont {{Webb}}},\ and\ \bibinfo {author}
  {\bibfnamefont {C.}~\bibnamefont {{Wolf}}},\ }\href
  {https://doi.org/10.1093/mnras/stz1748} {\bibfield  {journal} {\bibinfo
  {journal} {\mnras}\ }\textbf {\bibinfo {volume} {488}},\ \bibinfo {pages}
  {2989} (\bibinfo {year} {2019}{\natexlab{a}})},\ \Eprint
  {https://arxiv.org/abs/1905.02293} {arXiv:1905.02293 [astro-ph.HE]}
  \BibitemShut {NoStop}%
\bibitem [{\citenamefont {{CHIME/FRB Collaboration}}\ \emph
  {et~al.}(2018{\natexlab{a}})\citenamefont {{CHIME/FRB Collaboration}},
  \citenamefont {{Amiri}}, \citenamefont {{Bandura}}, \citenamefont {{Berger}},
  \citenamefont {{Bhardwaj}}, \citenamefont {{Boyce}}, \citenamefont {{Boyle}},
  \citenamefont {{Brar}}, \citenamefont {{Burhanpurkar}}, \citenamefont
  {{Chawla}}, \citenamefont {{Chowdhury}}, \citenamefont {{Cliche}},
  \citenamefont {{Cranmer}}, \citenamefont {{Cubranic}}, \citenamefont
  {{Deng}}, \citenamefont {{Denman}}, \citenamefont {{Dobbs}}, \citenamefont
  {{Fandino}}, \citenamefont {{Fonseca}}, \citenamefont {{Gaensler}},
  \citenamefont {{Giri}}, \citenamefont {{Gilbert}}, \citenamefont {{Good}},
  \citenamefont {{Guliani}}, \citenamefont {{Halpern}}, \citenamefont
  {{Hinshaw}}, \citenamefont {{H{\"o}fer}}, \citenamefont {{Josephy}},
  \citenamefont {{Kaspi}}, \citenamefont {{Landecker}}, \citenamefont {{Lang}},
  \citenamefont {{Liao}}, \citenamefont {{Masui}}, \citenamefont
  {{Mena-Parra}}, \citenamefont {{Naidu}}, \citenamefont {{Newburgh}},
  \citenamefont {{Ng}}, \citenamefont {{Patel}}, \citenamefont {{Pen}},
  \citenamefont {{Pinsonneault-Marotte}}, \citenamefont {{Pleunis}},
  \citenamefont {{Rafiei Ravandi}}, \citenamefont {{Ransom}}, \citenamefont
  {{Renard}}, \citenamefont {{Scholz}}, \citenamefont {{Sigurdson}},
  \citenamefont {{Siegel}}, \citenamefont {{Smith}}, \citenamefont {{Stairs}},
  \citenamefont {{Tendulkar}}, \citenamefont {{Vanderlinde}},\ and\
  \citenamefont {{Wiebe}}}]{chimefrb_overview}%
  \BibitemOpen
  \bibfield  {author} {\bibinfo {author} {\bibnamefont {{CHIME/FRB
  Collaboration}}}, \bibinfo {author} {\bibfnamefont {M.}~\bibnamefont
  {{Amiri}}}, \bibinfo {author} {\bibfnamefont {K.}~\bibnamefont {{Bandura}}},
  \bibinfo {author} {\bibfnamefont {P.}~\bibnamefont {{Berger}}}, \bibinfo
  {author} {\bibfnamefont {M.}~\bibnamefont {{Bhardwaj}}}, \bibinfo {author}
  {\bibfnamefont {M.~M.}\ \bibnamefont {{Boyce}}}, \bibinfo {author}
  {\bibfnamefont {P.~J.}\ \bibnamefont {{Boyle}}}, \bibinfo {author}
  {\bibfnamefont {C.}~\bibnamefont {{Brar}}}, \bibinfo {author} {\bibfnamefont
  {M.}~\bibnamefont {{Burhanpurkar}}}, \bibinfo {author} {\bibfnamefont
  {P.}~\bibnamefont {{Chawla}}}, \bibinfo {author} {\bibfnamefont
  {J.}~\bibnamefont {{Chowdhury}}}, \bibinfo {author} {\bibfnamefont {J.~F.}\
  \bibnamefont {{Cliche}}}, \bibinfo {author} {\bibfnamefont {M.~D.}\
  \bibnamefont {{Cranmer}}}, \bibinfo {author} {\bibfnamefont {D.}~\bibnamefont
  {{Cubranic}}}, \bibinfo {author} {\bibfnamefont {M.}~\bibnamefont {{Deng}}},
  \bibinfo {author} {\bibfnamefont {N.}~\bibnamefont {{Denman}}}, \bibinfo
  {author} {\bibfnamefont {M.}~\bibnamefont {{Dobbs}}}, \bibinfo {author}
  {\bibfnamefont {M.}~\bibnamefont {{Fandino}}}, \bibinfo {author}
  {\bibfnamefont {E.}~\bibnamefont {{Fonseca}}}, \bibinfo {author}
  {\bibfnamefont {B.~M.}\ \bibnamefont {{Gaensler}}}, \bibinfo {author}
  {\bibfnamefont {U.}~\bibnamefont {{Giri}}}, \bibinfo {author} {\bibfnamefont
  {A.~J.}\ \bibnamefont {{Gilbert}}}, \bibinfo {author} {\bibfnamefont {D.~C.}\
  \bibnamefont {{Good}}}, \bibinfo {author} {\bibfnamefont {S.}~\bibnamefont
  {{Guliani}}}, \bibinfo {author} {\bibfnamefont {M.}~\bibnamefont
  {{Halpern}}}, \bibinfo {author} {\bibfnamefont {G.}~\bibnamefont
  {{Hinshaw}}}, \bibinfo {author} {\bibfnamefont {C.}~\bibnamefont
  {{H{\"o}fer}}}, \bibinfo {author} {\bibfnamefont {A.}~\bibnamefont
  {{Josephy}}}, \bibinfo {author} {\bibfnamefont {V.~M.}\ \bibnamefont
  {{Kaspi}}}, \bibinfo {author} {\bibfnamefont {T.~L.}\ \bibnamefont
  {{Landecker}}}, \bibinfo {author} {\bibfnamefont {D.}~\bibnamefont {{Lang}}},
  \bibinfo {author} {\bibfnamefont {H.}~\bibnamefont {{Liao}}}, \bibinfo
  {author} {\bibfnamefont {K.~W.}\ \bibnamefont {{Masui}}}, \bibinfo {author}
  {\bibfnamefont {J.}~\bibnamefont {{Mena-Parra}}}, \bibinfo {author}
  {\bibfnamefont {A.}~\bibnamefont {{Naidu}}}, \bibinfo {author} {\bibfnamefont
  {L.~B.}\ \bibnamefont {{Newburgh}}}, \bibinfo {author} {\bibfnamefont
  {C.}~\bibnamefont {{Ng}}}, \bibinfo {author} {\bibfnamefont {C.}~\bibnamefont
  {{Patel}}}, \bibinfo {author} {\bibfnamefont {U.~L.}\ \bibnamefont {{Pen}}},
  \bibinfo {author} {\bibfnamefont {T.}~\bibnamefont {{Pinsonneault-Marotte}}},
  \bibinfo {author} {\bibfnamefont {Z.}~\bibnamefont {{Pleunis}}}, \bibinfo
  {author} {\bibfnamefont {M.}~\bibnamefont {{Rafiei Ravandi}}}, \bibinfo
  {author} {\bibfnamefont {S.~M.}\ \bibnamefont {{Ransom}}}, \bibinfo {author}
  {\bibfnamefont {A.}~\bibnamefont {{Renard}}}, \bibinfo {author}
  {\bibfnamefont {P.}~\bibnamefont {{Scholz}}}, \bibinfo {author}
  {\bibfnamefont {K.}~\bibnamefont {{Sigurdson}}}, \bibinfo {author}
  {\bibfnamefont {S.~R.}\ \bibnamefont {{Siegel}}}, \bibinfo {author}
  {\bibfnamefont {K.~M.}\ \bibnamefont {{Smith}}}, \bibinfo {author}
  {\bibfnamefont {I.~H.}\ \bibnamefont {{Stairs}}}, \bibinfo {author}
  {\bibfnamefont {S.~P.}\ \bibnamefont {{Tendulkar}}}, \bibinfo {author}
  {\bibfnamefont {K.}~\bibnamefont {{Vanderlinde}}},\ and\ \bibinfo {author}
  {\bibfnamefont {D.~V.}\ \bibnamefont {{Wiebe}}},\ }\href
  {https://doi.org/10.3847/1538-4357/aad188} {\bibfield  {journal} {\bibinfo
  {journal} {\apj}\ }\textbf {\bibinfo {volume} {863}},\ \bibinfo {eid} {48}
  (\bibinfo {year} {2018}{\natexlab{a}})},\ \Eprint
  {https://arxiv.org/abs/1803.11235} {arXiv:1803.11235 [astro-ph.IM]}
  \BibitemShut {NoStop}%
\bibitem [{\citenamefont {{Mu{\~n}oz}}\ \emph
  {et~al.}(2016{\natexlab{b}})\citenamefont {{Mu{\~n}oz}}, \citenamefont
  {{Kovetz}}, \citenamefont {{Dai}},\ and\ \citenamefont
  {{Kamionkowski}}}]{munoz2016lensing}%
  \BibitemOpen
  \bibfield  {author} {\bibinfo {author} {\bibfnamefont {J.~B.}\ \bibnamefont
  {{Mu{\~n}oz}}}, \bibinfo {author} {\bibfnamefont {E.~D.}\ \bibnamefont
  {{Kovetz}}}, \bibinfo {author} {\bibfnamefont {L.}~\bibnamefont {{Dai}}},\
  and\ \bibinfo {author} {\bibfnamefont {M.}~\bibnamefont {{Kamionkowski}}},\
  }\href {https://doi.org/10.1103/PhysRevLett.117.091301} {\bibfield  {journal}
  {\bibinfo  {journal} {\prl}\ }\textbf {\bibinfo {volume} {117}},\ \bibinfo
  {eid} {091301} (\bibinfo {year} {2016}{\natexlab{b}})},\ \Eprint
  {https://arxiv.org/abs/1605.00008} {arXiv:1605.00008 [astro-ph.CO]}
  \BibitemShut {NoStop}%
\bibitem [{\citenamefont {{Sammons}}\ \emph {et~al.}(2020)\citenamefont
  {{Sammons}}, \citenamefont {{Macquart}}, \citenamefont {{Ekers}},
  \citenamefont {{Shannon}}, \citenamefont {{Cho}}, \citenamefont
  {{Prochaska}}, \citenamefont {{Deller}},\ and\ \citenamefont
  {{Day}}}]{first_constraint}%
  \BibitemOpen
  \bibfield  {author} {\bibinfo {author} {\bibfnamefont {M.~W.}\ \bibnamefont
  {{Sammons}}}, \bibinfo {author} {\bibfnamefont {J.-P.}\ \bibnamefont
  {{Macquart}}}, \bibinfo {author} {\bibfnamefont {R.~D.}\ \bibnamefont
  {{Ekers}}}, \bibinfo {author} {\bibfnamefont {R.~M.}\ \bibnamefont
  {{Shannon}}}, \bibinfo {author} {\bibfnamefont {H.}~\bibnamefont {{Cho}}},
  \bibinfo {author} {\bibfnamefont {J.~X.}\ \bibnamefont {{Prochaska}}},
  \bibinfo {author} {\bibfnamefont {A.~T.}\ \bibnamefont {{Deller}}},\ and\
  \bibinfo {author} {\bibfnamefont {C.~K.}\ \bibnamefont {{Day}}},\ }\href
  {https://doi.org/10.3847/1538-4357/aba7bb} {\bibfield  {journal} {\bibinfo
  {journal} {\apj}\ }\textbf {\bibinfo {volume} {900}},\ \bibinfo {eid} {122}
  (\bibinfo {year} {2020})},\ \Eprint {https://arxiv.org/abs/2002.12533}
  {arXiv:2002.12533 [astro-ph.CO]} \BibitemShut {NoStop}%
\bibitem [{\citenamefont {{Cordes}}\ \emph {et~al.}(2017)\citenamefont
  {{Cordes}}, \citenamefont {{Wasserman}}, \citenamefont {{Hessels}},
  \citenamefont {{Lazio}}, \citenamefont {{Chatterjee}},\ and\ \citenamefont
  {{Wharton}}}]{cordes2017lensing}%
  \BibitemOpen
  \bibfield  {author} {\bibinfo {author} {\bibfnamefont {J.~M.}\ \bibnamefont
  {{Cordes}}}, \bibinfo {author} {\bibfnamefont {I.}~\bibnamefont
  {{Wasserman}}}, \bibinfo {author} {\bibfnamefont {J.~W.~T.}\ \bibnamefont
  {{Hessels}}}, \bibinfo {author} {\bibfnamefont {T.~J.~W.}\ \bibnamefont
  {{Lazio}}}, \bibinfo {author} {\bibfnamefont {S.}~\bibnamefont
  {{Chatterjee}}},\ and\ \bibinfo {author} {\bibfnamefont {R.~S.}\ \bibnamefont
  {{Wharton}}},\ }\href {https://doi.org/10.3847/1538-4357/aa74da} {\bibfield
  {journal} {\bibinfo  {journal} {\apj}\ }\textbf {\bibinfo {volume} {842}},\
  \bibinfo {eid} {35} (\bibinfo {year} {2017})},\ \Eprint
  {https://arxiv.org/abs/1703.06580} {arXiv:1703.06580 [astro-ph.HE]}
  \BibitemShut {NoStop}%
\bibitem [{\citenamefont {{Li}}\ \emph {et~al.}(2018)\citenamefont {{Li}},
  \citenamefont {{Gao}}, \citenamefont {{Ding}}, \citenamefont {{Wang}},\ and\
  \citenamefont {{Zhang}}}]{li2018strongly}%
  \BibitemOpen
  \bibfield  {author} {\bibinfo {author} {\bibfnamefont {Z.-X.}\ \bibnamefont
  {{Li}}}, \bibinfo {author} {\bibfnamefont {H.}~\bibnamefont {{Gao}}},
  \bibinfo {author} {\bibfnamefont {X.-H.}\ \bibnamefont {{Ding}}}, \bibinfo
  {author} {\bibfnamefont {G.-J.}\ \bibnamefont {{Wang}}},\ and\ \bibinfo
  {author} {\bibfnamefont {B.}~\bibnamefont {{Zhang}}},\ }\href
  {https://doi.org/10.1038/s41467-018-06303-0} {\bibfield  {journal} {\bibinfo
  {journal} {Nature Communications}\ }\textbf {\bibinfo {volume} {9}},\
  \bibinfo {eid} {3833} (\bibinfo {year} {2018})},\ \Eprint
  {https://arxiv.org/abs/1708.06357} {arXiv:1708.06357 [astro-ph.CO]}
  \BibitemShut {NoStop}%
\bibitem [{\citenamefont {{Jow}}\ \emph
  {et~al.}(2020{\natexlab{b}})\citenamefont {{Jow}}, \citenamefont {{Foreman}},
  \citenamefont {{Pen}},\ and\ \citenamefont {{Zhu}}}]{jow2020wave}%
  \BibitemOpen
  \bibfield  {author} {\bibinfo {author} {\bibfnamefont {D.~L.}\ \bibnamefont
  {{Jow}}}, \bibinfo {author} {\bibfnamefont {S.}~\bibnamefont {{Foreman}}},
  \bibinfo {author} {\bibfnamefont {U.-L.}\ \bibnamefont {{Pen}}},\ and\
  \bibinfo {author} {\bibfnamefont {W.}~\bibnamefont {{Zhu}}},\ }\href
  {https://doi.org/10.1093/mnras/staa2230} {\bibfield  {journal} {\bibinfo
  {journal} {\mnras}\ }\textbf {\bibinfo {volume} {497}},\ \bibinfo {pages}
  {4956} (\bibinfo {year} {2020}{\natexlab{b}})},\ \Eprint
  {https://arxiv.org/abs/2002.01570} {arXiv:2002.01570 [astro-ph.HE]}
  \BibitemShut {NoStop}%
\bibitem [{\citenamefont {{Katz}}\ \emph {et~al.}(2018)\citenamefont {{Katz}},
  \citenamefont {{Kopp}}, \citenamefont {{Sibiryakov}},\ and\ \citenamefont
  {{Xue}}}]{katz2018femtolensing}%
  \BibitemOpen
  \bibfield  {author} {\bibinfo {author} {\bibfnamefont {A.}~\bibnamefont
  {{Katz}}}, \bibinfo {author} {\bibfnamefont {J.}~\bibnamefont {{Kopp}}},
  \bibinfo {author} {\bibfnamefont {S.}~\bibnamefont {{Sibiryakov}}},\ and\
  \bibinfo {author} {\bibfnamefont {W.}~\bibnamefont {{Xue}}},\ }\href
  {https://doi.org/10.1088/1475-7516/2018/12/005} {\bibfield  {journal}
  {\bibinfo  {journal} {\jcap}\ }\textbf {\bibinfo {volume} {2018}},\ \bibinfo
  {eid} {005} (\bibinfo {year} {2018})},\ \Eprint
  {https://arxiv.org/abs/1807.11495} {arXiv:1807.11495 [astro-ph.CO]}
  \BibitemShut {NoStop}%
\bibitem [{\citenamefont {{Oguri}}(2019)}]{oguri2019strong}%
  \BibitemOpen
  \bibfield  {author} {\bibinfo {author} {\bibfnamefont {M.}~\bibnamefont
  {{Oguri}}},\ }\href {https://doi.org/10.1088/1361-6633/ab4fc5} {\bibfield
  {journal} {\bibinfo  {journal} {Reports on Progress in Physics}\ }\textbf
  {\bibinfo {volume} {82}},\ \bibinfo {eid} {126901} (\bibinfo {year}
  {2019})},\ \Eprint {https://arxiv.org/abs/1907.06830} {arXiv:1907.06830
  [astro-ph.CO]} \BibitemShut {NoStop}%
\bibitem [{\citenamefont {{Dai}}\ and\ \citenamefont
  {{Lu}}(2017)}]{dai2017probing}%
  \BibitemOpen
  \bibfield  {author} {\bibinfo {author} {\bibfnamefont {L.}~\bibnamefont
  {{Dai}}}\ and\ \bibinfo {author} {\bibfnamefont {W.}~\bibnamefont {{Lu}}},\
  }\href {https://doi.org/10.3847/1538-4357/aa8873} {\bibfield  {journal}
  {\bibinfo  {journal} {\apj}\ }\textbf {\bibinfo {volume} {847}},\ \bibinfo
  {eid} {19} (\bibinfo {year} {2017})},\ \Eprint
  {https://arxiv.org/abs/1706.06103} {arXiv:1706.06103 [astro-ph.HE]}
  \BibitemShut {NoStop}%
\bibitem [{\citenamefont {{Pearson}}\ \emph {et~al.}(2020)\citenamefont
  {{Pearson}}, \citenamefont {{Trendafilova}},\ and\ \citenamefont
  {{Meyers}}}]{pearson2020searching}%
  \BibitemOpen
  \bibfield  {author} {\bibinfo {author} {\bibfnamefont {N.}~\bibnamefont
  {{Pearson}}}, \bibinfo {author} {\bibfnamefont {C.}~\bibnamefont
  {{Trendafilova}}},\ and\ \bibinfo {author} {\bibfnamefont {J.}~\bibnamefont
  {{Meyers}}},\ }\href@noop {} {\bibfield  {journal} {\bibinfo  {journal}
  {arXiv e-prints}\ ,\ \bibinfo {eid} {arXiv:2009.11252}} (\bibinfo {year}
  {2020})},\ \Eprint {https://arxiv.org/abs/2009.11252} {arXiv:2009.11252
  [astro-ph.CO]} \BibitemShut {NoStop}%
\bibitem [{\citenamefont {{Wucknitz}}\ \emph {et~al.}(2021)\citenamefont
  {{Wucknitz}}, \citenamefont {{Spitler}},\ and\ \citenamefont
  {{Pen}}}]{wucknitz2021cosmology}%
  \BibitemOpen
  \bibfield  {author} {\bibinfo {author} {\bibfnamefont {O.}~\bibnamefont
  {{Wucknitz}}}, \bibinfo {author} {\bibfnamefont {L.~G.}\ \bibnamefont
  {{Spitler}}},\ and\ \bibinfo {author} {\bibfnamefont {U.~L.}\ \bibnamefont
  {{Pen}}},\ }\href {https://doi.org/10.1051/0004-6361/202038248} {\bibfield
  {journal} {\bibinfo  {journal} {\aap}\ }\textbf {\bibinfo {volume} {645}},\
  \bibinfo {eid} {A44} (\bibinfo {year} {2021})},\ \Eprint
  {https://arxiv.org/abs/2004.11643} {arXiv:2004.11643 [astro-ph.CO]}
  \BibitemShut {NoStop}%
\bibitem [{\citenamefont {{Paynter}}\ \emph {et~al.}(2021)\citenamefont
  {{Paynter}}, \citenamefont {{Webster}},\ and\ \citenamefont
  {{Thrane}}}]{paynter2021evidence}%
  \BibitemOpen
  \bibfield  {author} {\bibinfo {author} {\bibfnamefont {J.}~\bibnamefont
  {{Paynter}}}, \bibinfo {author} {\bibfnamefont {R.}~\bibnamefont
  {{Webster}}},\ and\ \bibinfo {author} {\bibfnamefont {E.}~\bibnamefont
  {{Thrane}}},\ }\bibfield  {journal} {\bibinfo  {journal} {Nature Astronomy}\
  }\href {https://doi.org/10.1038/s41550-021-01307-1}
  {10.1038/s41550-021-01307-1} (\bibinfo {year} {2021}),\ \Eprint
  {https://arxiv.org/abs/2103.15414} {arXiv:2103.15414 [astro-ph.HE]}
  \BibitemShut {NoStop}%
\bibitem [{\citenamefont {{The CHIME Collaboration}}\ \emph
  {et~al.}(2022)\citenamefont {{The CHIME Collaboration}}, \citenamefont
  {{Amiri}}, \citenamefont {{Bandura}}, \citenamefont {{Boskovic}},
  \citenamefont {{Chen}}, \citenamefont {{Cliche}}, \citenamefont {{Deng}},
  \citenamefont {{Denman}}, \citenamefont {{Dobbs}}, \citenamefont {{Fandino}},
  \citenamefont {{Foreman}}, \citenamefont {{Halpern}}, \citenamefont
  {{Hanna}}, \citenamefont {{Hill}}, \citenamefont {{Hinshaw}}, \citenamefont
  {{H{\"o}fer}}, \citenamefont {{Kania}}, \citenamefont {{Klages}},
  \citenamefont {{Landecker}}, \citenamefont {{MacEachern}}, \citenamefont
  {{Masui}}, \citenamefont {{Mena-Parra}}, \citenamefont {{Milutinovic}},
  \citenamefont {{Mirhosseini}}, \citenamefont {{Newburgh}}, \citenamefont
  {{Nitsche}}, \citenamefont {{Ordog}}, \citenamefont {{Pen}}, \citenamefont
  {{Pinsonneault-Marotte}}, \citenamefont {{Polzin}}, \citenamefont {{Reda}},
  \citenamefont {{Renard}}, \citenamefont {{Shaw}}, \citenamefont {{Siegel}},
  \citenamefont {{Singh}}, \citenamefont {{Smegal}}, \citenamefont
  {{Tretyakov}}, \citenamefont {{Van Gassen}}, \citenamefont {{Vanderlinde}},
  \citenamefont {{Wang}}, \citenamefont {{Wiebe}}, \citenamefont {{Willis}},\
  and\ \citenamefont {{Wulf}}}]{CHIMECosmologyOverview}%
  \BibitemOpen
  \bibfield  {author} {\bibinfo {author} {\bibnamefont {{The CHIME
  Collaboration}}}, \bibinfo {author} {\bibfnamefont {M.}~\bibnamefont
  {{Amiri}}}, \bibinfo {author} {\bibfnamefont {K.}~\bibnamefont {{Bandura}}},
  \bibinfo {author} {\bibfnamefont {A.}~\bibnamefont {{Boskovic}}}, \bibinfo
  {author} {\bibfnamefont {T.}~\bibnamefont {{Chen}}}, \bibinfo {author}
  {\bibfnamefont {J.-F.}\ \bibnamefont {{Cliche}}}, \bibinfo {author}
  {\bibfnamefont {M.}~\bibnamefont {{Deng}}}, \bibinfo {author} {\bibfnamefont
  {N.}~\bibnamefont {{Denman}}}, \bibinfo {author} {\bibfnamefont
  {M.}~\bibnamefont {{Dobbs}}}, \bibinfo {author} {\bibfnamefont
  {M.}~\bibnamefont {{Fandino}}}, \bibinfo {author} {\bibfnamefont
  {S.}~\bibnamefont {{Foreman}}}, \bibinfo {author} {\bibfnamefont
  {M.}~\bibnamefont {{Halpern}}}, \bibinfo {author} {\bibfnamefont
  {D.}~\bibnamefont {{Hanna}}}, \bibinfo {author} {\bibfnamefont {A.~S.}\
  \bibnamefont {{Hill}}}, \bibinfo {author} {\bibfnamefont {G.}~\bibnamefont
  {{Hinshaw}}}, \bibinfo {author} {\bibfnamefont {C.}~\bibnamefont
  {{H{\"o}fer}}}, \bibinfo {author} {\bibfnamefont {J.}~\bibnamefont
  {{Kania}}}, \bibinfo {author} {\bibfnamefont {P.}~\bibnamefont {{Klages}}},
  \bibinfo {author} {\bibfnamefont {T.~L.}\ \bibnamefont {{Landecker}}},
  \bibinfo {author} {\bibfnamefont {J.}~\bibnamefont {{MacEachern}}}, \bibinfo
  {author} {\bibfnamefont {K.}~\bibnamefont {{Masui}}}, \bibinfo {author}
  {\bibfnamefont {J.}~\bibnamefont {{Mena-Parra}}}, \bibinfo {author}
  {\bibfnamefont {N.}~\bibnamefont {{Milutinovic}}}, \bibinfo {author}
  {\bibfnamefont {A.}~\bibnamefont {{Mirhosseini}}}, \bibinfo {author}
  {\bibfnamefont {L.}~\bibnamefont {{Newburgh}}}, \bibinfo {author}
  {\bibfnamefont {R.}~\bibnamefont {{Nitsche}}}, \bibinfo {author}
  {\bibfnamefont {A.}~\bibnamefont {{Ordog}}}, \bibinfo {author} {\bibfnamefont
  {U.-L.}\ \bibnamefont {{Pen}}}, \bibinfo {author} {\bibfnamefont
  {T.}~\bibnamefont {{Pinsonneault-Marotte}}}, \bibinfo {author} {\bibfnamefont
  {A.}~\bibnamefont {{Polzin}}}, \bibinfo {author} {\bibfnamefont
  {A.}~\bibnamefont {{Reda}}}, \bibinfo {author} {\bibfnamefont
  {A.}~\bibnamefont {{Renard}}}, \bibinfo {author} {\bibfnamefont {J.~R.}\
  \bibnamefont {{Shaw}}}, \bibinfo {author} {\bibfnamefont {S.~R.}\
  \bibnamefont {{Siegel}}}, \bibinfo {author} {\bibfnamefont {S.}~\bibnamefont
  {{Singh}}}, \bibinfo {author} {\bibfnamefont {R.}~\bibnamefont {{Smegal}}},
  \bibinfo {author} {\bibfnamefont {I.}~\bibnamefont {{Tretyakov}}}, \bibinfo
  {author} {\bibfnamefont {K.}~\bibnamefont {{Van Gassen}}}, \bibinfo {author}
  {\bibfnamefont {K.}~\bibnamefont {{Vanderlinde}}}, \bibinfo {author}
  {\bibfnamefont {H.}~\bibnamefont {{Wang}}}, \bibinfo {author} {\bibfnamefont
  {D.~V.}\ \bibnamefont {{Wiebe}}}, \bibinfo {author} {\bibfnamefont {J.~S.}\
  \bibnamefont {{Willis}}},\ and\ \bibinfo {author} {\bibfnamefont
  {D.}~\bibnamefont {{Wulf}}},\ }\href@noop {} {\bibfield  {journal} {\bibinfo
  {journal} {arXiv e-prints}\ ,\ \bibinfo {eid} {arXiv:2201.07869}} (\bibinfo
  {year} {2022})},\ \Eprint {https://arxiv.org/abs/2201.07869}
  {arXiv:2201.07869 [astro-ph.IM]} \BibitemShut {NoStop}%
\bibitem [{\citenamefont {{Michilli}}\ \emph {et~al.}(2021)\citenamefont
  {{Michilli}}, \citenamefont {{Masui}}, \citenamefont {{Mckinven}},
  \citenamefont {{Cubranic}}, \citenamefont {{Bruneault}}, \citenamefont
  {{Brar}}, \citenamefont {{Patel}}, \citenamefont {{Boyle}}, \citenamefont
  {{Stairs}}, \citenamefont {{Renard}}, \citenamefont {{Bandura}},
  \citenamefont {{Berger}}, \citenamefont {{Breitman}}, \citenamefont
  {{Cassanelli}}, \citenamefont {{Dobbs}}, \citenamefont {{Kaspi}},
  \citenamefont {{Leung}}, \citenamefont {{Mena-Parra}}, \citenamefont
  {{Pleunis}}, \citenamefont {{Russell}}, \citenamefont {{Scholz}},
  \citenamefont {{Siegel}}, \citenamefont {{Tendulkar}},\ and\ \citenamefont
  {{Vanderlinde}}}]{baseband_paper}%
  \BibitemOpen
  \bibfield  {author} {\bibinfo {author} {\bibfnamefont {D.}~\bibnamefont
  {{Michilli}}}, \bibinfo {author} {\bibfnamefont {K.~W.}\ \bibnamefont
  {{Masui}}}, \bibinfo {author} {\bibfnamefont {R.}~\bibnamefont {{Mckinven}}},
  \bibinfo {author} {\bibfnamefont {D.}~\bibnamefont {{Cubranic}}}, \bibinfo
  {author} {\bibfnamefont {M.}~\bibnamefont {{Bruneault}}}, \bibinfo {author}
  {\bibfnamefont {C.}~\bibnamefont {{Brar}}}, \bibinfo {author} {\bibfnamefont
  {C.}~\bibnamefont {{Patel}}}, \bibinfo {author} {\bibfnamefont {P.~J.}\
  \bibnamefont {{Boyle}}}, \bibinfo {author} {\bibfnamefont {I.~H.}\
  \bibnamefont {{Stairs}}}, \bibinfo {author} {\bibfnamefont {A.}~\bibnamefont
  {{Renard}}}, \bibinfo {author} {\bibfnamefont {K.}~\bibnamefont {{Bandura}}},
  \bibinfo {author} {\bibfnamefont {S.}~\bibnamefont {{Berger}}}, \bibinfo
  {author} {\bibfnamefont {D.}~\bibnamefont {{Breitman}}}, \bibinfo {author}
  {\bibfnamefont {T.}~\bibnamefont {{Cassanelli}}}, \bibinfo {author}
  {\bibfnamefont {M.}~\bibnamefont {{Dobbs}}}, \bibinfo {author} {\bibfnamefont
  {V.~M.}\ \bibnamefont {{Kaspi}}}, \bibinfo {author} {\bibfnamefont
  {C.}~\bibnamefont {{Leung}}}, \bibinfo {author} {\bibfnamefont
  {J.}~\bibnamefont {{Mena-Parra}}}, \bibinfo {author} {\bibfnamefont
  {Z.}~\bibnamefont {{Pleunis}}}, \bibinfo {author} {\bibfnamefont
  {L.}~\bibnamefont {{Russell}}}, \bibinfo {author} {\bibfnamefont
  {P.}~\bibnamefont {{Scholz}}}, \bibinfo {author} {\bibfnamefont {S.~R.}\
  \bibnamefont {{Siegel}}}, \bibinfo {author} {\bibfnamefont {S.~P.}\
  \bibnamefont {{Tendulkar}}},\ and\ \bibinfo {author} {\bibfnamefont
  {K.}~\bibnamefont {{Vanderlinde}}},\ }\href
  {https://doi.org/10.3847/1538-4357/abe626} {\bibfield  {journal} {\bibinfo
  {journal} {\apj}\ }\textbf {\bibinfo {volume} {910}},\ \bibinfo {eid} {147}
  (\bibinfo {year} {2021})},\ \Eprint {https://arxiv.org/abs/2010.06748}
  {arXiv:2010.06748 [astro-ph.HE]} \BibitemShut {NoStop}%
\bibitem [{\citenamefont {Bandura}\ \emph {et~al.}(2016)\citenamefont
  {Bandura}, \citenamefont {Bender}, \citenamefont {Cliche}, \citenamefont
  {de~Haan}, \citenamefont {Dobbs}, \citenamefont {Gilbert}, \citenamefont
  {Griffin}, \citenamefont {Hsyu}, \citenamefont {Ittah}, \citenamefont
  {Parra}, \citenamefont {Montgomery}, \citenamefont {Pinsonneault-Marotte},
  \citenamefont {Siegel}, \citenamefont {Smecher}, \citenamefont {Tang},
  \citenamefont {Vanderlinde},\ and\ \citenamefont
  {Whitehorn}}]{FEngineOverview}%
  \BibitemOpen
  \bibfield  {author} {\bibinfo {author} {\bibfnamefont {K.}~\bibnamefont
  {Bandura}}, \bibinfo {author} {\bibfnamefont {A.~N.}\ \bibnamefont {Bender}},
  \bibinfo {author} {\bibfnamefont {J.~F.}\ \bibnamefont {Cliche}}, \bibinfo
  {author} {\bibfnamefont {T.}~\bibnamefont {de~Haan}}, \bibinfo {author}
  {\bibfnamefont {M.~A.}\ \bibnamefont {Dobbs}}, \bibinfo {author}
  {\bibfnamefont {A.~J.}\ \bibnamefont {Gilbert}}, \bibinfo {author}
  {\bibfnamefont {S.}~\bibnamefont {Griffin}}, \bibinfo {author} {\bibfnamefont
  {G.}~\bibnamefont {Hsyu}}, \bibinfo {author} {\bibfnamefont {D.}~\bibnamefont
  {Ittah}}, \bibinfo {author} {\bibfnamefont {J.~M.}\ \bibnamefont {Parra}},
  \bibinfo {author} {\bibfnamefont {J.}~\bibnamefont {Montgomery}}, \bibinfo
  {author} {\bibfnamefont {T.}~\bibnamefont {Pinsonneault-Marotte}}, \bibinfo
  {author} {\bibfnamefont {S.}~\bibnamefont {Siegel}}, \bibinfo {author}
  {\bibfnamefont {G.}~\bibnamefont {Smecher}}, \bibinfo {author} {\bibfnamefont
  {Q.~Y.}\ \bibnamefont {Tang}}, \bibinfo {author} {\bibfnamefont
  {K.}~\bibnamefont {Vanderlinde}},\ and\ \bibinfo {author} {\bibfnamefont
  {N.}~\bibnamefont {Whitehorn}},\ }\href
  {https://doi.org/10.1142/S2251171716410051} {\bibfield  {journal} {\bibinfo
  {journal} {Journal of Astronomical Instrumentation}\ }\textbf {\bibinfo
  {volume} {05}},\ \bibinfo {pages} {1641005} (\bibinfo {year} {2016})},\
  \Eprint {https://arxiv.org/abs/https://doi.org/10.1142/S2251171716410051}
  {https://doi.org/10.1142/S2251171716410051} \BibitemShut {NoStop}%
\bibitem [{\citenamefont {{CHIME/FRB Collaboration}}\ \emph
  {et~al.}(2018{\natexlab{b}})\citenamefont {{CHIME/FRB Collaboration}},
  \citenamefont {{Amiri}}, \citenamefont {{Bandura}}, \citenamefont {{Berger}},
  \citenamefont {{Bhardwaj}}, \citenamefont {{Boyce}}, \citenamefont {{Boyle}},
  \citenamefont {{Brar}}, \citenamefont {{Burhanpurkar}}, \citenamefont
  {{Chawla}}, \citenamefont {{Chowdhury}}, \citenamefont {{Cliche}},
  \citenamefont {{Cranmer}}, \citenamefont {{Cubranic}}, \citenamefont
  {{Deng}}, \citenamefont {{Denman}}, \citenamefont {{Dobbs}}, \citenamefont
  {{Fandino}}, \citenamefont {{Fonseca}}, \citenamefont {{Gaensler}},
  \citenamefont {{Giri}}, \citenamefont {{Gilbert}}, \citenamefont {{Good}},
  \citenamefont {{Guliani}}, \citenamefont {{Halpern}}, \citenamefont
  {{Hinshaw}}, \citenamefont {{H{\"o}fer}}, \citenamefont {{Josephy}},
  \citenamefont {{Kaspi}}, \citenamefont {{Landecker}}, \citenamefont {{Lang}},
  \citenamefont {{Liao}}, \citenamefont {{Masui}}, \citenamefont
  {{Mena-Parra}}, \citenamefont {{Naidu}}, \citenamefont {{Newburgh}},
  \citenamefont {{Ng}}, \citenamefont {{Patel}}, \citenamefont {{Pen}},
  \citenamefont {{Pinsonneault-Marotte}}, \citenamefont {{Pleunis}},
  \citenamefont {{Rafiei Ravandi}}, \citenamefont {{Ransom}}, \citenamefont
  {{Renard}}, \citenamefont {{Scholz}}, \citenamefont {{Sigurdson}},
  \citenamefont {{Siegel}}, \citenamefont {{Smith}}, \citenamefont {{Stairs}},
  \citenamefont {{Tendulkar}}, \citenamefont {{Vand erlinde}},\ and\
  \citenamefont {{Wiebe}}}]{FRBSystemOverview}%
  \BibitemOpen
  \bibfield  {author} {\bibinfo {author} {\bibnamefont {{CHIME/FRB
  Collaboration}}}, \bibinfo {author} {\bibfnamefont {M.}~\bibnamefont
  {{Amiri}}}, \bibinfo {author} {\bibfnamefont {K.}~\bibnamefont {{Bandura}}},
  \bibinfo {author} {\bibfnamefont {P.}~\bibnamefont {{Berger}}}, \bibinfo
  {author} {\bibfnamefont {M.}~\bibnamefont {{Bhardwaj}}}, \bibinfo {author}
  {\bibfnamefont {M.~M.}\ \bibnamefont {{Boyce}}}, \bibinfo {author}
  {\bibfnamefont {P.~J.}\ \bibnamefont {{Boyle}}}, \bibinfo {author}
  {\bibfnamefont {C.}~\bibnamefont {{Brar}}}, \bibinfo {author} {\bibfnamefont
  {M.}~\bibnamefont {{Burhanpurkar}}}, \bibinfo {author} {\bibfnamefont
  {P.}~\bibnamefont {{Chawla}}}, \bibinfo {author} {\bibfnamefont
  {J.}~\bibnamefont {{Chowdhury}}}, \bibinfo {author} {\bibfnamefont {J.~F.}\
  \bibnamefont {{Cliche}}}, \bibinfo {author} {\bibfnamefont {M.~D.}\
  \bibnamefont {{Cranmer}}}, \bibinfo {author} {\bibfnamefont {D.}~\bibnamefont
  {{Cubranic}}}, \bibinfo {author} {\bibfnamefont {M.}~\bibnamefont {{Deng}}},
  \bibinfo {author} {\bibfnamefont {N.}~\bibnamefont {{Denman}}}, \bibinfo
  {author} {\bibfnamefont {M.}~\bibnamefont {{Dobbs}}}, \bibinfo {author}
  {\bibfnamefont {M.}~\bibnamefont {{Fandino}}}, \bibinfo {author}
  {\bibfnamefont {E.}~\bibnamefont {{Fonseca}}}, \bibinfo {author}
  {\bibfnamefont {B.~M.}\ \bibnamefont {{Gaensler}}}, \bibinfo {author}
  {\bibfnamefont {U.}~\bibnamefont {{Giri}}}, \bibinfo {author} {\bibfnamefont
  {A.~J.}\ \bibnamefont {{Gilbert}}}, \bibinfo {author} {\bibfnamefont {D.~C.}\
  \bibnamefont {{Good}}}, \bibinfo {author} {\bibfnamefont {S.}~\bibnamefont
  {{Guliani}}}, \bibinfo {author} {\bibfnamefont {M.}~\bibnamefont
  {{Halpern}}}, \bibinfo {author} {\bibfnamefont {G.}~\bibnamefont
  {{Hinshaw}}}, \bibinfo {author} {\bibfnamefont {C.}~\bibnamefont
  {{H{\"o}fer}}}, \bibinfo {author} {\bibfnamefont {A.}~\bibnamefont
  {{Josephy}}}, \bibinfo {author} {\bibfnamefont {V.~M.}\ \bibnamefont
  {{Kaspi}}}, \bibinfo {author} {\bibfnamefont {T.~L.}\ \bibnamefont
  {{Landecker}}}, \bibinfo {author} {\bibfnamefont {D.}~\bibnamefont {{Lang}}},
  \bibinfo {author} {\bibfnamefont {H.}~\bibnamefont {{Liao}}}, \bibinfo
  {author} {\bibfnamefont {K.~W.}\ \bibnamefont {{Masui}}}, \bibinfo {author}
  {\bibfnamefont {J.}~\bibnamefont {{Mena-Parra}}}, \bibinfo {author}
  {\bibfnamefont {A.}~\bibnamefont {{Naidu}}}, \bibinfo {author} {\bibfnamefont
  {L.~B.}\ \bibnamefont {{Newburgh}}}, \bibinfo {author} {\bibfnamefont
  {C.}~\bibnamefont {{Ng}}}, \bibinfo {author} {\bibfnamefont {C.}~\bibnamefont
  {{Patel}}}, \bibinfo {author} {\bibfnamefont {U.~L.}\ \bibnamefont {{Pen}}},
  \bibinfo {author} {\bibfnamefont {T.}~\bibnamefont {{Pinsonneault-Marotte}}},
  \bibinfo {author} {\bibfnamefont {Z.}~\bibnamefont {{Pleunis}}}, \bibinfo
  {author} {\bibfnamefont {M.}~\bibnamefont {{Rafiei Ravandi}}}, \bibinfo
  {author} {\bibfnamefont {S.~M.}\ \bibnamefont {{Ransom}}}, \bibinfo {author}
  {\bibfnamefont {A.}~\bibnamefont {{Renard}}}, \bibinfo {author}
  {\bibfnamefont {P.}~\bibnamefont {{Scholz}}}, \bibinfo {author}
  {\bibfnamefont {K.}~\bibnamefont {{Sigurdson}}}, \bibinfo {author}
  {\bibfnamefont {S.~R.}\ \bibnamefont {{Siegel}}}, \bibinfo {author}
  {\bibfnamefont {K.~M.}\ \bibnamefont {{Smith}}}, \bibinfo {author}
  {\bibfnamefont {I.~H.}\ \bibnamefont {{Stairs}}}, \bibinfo {author}
  {\bibfnamefont {S.~P.}\ \bibnamefont {{Tendulkar}}}, \bibinfo {author}
  {\bibfnamefont {K.}~\bibnamefont {{Vand erlinde}}},\ and\ \bibinfo {author}
  {\bibfnamefont {D.~V.}\ \bibnamefont {{Wiebe}}},\ }\href
  {https://doi.org/10.3847/1538-4357/aad188} {\bibfield  {journal} {\bibinfo
  {journal} {\apj}\ }\textbf {\bibinfo {volume} {863}},\ \bibinfo {eid} {48}
  (\bibinfo {year} {2018}{\natexlab{b}})},\ \Eprint
  {https://arxiv.org/abs/1803.11235} {arXiv:1803.11235 [astro-ph.IM]}
  \BibitemShut {NoStop}%
\bibitem [{\citenamefont {{CHIME/Pulsar Collaboration}}\ \emph
  {et~al.}(2021)\citenamefont {{CHIME/Pulsar Collaboration}}, \citenamefont
  {{Amiri}}, \citenamefont {{Bandura}}, \citenamefont {{Boyle}}, \citenamefont
  {{Brar}}, \citenamefont {{Cliche}}, \citenamefont {{Crowter}}, \citenamefont
  {{Cubranic}}, \citenamefont {{Demorest}}, \citenamefont {{Denman}},
  \citenamefont {{Dobbs}}, \citenamefont {{Dong}}, \citenamefont {{Fandino}},
  \citenamefont {{Fonseca}}, \citenamefont {{Good}}, \citenamefont {{Halpern}},
  \citenamefont {{Hill}}, \citenamefont {{H{\"o}fer}}, \citenamefont {{Kaspi}},
  \citenamefont {{Landecker}}, \citenamefont {{Leung}}, \citenamefont {{Lin}},
  \citenamefont {{Luo}}, \citenamefont {{Masui}}, \citenamefont {{McKee}},
  \citenamefont {{Mena-Parra}}, \citenamefont {{Meyers}}, \citenamefont
  {{Michilli}}, \citenamefont {{Naidu}}, \citenamefont {{Newburgh}},
  \citenamefont {{Ng}}, \citenamefont {{Patel}}, \citenamefont
  {{Pinsonneault-Marotte}}, \citenamefont {{Ransom}}, \citenamefont {{Renard}},
  \citenamefont {{Scholz}}, \citenamefont {{Shaw}}, \citenamefont {{Sikora}},
  \citenamefont {{Stairs}}, \citenamefont {{Tan}}, \citenamefont {{Tendulkar}},
  \citenamefont {{Tretyakov}}, \citenamefont {{Vanderlinde}}, \citenamefont
  {{Wang}},\ and\ \citenamefont {{Wang}}}]{CHIMEPulsarOverview}%
  \BibitemOpen
  \bibfield  {author} {\bibinfo {author} {\bibnamefont {{CHIME/Pulsar
  Collaboration}}}, \bibinfo {author} {\bibfnamefont {M.}~\bibnamefont
  {{Amiri}}}, \bibinfo {author} {\bibfnamefont {K.~M.}\ \bibnamefont
  {{Bandura}}}, \bibinfo {author} {\bibfnamefont {P.~J.}\ \bibnamefont
  {{Boyle}}}, \bibinfo {author} {\bibfnamefont {C.}~\bibnamefont {{Brar}}},
  \bibinfo {author} {\bibfnamefont {J.~F.}\ \bibnamefont {{Cliche}}}, \bibinfo
  {author} {\bibfnamefont {K.}~\bibnamefont {{Crowter}}}, \bibinfo {author}
  {\bibfnamefont {D.}~\bibnamefont {{Cubranic}}}, \bibinfo {author}
  {\bibfnamefont {P.~B.}\ \bibnamefont {{Demorest}}}, \bibinfo {author}
  {\bibfnamefont {N.~T.}\ \bibnamefont {{Denman}}}, \bibinfo {author}
  {\bibfnamefont {M.}~\bibnamefont {{Dobbs}}}, \bibinfo {author} {\bibfnamefont
  {F.~Q.}\ \bibnamefont {{Dong}}}, \bibinfo {author} {\bibfnamefont
  {M.}~\bibnamefont {{Fandino}}}, \bibinfo {author} {\bibfnamefont
  {E.}~\bibnamefont {{Fonseca}}}, \bibinfo {author} {\bibfnamefont {D.~C.}\
  \bibnamefont {{Good}}}, \bibinfo {author} {\bibfnamefont {M.}~\bibnamefont
  {{Halpern}}}, \bibinfo {author} {\bibfnamefont {A.~S.}\ \bibnamefont
  {{Hill}}}, \bibinfo {author} {\bibfnamefont {C.}~\bibnamefont {{H{\"o}fer}}},
  \bibinfo {author} {\bibfnamefont {V.~M.}\ \bibnamefont {{Kaspi}}}, \bibinfo
  {author} {\bibfnamefont {T.~L.}\ \bibnamefont {{Landecker}}}, \bibinfo
  {author} {\bibfnamefont {C.}~\bibnamefont {{Leung}}}, \bibinfo {author}
  {\bibfnamefont {H.~H.}\ \bibnamefont {{Lin}}}, \bibinfo {author}
  {\bibfnamefont {J.}~\bibnamefont {{Luo}}}, \bibinfo {author} {\bibfnamefont
  {K.~W.}\ \bibnamefont {{Masui}}}, \bibinfo {author} {\bibfnamefont {J.~W.}\
  \bibnamefont {{McKee}}}, \bibinfo {author} {\bibfnamefont {J.}~\bibnamefont
  {{Mena-Parra}}}, \bibinfo {author} {\bibfnamefont {B.~W.}\ \bibnamefont
  {{Meyers}}}, \bibinfo {author} {\bibfnamefont {D.}~\bibnamefont
  {{Michilli}}}, \bibinfo {author} {\bibfnamefont {A.}~\bibnamefont {{Naidu}}},
  \bibinfo {author} {\bibfnamefont {L.}~\bibnamefont {{Newburgh}}}, \bibinfo
  {author} {\bibfnamefont {C.}~\bibnamefont {{Ng}}}, \bibinfo {author}
  {\bibfnamefont {C.}~\bibnamefont {{Patel}}}, \bibinfo {author} {\bibfnamefont
  {T.}~\bibnamefont {{Pinsonneault-Marotte}}}, \bibinfo {author} {\bibfnamefont
  {S.~M.}\ \bibnamefont {{Ransom}}}, \bibinfo {author} {\bibfnamefont
  {A.}~\bibnamefont {{Renard}}}, \bibinfo {author} {\bibfnamefont
  {P.}~\bibnamefont {{Scholz}}}, \bibinfo {author} {\bibfnamefont {J.~R.}\
  \bibnamefont {{Shaw}}}, \bibinfo {author} {\bibfnamefont {A.~E.}\
  \bibnamefont {{Sikora}}}, \bibinfo {author} {\bibfnamefont {I.~H.}\
  \bibnamefont {{Stairs}}}, \bibinfo {author} {\bibfnamefont {C.~M.}\
  \bibnamefont {{Tan}}}, \bibinfo {author} {\bibfnamefont {S.~P.}\ \bibnamefont
  {{Tendulkar}}}, \bibinfo {author} {\bibfnamefont {I.}~\bibnamefont
  {{Tretyakov}}}, \bibinfo {author} {\bibfnamefont {K.}~\bibnamefont
  {{Vanderlinde}}}, \bibinfo {author} {\bibfnamefont {H.}~\bibnamefont
  {{Wang}}},\ and\ \bibinfo {author} {\bibfnamefont {X.}~\bibnamefont
  {{Wang}}},\ }\href {https://doi.org/10.3847/1538-4365/abfdcb} {\bibfield
  {journal} {\bibinfo  {journal} {\apjs}\ }\textbf {\bibinfo {volume} {255}},\
  \bibinfo {eid} {5} (\bibinfo {year} {2021})},\ \Eprint
  {https://arxiv.org/abs/2008.05681} {arXiv:2008.05681 [astro-ph.IM]}
  \BibitemShut {NoStop}%
\bibitem [{\citenamefont {Ng}\ \emph {et~al.}(2017)\citenamefont {Ng},
  \citenamefont {Vanderlinde}, \citenamefont {Paradise}, \citenamefont
  {Klages}, \citenamefont {Masui}, \citenamefont {Smith}, \citenamefont
  {Bandura}, \citenamefont {Boyle}, \citenamefont {Dobbs}, \citenamefont
  {Kaspi} \emph {et~al.}}]{CHIMEBeamforming}%
  \BibitemOpen
  \bibfield  {author} {\bibinfo {author} {\bibfnamefont {C.}~\bibnamefont
  {Ng}}, \bibinfo {author} {\bibfnamefont {K.}~\bibnamefont {Vanderlinde}},
  \bibinfo {author} {\bibfnamefont {A.}~\bibnamefont {Paradise}}, \bibinfo
  {author} {\bibfnamefont {P.}~\bibnamefont {Klages}}, \bibinfo {author}
  {\bibfnamefont {K.}~\bibnamefont {Masui}}, \bibinfo {author} {\bibfnamefont
  {K.}~\bibnamefont {Smith}}, \bibinfo {author} {\bibfnamefont
  {K.}~\bibnamefont {Bandura}}, \bibinfo {author} {\bibfnamefont {P.~J.}\
  \bibnamefont {Boyle}}, \bibinfo {author} {\bibfnamefont {M.}~\bibnamefont
  {Dobbs}}, \bibinfo {author} {\bibfnamefont {V.}~\bibnamefont {Kaspi}}, \emph
  {et~al.},\ }in\ \href@noop {} {\emph {\bibinfo {booktitle} {2017 XXXIInd
  General Assembly and Scientific Symposium of the International Union of Radio
  Science (URSI GASS)}}}\ (\bibinfo {organization} {IEEE},\ \bibinfo {year}
  {2017})\ pp.\ \bibinfo {pages} {1--4}\BibitemShut {NoStop}%
\bibitem [{\citenamefont {{Masui}}\ \emph {et~al.}(2017)\citenamefont
  {{Masui}}, \citenamefont {{Shaw}}, \citenamefont {{Ng}}, \citenamefont
  {{Smith}}, \citenamefont {{Vanderlinde}},\ and\ \citenamefont
  {{Paradise}}}]{beamforming}%
  \BibitemOpen
  \bibfield  {author} {\bibinfo {author} {\bibfnamefont {K.~W.}\ \bibnamefont
  {{Masui}}}, \bibinfo {author} {\bibfnamefont {J.~R.}\ \bibnamefont {{Shaw}}},
  \bibinfo {author} {\bibfnamefont {C.}~\bibnamefont {{Ng}}}, \bibinfo {author}
  {\bibfnamefont {K.~M.}\ \bibnamefont {{Smith}}}, \bibinfo {author}
  {\bibfnamefont {K.}~\bibnamefont {{Vanderlinde}}},\ and\ \bibinfo {author}
  {\bibfnamefont {A.}~\bibnamefont {{Paradise}}},\ }\href@noop {} {\bibfield
  {journal} {\bibinfo  {journal} {arXiv e-prints}\ ,\ \bibinfo {eid}
  {arXiv:1710.08591}} (\bibinfo {year} {2017})},\ \Eprint
  {https://arxiv.org/abs/1710.08591} {arXiv:1710.08591 [astro-ph.IM]}
  \BibitemShut {NoStop}%
\bibitem [{\citenamefont {Hankins}\ and\ \citenamefont
  {Rickett}(1975)}]{hankins_pulsar_signal_process}%
  \BibitemOpen
  \bibfield  {author} {\bibinfo {author} {\bibfnamefont {T.~H.}\ \bibnamefont
  {Hankins}}\ and\ \bibinfo {author} {\bibfnamefont {B.~J.}\ \bibnamefont
  {Rickett}},\ }\href {https://doi.org/10.1016/b978-0-12-460814-6.50007-3}
  {\bibfield  {journal} {\bibinfo  {journal} {Methods in Computational Physics:
  Advances in Research and Applications Radio Astronomy}\ ,\ \bibinfo {pages}
  {55–129}} (\bibinfo {year} {1975})}\BibitemShut {NoStop}%
\bibitem [{\citenamefont {{Lorimer}}\ and\ \citenamefont
  {{Kramer}}(2004)}]{lorimer2004handbook}%
  \BibitemOpen
  \bibfield  {author} {\bibinfo {author} {\bibfnamefont {D.~R.}\ \bibnamefont
  {{Lorimer}}}\ and\ \bibinfo {author} {\bibfnamefont {M.}~\bibnamefont
  {{Kramer}}},\ }\href@noop {} {\emph {\bibinfo {title} {{Handbook of Pulsar
  Astronomy}}}},\ Vol.~\bibinfo {volume} {4}\ (\bibinfo  {publisher} {Cambridge
  University Press},\ \bibinfo {year} {2004})\BibitemShut {NoStop}%
\bibitem [{\citenamefont {{Farah}}\ \emph
  {et~al.}(2019{\natexlab{b}})\citenamefont {{Farah}}, \citenamefont {{Flynn}},
  \citenamefont {{Bailes}}, \citenamefont {{Jameson}}, \citenamefont
  {{Bateman}}, \citenamefont {{Campbell-Wilson}}, \citenamefont {{Day}},
  \citenamefont {{Deller}}, \citenamefont {{Green}}, \citenamefont {{Gupta}},
  \citenamefont {{Hunstead}}, \citenamefont {{Lower}}, \citenamefont
  {{Os{\l}owski}}, \citenamefont {{Parthasarathy}}, \citenamefont {{Price}},
  \citenamefont {{Ravi}}, \citenamefont {{Shannon}}, \citenamefont
  {{Sutherland}}, \citenamefont {{Temby}}, \citenamefont {{Krishnan}},
  \citenamefont {{Caleb}}, \citenamefont {{Chang}}, \citenamefont {{Cruces}},
  \citenamefont {{Roy}}, \citenamefont {{Morello}}, \citenamefont {{Onken}},
  \citenamefont {{Stappers}}, \citenamefont {{Webb}},\ and\ \citenamefont
  {{Wolf}}}]{2019MNRAS.488.2989F}%
  \BibitemOpen
  \bibfield  {author} {\bibinfo {author} {\bibfnamefont {W.}~\bibnamefont
  {{Farah}}}, \bibinfo {author} {\bibfnamefont {C.}~\bibnamefont {{Flynn}}},
  \bibinfo {author} {\bibfnamefont {M.}~\bibnamefont {{Bailes}}}, \bibinfo
  {author} {\bibfnamefont {A.}~\bibnamefont {{Jameson}}}, \bibinfo {author}
  {\bibfnamefont {T.}~\bibnamefont {{Bateman}}}, \bibinfo {author}
  {\bibfnamefont {D.}~\bibnamefont {{Campbell-Wilson}}}, \bibinfo {author}
  {\bibfnamefont {C.~K.}\ \bibnamefont {{Day}}}, \bibinfo {author}
  {\bibfnamefont {A.~T.}\ \bibnamefont {{Deller}}}, \bibinfo {author}
  {\bibfnamefont {A.~J.}\ \bibnamefont {{Green}}}, \bibinfo {author}
  {\bibfnamefont {V.}~\bibnamefont {{Gupta}}}, \bibinfo {author} {\bibfnamefont
  {R.}~\bibnamefont {{Hunstead}}}, \bibinfo {author} {\bibfnamefont {M.~E.}\
  \bibnamefont {{Lower}}}, \bibinfo {author} {\bibfnamefont {S.}~\bibnamefont
  {{Os{\l}owski}}}, \bibinfo {author} {\bibfnamefont {A.}~\bibnamefont
  {{Parthasarathy}}}, \bibinfo {author} {\bibfnamefont {D.~C.}\ \bibnamefont
  {{Price}}}, \bibinfo {author} {\bibfnamefont {V.}~\bibnamefont {{Ravi}}},
  \bibinfo {author} {\bibfnamefont {R.~M.}\ \bibnamefont {{Shannon}}}, \bibinfo
  {author} {\bibfnamefont {A.}~\bibnamefont {{Sutherland}}}, \bibinfo {author}
  {\bibfnamefont {D.}~\bibnamefont {{Temby}}}, \bibinfo {author} {\bibfnamefont
  {V.~V.}\ \bibnamefont {{Krishnan}}}, \bibinfo {author} {\bibfnamefont
  {M.}~\bibnamefont {{Caleb}}}, \bibinfo {author} {\bibfnamefont {S.~W.}\
  \bibnamefont {{Chang}}}, \bibinfo {author} {\bibfnamefont {M.}~\bibnamefont
  {{Cruces}}}, \bibinfo {author} {\bibfnamefont {J.}~\bibnamefont {{Roy}}},
  \bibinfo {author} {\bibfnamefont {V.}~\bibnamefont {{Morello}}}, \bibinfo
  {author} {\bibfnamefont {C.~A.}\ \bibnamefont {{Onken}}}, \bibinfo {author}
  {\bibfnamefont {B.~W.}\ \bibnamefont {{Stappers}}}, \bibinfo {author}
  {\bibfnamefont {S.}~\bibnamefont {{Webb}}},\ and\ \bibinfo {author}
  {\bibfnamefont {C.}~\bibnamefont {{Wolf}}},\ }\href
  {https://doi.org/10.1093/mnras/stz1748} {\bibfield  {journal} {\bibinfo
  {journal} {\mnras}\ }\textbf {\bibinfo {volume} {488}},\ \bibinfo {pages}
  {2989} (\bibinfo {year} {2019}{\natexlab{b}})},\ \Eprint
  {https://arxiv.org/abs/1905.02293} {arXiv:1905.02293 [astro-ph.HE]}
  \BibitemShut {NoStop}%
\bibitem [{\citenamefont {{Cho}}\ \emph
  {et~al.}(2020{\natexlab{b}})\citenamefont {{Cho}}, \citenamefont
  {{Macquart}}, \citenamefont {{Shannon}}, \citenamefont {{Deller}},
  \citenamefont {{Morrison}}, \citenamefont {{Ekers}}, \citenamefont
  {{Bannister}}, \citenamefont {{Farah}}, \citenamefont {{Qiu}}, \citenamefont
  {{Sammons}}, \citenamefont {{Bailes}}, \citenamefont {{Bhandari}},
  \citenamefont {{Day}}, \citenamefont {{James}}, \citenamefont {{Phillips}},
  \citenamefont {{Prochaska}},\ and\ \citenamefont
  {{Tuthill}}}]{2020ApJ...891L..38C}%
  \BibitemOpen
  \bibfield  {author} {\bibinfo {author} {\bibfnamefont {H.}~\bibnamefont
  {{Cho}}}, \bibinfo {author} {\bibfnamefont {J.-P.}\ \bibnamefont
  {{Macquart}}}, \bibinfo {author} {\bibfnamefont {R.~M.}\ \bibnamefont
  {{Shannon}}}, \bibinfo {author} {\bibfnamefont {A.~T.}\ \bibnamefont
  {{Deller}}}, \bibinfo {author} {\bibfnamefont {I.~S.}\ \bibnamefont
  {{Morrison}}}, \bibinfo {author} {\bibfnamefont {R.~D.}\ \bibnamefont
  {{Ekers}}}, \bibinfo {author} {\bibfnamefont {K.~W.}\ \bibnamefont
  {{Bannister}}}, \bibinfo {author} {\bibfnamefont {W.}~\bibnamefont
  {{Farah}}}, \bibinfo {author} {\bibfnamefont {H.}~\bibnamefont {{Qiu}}},
  \bibinfo {author} {\bibfnamefont {M.~W.}\ \bibnamefont {{Sammons}}}, \bibinfo
  {author} {\bibfnamefont {M.}~\bibnamefont {{Bailes}}}, \bibinfo {author}
  {\bibfnamefont {S.}~\bibnamefont {{Bhandari}}}, \bibinfo {author}
  {\bibfnamefont {C.~K.}\ \bibnamefont {{Day}}}, \bibinfo {author}
  {\bibfnamefont {C.~W.}\ \bibnamefont {{James}}}, \bibinfo {author}
  {\bibfnamefont {C.~J.}\ \bibnamefont {{Phillips}}}, \bibinfo {author}
  {\bibfnamefont {J.~X.}\ \bibnamefont {{Prochaska}}},\ and\ \bibinfo {author}
  {\bibfnamefont {J.}~\bibnamefont {{Tuthill}}},\ }\href
  {https://doi.org/10.3847/2041-8213/ab7824} {\bibfield  {journal} {\bibinfo
  {journal} {\apjl}\ }\textbf {\bibinfo {volume} {891}},\ \bibinfo {eid} {L38}
  (\bibinfo {year} {2020}{\natexlab{b}})},\ \Eprint
  {https://arxiv.org/abs/2002.12539} {arXiv:2002.12539 [astro-ph.HE]}
  \BibitemShut {NoStop}%
\bibitem [{\citenamefont {{Masui}}\ \emph {et~al.}(2015)\citenamefont
  {{Masui}}, \citenamefont {{Lin}}, \citenamefont {{Sievers}}, \citenamefont
  {{Anderson}}, \citenamefont {{Chang}}, \citenamefont {{Chen}}, \citenamefont
  {{Ganguly}}, \citenamefont {{Jarvis}}, \citenamefont {{Kuo}}, \citenamefont
  {{Li}}, \citenamefont {{Liao}}, \citenamefont {{McLaughlin}}, \citenamefont
  {{Pen}}, \citenamefont {{Peterson}}, \citenamefont {{Roman}}, \citenamefont
  {{Timbie}}, \citenamefont {{Voytek}},\ and\ \citenamefont
  {{Yadav}}}]{masui_scattering}%
  \BibitemOpen
  \bibfield  {author} {\bibinfo {author} {\bibfnamefont {K.}~\bibnamefont
  {{Masui}}}, \bibinfo {author} {\bibfnamefont {H.-H.}\ \bibnamefont {{Lin}}},
  \bibinfo {author} {\bibfnamefont {J.}~\bibnamefont {{Sievers}}}, \bibinfo
  {author} {\bibfnamefont {C.~J.}\ \bibnamefont {{Anderson}}}, \bibinfo
  {author} {\bibfnamefont {T.-C.}\ \bibnamefont {{Chang}}}, \bibinfo {author}
  {\bibfnamefont {X.}~\bibnamefont {{Chen}}}, \bibinfo {author} {\bibfnamefont
  {A.}~\bibnamefont {{Ganguly}}}, \bibinfo {author} {\bibfnamefont
  {M.}~\bibnamefont {{Jarvis}}}, \bibinfo {author} {\bibfnamefont {C.-Y.}\
  \bibnamefont {{Kuo}}}, \bibinfo {author} {\bibfnamefont {Y.-C.}\ \bibnamefont
  {{Li}}}, \bibinfo {author} {\bibfnamefont {Y.-W.}\ \bibnamefont {{Liao}}},
  \bibinfo {author} {\bibfnamefont {M.}~\bibnamefont {{McLaughlin}}}, \bibinfo
  {author} {\bibfnamefont {U.-L.}\ \bibnamefont {{Pen}}}, \bibinfo {author}
  {\bibfnamefont {J.~B.}\ \bibnamefont {{Peterson}}}, \bibinfo {author}
  {\bibfnamefont {A.}~\bibnamefont {{Roman}}}, \bibinfo {author} {\bibfnamefont
  {P.~T.}\ \bibnamefont {{Timbie}}}, \bibinfo {author} {\bibfnamefont
  {T.}~\bibnamefont {{Voytek}}},\ and\ \bibinfo {author} {\bibfnamefont
  {J.~K.}\ \bibnamefont {{Yadav}}},\ }\href
  {https://doi.org/10.1038/nature15769} {\bibfield  {journal} {\bibinfo
  {journal} {\nat}\ }\textbf {\bibinfo {volume} {528}},\ \bibinfo {pages} {523}
  (\bibinfo {year} {2015})},\ \Eprint {https://arxiv.org/abs/1512.00529}
  {arXiv:1512.00529 [astro-ph.HE]} \BibitemShut {NoStop}%
\bibitem [{\citenamefont {{Macquart}}\ \emph {et~al.}(2019)\citenamefont
  {{Macquart}}, \citenamefont {{Shannon}}, \citenamefont {{Bannister}},
  \citenamefont {{James}}, \citenamefont {{Ekers}},\ and\ \citenamefont
  {{Bunton}}}]{macquart2019spectral}%
  \BibitemOpen
  \bibfield  {author} {\bibinfo {author} {\bibfnamefont {J.~P.}\ \bibnamefont
  {{Macquart}}}, \bibinfo {author} {\bibfnamefont {R.~M.}\ \bibnamefont
  {{Shannon}}}, \bibinfo {author} {\bibfnamefont {K.~W.}\ \bibnamefont
  {{Bannister}}}, \bibinfo {author} {\bibfnamefont {C.~W.}\ \bibnamefont
  {{James}}}, \bibinfo {author} {\bibfnamefont {R.~D.}\ \bibnamefont
  {{Ekers}}},\ and\ \bibinfo {author} {\bibfnamefont {J.~D.}\ \bibnamefont
  {{Bunton}}},\ }\href {https://doi.org/10.3847/2041-8213/ab03d6} {\bibfield
  {journal} {\bibinfo  {journal} {\apjl}\ }\textbf {\bibinfo {volume} {872}},\
  \bibinfo {eid} {L19} (\bibinfo {year} {2019})},\ \Eprint
  {https://arxiv.org/abs/1810.04353} {arXiv:1810.04353 [astro-ph.HE]}
  \BibitemShut {NoStop}%
\bibitem [{\citenamefont {{Schoen}}\ \emph {et~al.}(2021)\citenamefont
  {{Schoen}}, \citenamefont {{Leung}}, \citenamefont {{Masui}}, \citenamefont
  {{Michilli}}, \citenamefont {{Chawla}}, \citenamefont {{Pearlman}},
  \citenamefont {{Shin}}, \citenamefont {{Stock}},\ and\ \citenamefont
  {{CHIME/FRB Collaboration}}}]{schoen2020scintillation}%
  \BibitemOpen
  \bibfield  {author} {\bibinfo {author} {\bibfnamefont {E.}~\bibnamefont
  {{Schoen}}}, \bibinfo {author} {\bibfnamefont {C.}~\bibnamefont {{Leung}}},
  \bibinfo {author} {\bibfnamefont {K.}~\bibnamefont {{Masui}}}, \bibinfo
  {author} {\bibfnamefont {D.}~\bibnamefont {{Michilli}}}, \bibinfo {author}
  {\bibfnamefont {P.}~\bibnamefont {{Chawla}}}, \bibinfo {author}
  {\bibfnamefont {A.~B.}\ \bibnamefont {{Pearlman}}}, \bibinfo {author}
  {\bibfnamefont {K.}~\bibnamefont {{Shin}}}, \bibinfo {author} {\bibfnamefont
  {A.}~\bibnamefont {{Stock}}},\ and\ \bibinfo {author} {\bibnamefont
  {{CHIME/FRB Collaboration}}},\ }\href
  {https://doi.org/10.3847/2515-5172/ac3af9} {\bibfield  {journal} {\bibinfo
  {journal} {Research Notes of the American Astronomical Society}\ }\textbf
  {\bibinfo {volume} {5}},\ \bibinfo {eid} {271} (\bibinfo {year} {2021})},\
  \Eprint {https://arxiv.org/abs/2111.08753} {arXiv:2111.08753 [astro-ph.HE]}
  \BibitemShut {NoStop}%
\bibitem [{\citenamefont {{The CHIME/FRB Collaboration}}\ \emph
  {et~al.}(2021)\citenamefont {{The CHIME/FRB Collaboration}}, \citenamefont
  {{:}}, \citenamefont {{Amiri}}, \citenamefont {{Andersen}}, \citenamefont
  {{Bandura}}, \citenamefont {{Berger}}, \citenamefont {{Bhardwaj}},
  \citenamefont {{Boyce}}, \citenamefont {{Boyle}}, \citenamefont {{Brar}},
  \citenamefont {{Breitman}}, \citenamefont {{Cassanelli}}, \citenamefont
  {{Chawla}}, \citenamefont {{Chen}}, \citenamefont {{Cliche}}, \citenamefont
  {{Cook}}, \citenamefont {{Cubranic}}, \citenamefont {{Curtin}}, \citenamefont
  {{Deng}}, \citenamefont {{Dobbs}}, \citenamefont {{Fengqiu}}, \citenamefont
  {{Dong}}, \citenamefont {{Eadie}}, \citenamefont {{Fandino}}, \citenamefont
  {{Fonseca}}, \citenamefont {{Gaensler}}, \citenamefont {{Giri}},
  \citenamefont {{Good}}, \citenamefont {{Halpern}}, \citenamefont {{Hill}},
  \citenamefont {{Hinshaw}}, \citenamefont {{Josephy}}, \citenamefont
  {{Kaczmarek}}, \citenamefont {{Kader}}, \citenamefont {{Kania}},
  \citenamefont {{Kaspi}}, \citenamefont {{Landecker}}, \citenamefont {{Lang}},
  \citenamefont {{Leung}}, \citenamefont {{Li}}, \citenamefont {{Lin}},
  \citenamefont {{Masui}}, \citenamefont {{Mckinven}}, \citenamefont
  {{Mena-Parra}}, \citenamefont {{Merryfield}}, \citenamefont {{Meyers}},
  \citenamefont {{Michilli}}, \citenamefont {{Milutinovic}}, \citenamefont
  {{Mirhosseini}}, \citenamefont {{M{\"u}nchmeyer}}, \citenamefont {{Naidu}},
  \citenamefont {{Newburgh}}, \citenamefont {{Ng}}, \citenamefont {{Patel}},
  \citenamefont {{Pen}}, \citenamefont {{Petroff}}, \citenamefont
  {{Pinsonneault-Marotte}}, \citenamefont {{Pleunis}}, \citenamefont
  {{Rafiei-Ravandi}}, \citenamefont {{Rahman}}, \citenamefont {{Ransom}},
  \citenamefont {{Renard}}, \citenamefont {{Sanghavi}}, \citenamefont
  {{Scholz}}, \citenamefont {{Shaw}}, \citenamefont {{Shin}}, \citenamefont
  {{Siegel}}, \citenamefont {{Sikora}}, \citenamefont {{Singh}}, \citenamefont
  {{Smith}}, \citenamefont {{Stairs}}, \citenamefont {{Tan}}, \citenamefont
  {{Tendulkar}}, \citenamefont {{Vanderlinde}}, \citenamefont {{Wang}},
  \citenamefont {{Wulf}},\ and\ \citenamefont {{Zwaniga}}}]{chimefrbcatalog1}%
  \BibitemOpen
  \bibfield  {author} {\bibinfo {author} {\bibnamefont {{The CHIME/FRB
  Collaboration}}}, \bibinfo {author} {\bibnamefont {{:}}}, \bibinfo {author}
  {\bibfnamefont {M.}~\bibnamefont {{Amiri}}}, \bibinfo {author} {\bibfnamefont
  {B.~C.}\ \bibnamefont {{Andersen}}}, \bibinfo {author} {\bibfnamefont
  {K.}~\bibnamefont {{Bandura}}}, \bibinfo {author} {\bibfnamefont
  {S.}~\bibnamefont {{Berger}}}, \bibinfo {author} {\bibfnamefont
  {M.}~\bibnamefont {{Bhardwaj}}}, \bibinfo {author} {\bibfnamefont {M.~M.}\
  \bibnamefont {{Boyce}}}, \bibinfo {author} {\bibfnamefont {P.~J.}\
  \bibnamefont {{Boyle}}}, \bibinfo {author} {\bibfnamefont {C.}~\bibnamefont
  {{Brar}}}, \bibinfo {author} {\bibfnamefont {D.}~\bibnamefont {{Breitman}}},
  \bibinfo {author} {\bibfnamefont {T.}~\bibnamefont {{Cassanelli}}}, \bibinfo
  {author} {\bibfnamefont {P.}~\bibnamefont {{Chawla}}}, \bibinfo {author}
  {\bibfnamefont {T.}~\bibnamefont {{Chen}}}, \bibinfo {author} {\bibfnamefont
  {J.~F.}\ \bibnamefont {{Cliche}}}, \bibinfo {author} {\bibfnamefont
  {A.}~\bibnamefont {{Cook}}}, \bibinfo {author} {\bibfnamefont
  {D.}~\bibnamefont {{Cubranic}}}, \bibinfo {author} {\bibfnamefont {A.~P.}\
  \bibnamefont {{Curtin}}}, \bibinfo {author} {\bibfnamefont {M.}~\bibnamefont
  {{Deng}}}, \bibinfo {author} {\bibfnamefont {M.}~\bibnamefont {{Dobbs}}},
  \bibinfo {author} {\bibnamefont {{Fengqiu}}}, \bibinfo {author} {\bibnamefont
  {{Dong}}}, \bibinfo {author} {\bibfnamefont {G.}~\bibnamefont {{Eadie}}},
  \bibinfo {author} {\bibfnamefont {M.}~\bibnamefont {{Fandino}}}, \bibinfo
  {author} {\bibfnamefont {E.}~\bibnamefont {{Fonseca}}}, \bibinfo {author}
  {\bibfnamefont {B.~M.}\ \bibnamefont {{Gaensler}}}, \bibinfo {author}
  {\bibfnamefont {U.}~\bibnamefont {{Giri}}}, \bibinfo {author} {\bibfnamefont
  {D.~C.}\ \bibnamefont {{Good}}}, \bibinfo {author} {\bibfnamefont
  {M.}~\bibnamefont {{Halpern}}}, \bibinfo {author} {\bibfnamefont {A.~S.}\
  \bibnamefont {{Hill}}}, \bibinfo {author} {\bibfnamefont {G.}~\bibnamefont
  {{Hinshaw}}}, \bibinfo {author} {\bibfnamefont {A.}~\bibnamefont
  {{Josephy}}}, \bibinfo {author} {\bibfnamefont {J.~F.}\ \bibnamefont
  {{Kaczmarek}}}, \bibinfo {author} {\bibfnamefont {Z.}~\bibnamefont
  {{Kader}}}, \bibinfo {author} {\bibfnamefont {J.~W.}\ \bibnamefont
  {{Kania}}}, \bibinfo {author} {\bibfnamefont {V.~M.}\ \bibnamefont
  {{Kaspi}}}, \bibinfo {author} {\bibfnamefont {T.~L.}\ \bibnamefont
  {{Landecker}}}, \bibinfo {author} {\bibfnamefont {D.}~\bibnamefont {{Lang}}},
  \bibinfo {author} {\bibfnamefont {C.}~\bibnamefont {{Leung}}}, \bibinfo
  {author} {\bibfnamefont {D.}~\bibnamefont {{Li}}}, \bibinfo {author}
  {\bibfnamefont {H.-H.}\ \bibnamefont {{Lin}}}, \bibinfo {author}
  {\bibfnamefont {K.~W.}\ \bibnamefont {{Masui}}}, \bibinfo {author}
  {\bibfnamefont {R.}~\bibnamefont {{Mckinven}}}, \bibinfo {author}
  {\bibfnamefont {J.}~\bibnamefont {{Mena-Parra}}}, \bibinfo {author}
  {\bibfnamefont {M.}~\bibnamefont {{Merryfield}}}, \bibinfo {author}
  {\bibfnamefont {B.~W.}\ \bibnamefont {{Meyers}}}, \bibinfo {author}
  {\bibfnamefont {D.}~\bibnamefont {{Michilli}}}, \bibinfo {author}
  {\bibfnamefont {N.}~\bibnamefont {{Milutinovic}}}, \bibinfo {author}
  {\bibfnamefont {A.}~\bibnamefont {{Mirhosseini}}}, \bibinfo {author}
  {\bibfnamefont {M.}~\bibnamefont {{M{\"u}nchmeyer}}}, \bibinfo {author}
  {\bibfnamefont {A.}~\bibnamefont {{Naidu}}}, \bibinfo {author} {\bibfnamefont
  {L.}~\bibnamefont {{Newburgh}}}, \bibinfo {author} {\bibfnamefont
  {C.}~\bibnamefont {{Ng}}}, \bibinfo {author} {\bibfnamefont {C.}~\bibnamefont
  {{Patel}}}, \bibinfo {author} {\bibfnamefont {U.-L.}\ \bibnamefont {{Pen}}},
  \bibinfo {author} {\bibfnamefont {E.}~\bibnamefont {{Petroff}}}, \bibinfo
  {author} {\bibfnamefont {T.}~\bibnamefont {{Pinsonneault-Marotte}}}, \bibinfo
  {author} {\bibfnamefont {Z.}~\bibnamefont {{Pleunis}}}, \bibinfo {author}
  {\bibfnamefont {M.}~\bibnamefont {{Rafiei-Ravandi}}}, \bibinfo {author}
  {\bibfnamefont {M.}~\bibnamefont {{Rahman}}}, \bibinfo {author}
  {\bibfnamefont {S.~M.}\ \bibnamefont {{Ransom}}}, \bibinfo {author}
  {\bibfnamefont {A.}~\bibnamefont {{Renard}}}, \bibinfo {author}
  {\bibfnamefont {P.}~\bibnamefont {{Sanghavi}}}, \bibinfo {author}
  {\bibfnamefont {P.}~\bibnamefont {{Scholz}}}, \bibinfo {author}
  {\bibfnamefont {J.~R.}\ \bibnamefont {{Shaw}}}, \bibinfo {author}
  {\bibfnamefont {K.}~\bibnamefont {{Shin}}}, \bibinfo {author} {\bibfnamefont
  {S.~R.}\ \bibnamefont {{Siegel}}}, \bibinfo {author} {\bibfnamefont {A.~E.}\
  \bibnamefont {{Sikora}}}, \bibinfo {author} {\bibfnamefont {S.}~\bibnamefont
  {{Singh}}}, \bibinfo {author} {\bibfnamefont {K.~M.}\ \bibnamefont
  {{Smith}}}, \bibinfo {author} {\bibfnamefont {I.}~\bibnamefont {{Stairs}}},
  \bibinfo {author} {\bibfnamefont {C.~M.}\ \bibnamefont {{Tan}}}, \bibinfo
  {author} {\bibfnamefont {S.~P.}\ \bibnamefont {{Tendulkar}}}, \bibinfo
  {author} {\bibfnamefont {K.}~\bibnamefont {{Vanderlinde}}}, \bibinfo {author}
  {\bibfnamefont {H.}~\bibnamefont {{Wang}}}, \bibinfo {author} {\bibfnamefont
  {D.}~\bibnamefont {{Wulf}}},\ and\ \bibinfo {author} {\bibfnamefont {A.~V.}\
  \bibnamefont {{Zwaniga}}},\ }\href@noop {} {\bibfield  {journal} {\bibinfo
  {journal} {arXiv e-prints}\ ,\ \bibinfo {eid} {arXiv:2106.04352}} (\bibinfo
  {year} {2021})},\ \Eprint {https://arxiv.org/abs/2106.04352}
  {arXiv:2106.04352 [astro-ph.HE]} \BibitemShut {NoStop}%
\bibitem [{\citenamefont {{Leung}}\ \emph {et~al.}(2021)\citenamefont
  {{Leung}}, \citenamefont {{Mena-Parra}}, \citenamefont {{Masui}},
  \citenamefont {{Bandura}}, \citenamefont {{Bhardwaj}}, \citenamefont
  {{Boyle}}, \citenamefont {{Brar}}, \citenamefont {{Bruneault}}, \citenamefont
  {{Cassanelli}}, \citenamefont {{Cubranic}}, \citenamefont {{Kaczmarek}},
  \citenamefont {{Kaspi}}, \citenamefont {{Landecker}}, \citenamefont
  {{Michilli}}, \citenamefont {{Milutinovic}}, \citenamefont {{Patel}},
  \citenamefont {{Pleunis}}, \citenamefont {{Rahman}}, \citenamefont
  {{Renard}}, \citenamefont {{Sanghavi}}, \citenamefont {{Stairs}},
  \citenamefont {{Scholz}}, \citenamefont {{Vanderlinde}},\ and\ \citenamefont
  {{Chime/Frb Collaboration}}}]{leung2021synoptic}%
  \BibitemOpen
  \bibfield  {author} {\bibinfo {author} {\bibfnamefont {C.}~\bibnamefont
  {{Leung}}}, \bibinfo {author} {\bibfnamefont {J.}~\bibnamefont
  {{Mena-Parra}}}, \bibinfo {author} {\bibfnamefont {K.}~\bibnamefont
  {{Masui}}}, \bibinfo {author} {\bibfnamefont {K.}~\bibnamefont {{Bandura}}},
  \bibinfo {author} {\bibfnamefont {M.}~\bibnamefont {{Bhardwaj}}}, \bibinfo
  {author} {\bibfnamefont {P.~J.}\ \bibnamefont {{Boyle}}}, \bibinfo {author}
  {\bibfnamefont {C.}~\bibnamefont {{Brar}}}, \bibinfo {author} {\bibfnamefont
  {M.}~\bibnamefont {{Bruneault}}}, \bibinfo {author} {\bibfnamefont
  {T.}~\bibnamefont {{Cassanelli}}}, \bibinfo {author} {\bibfnamefont
  {D.}~\bibnamefont {{Cubranic}}}, \bibinfo {author} {\bibfnamefont {J.~F.}\
  \bibnamefont {{Kaczmarek}}}, \bibinfo {author} {\bibfnamefont
  {V.}~\bibnamefont {{Kaspi}}}, \bibinfo {author} {\bibfnamefont
  {T.}~\bibnamefont {{Landecker}}}, \bibinfo {author} {\bibfnamefont
  {D.}~\bibnamefont {{Michilli}}}, \bibinfo {author} {\bibfnamefont
  {N.}~\bibnamefont {{Milutinovic}}}, \bibinfo {author} {\bibfnamefont
  {C.}~\bibnamefont {{Patel}}}, \bibinfo {author} {\bibfnamefont
  {Z.}~\bibnamefont {{Pleunis}}}, \bibinfo {author} {\bibfnamefont
  {M.}~\bibnamefont {{Rahman}}}, \bibinfo {author} {\bibfnamefont
  {A.}~\bibnamefont {{Renard}}}, \bibinfo {author} {\bibfnamefont
  {P.}~\bibnamefont {{Sanghavi}}}, \bibinfo {author} {\bibfnamefont {I.~H.}\
  \bibnamefont {{Stairs}}}, \bibinfo {author} {\bibfnamefont {P.}~\bibnamefont
  {{Scholz}}}, \bibinfo {author} {\bibfnamefont {K.}~\bibnamefont
  {{Vanderlinde}}},\ and\ \bibinfo {author} {\bibnamefont {{Chime/Frb
  Collaboration}}},\ }\href {https://doi.org/10.3847/1538-3881/abd174}
  {\bibfield  {journal} {\bibinfo  {journal} {\aj}\ }\textbf {\bibinfo {volume}
  {161}},\ \bibinfo {eid} {81} (\bibinfo {year} {2021})},\ \Eprint
  {https://arxiv.org/abs/2008.11738} {arXiv:2008.11738 [astro-ph.IM]}
  \BibitemShut {NoStop}%
\bibitem [{\citenamefont {{Michilli}}\ \emph {et~al.}(2020)\citenamefont
  {{Michilli}}, \citenamefont {{Masui}}, \citenamefont {{Mckinven}},
  \citenamefont {{Cubranic}}, \citenamefont {{Bruneault}}, \citenamefont
  {{Brar}}, \citenamefont {{Patel}}, \citenamefont {{Boyle}}, \citenamefont
  {{Stairs}}, \citenamefont {{Renard}}, \citenamefont {{Bandura}},
  \citenamefont {{Berger}}, \citenamefont {{Breitman}}, \citenamefont
  {{Cassanelli}}, \citenamefont {{Dobbs}}, \citenamefont {{Kaspi}},
  \citenamefont {{Leung}}, \citenamefont {{Mena-Parra}}, \citenamefont
  {{Pleunis}}, \citenamefont {{Russell}}, \citenamefont {{Scholz}},
  \citenamefont {{Siegel}}, \citenamefont {{Tendulkar}},\ and\ \citenamefont
  {{Vand erlinde}}}]{michilli2020analysis}%
  \BibitemOpen
  \bibfield  {author} {\bibinfo {author} {\bibfnamefont {D.}~\bibnamefont
  {{Michilli}}}, \bibinfo {author} {\bibfnamefont {K.~W.}\ \bibnamefont
  {{Masui}}}, \bibinfo {author} {\bibfnamefont {R.}~\bibnamefont {{Mckinven}}},
  \bibinfo {author} {\bibfnamefont {D.}~\bibnamefont {{Cubranic}}}, \bibinfo
  {author} {\bibfnamefont {M.}~\bibnamefont {{Bruneault}}}, \bibinfo {author}
  {\bibfnamefont {C.}~\bibnamefont {{Brar}}}, \bibinfo {author} {\bibfnamefont
  {C.}~\bibnamefont {{Patel}}}, \bibinfo {author} {\bibfnamefont {P.~J.}\
  \bibnamefont {{Boyle}}}, \bibinfo {author} {\bibfnamefont {I.~H.}\
  \bibnamefont {{Stairs}}}, \bibinfo {author} {\bibfnamefont {A.}~\bibnamefont
  {{Renard}}}, \bibinfo {author} {\bibfnamefont {K.}~\bibnamefont {{Bandura}}},
  \bibinfo {author} {\bibfnamefont {S.}~\bibnamefont {{Berger}}}, \bibinfo
  {author} {\bibfnamefont {D.}~\bibnamefont {{Breitman}}}, \bibinfo {author}
  {\bibfnamefont {T.}~\bibnamefont {{Cassanelli}}}, \bibinfo {author}
  {\bibfnamefont {M.}~\bibnamefont {{Dobbs}}}, \bibinfo {author} {\bibfnamefont
  {V.~M.}\ \bibnamefont {{Kaspi}}}, \bibinfo {author} {\bibfnamefont
  {C.}~\bibnamefont {{Leung}}}, \bibinfo {author} {\bibfnamefont
  {J.}~\bibnamefont {{Mena-Parra}}}, \bibinfo {author} {\bibfnamefont
  {Z.}~\bibnamefont {{Pleunis}}}, \bibinfo {author} {\bibfnamefont
  {L.}~\bibnamefont {{Russell}}}, \bibinfo {author} {\bibfnamefont
  {P.}~\bibnamefont {{Scholz}}}, \bibinfo {author} {\bibfnamefont {S.~R.}\
  \bibnamefont {{Siegel}}}, \bibinfo {author} {\bibfnamefont {S.~P.}\
  \bibnamefont {{Tendulkar}}},\ and\ \bibinfo {author} {\bibfnamefont
  {K.}~\bibnamefont {{Vand erlinde}}},\ }\href@noop {} {\bibfield  {journal}
  {\bibinfo  {journal} {arXiv e-prints}\ ,\ \bibinfo {eid} {arXiv:2010.06748}}
  (\bibinfo {year} {2020})},\ \Eprint {https://arxiv.org/abs/2010.06748}
  {arXiv:2010.06748 [astro-ph.HE]} \BibitemShut {NoStop}%
\bibitem [{\citenamefont {{Kader}}(2022)}]{kadermscthesis}%
  \BibitemOpen
  \bibfield  {author} {\bibinfo {author} {\bibfnamefont {Z.}~\bibnamefont
  {{Kader}}},\ }\emph {\bibinfo {title} {A High Time Resolution Search for
  Gravitationally Lensed Fast Radio Bursts using the CHIME telescope}},\
  \href@noop {} {Master's thesis},\ \bibinfo  {school} {McGill University}
  (\bibinfo {year} {2022})\BibitemShut {NoStop}%
\bibitem [{\citenamefont {{Cordes}}(2004)}]{cordes2001new}%
  \BibitemOpen
  \bibfield  {author} {\bibinfo {author} {\bibfnamefont {J.~M.}\ \bibnamefont
  {{Cordes}}},\ }in\ \href@noop {} {\emph {\bibinfo {booktitle} {Milky Way
  Surveys: The Structure and Evolution of our Galaxy}}},\ \bibinfo {series}
  {Astronomical Society of the Pacific Conference Series}, Vol.\ \bibinfo
  {volume} {317},\ \bibinfo {editor} {edited by\ \bibinfo {editor}
  {\bibfnamefont {D.}~\bibnamefont {{Clemens}}}, \bibinfo {editor}
  {\bibfnamefont {R.}~\bibnamefont {{Shah}}},\ and\ \bibinfo {editor}
  {\bibfnamefont {T.}~\bibnamefont {{Brainerd}}}}\ (\bibinfo {year} {2004})\
  p.\ \bibinfo {pages} {211}\BibitemShut {NoStop}%
\end{thebibliography}%

\end{document}